\documentclass[11pt,a4paper]{article}

\usepackage{ILD}

\usepackage[symbol]{footmisc}
\usepackage{feynmf}
\usepackage{textpos}
\usepackage{lineno}
\usepackage{tabularx}
\usepackage{footnote}
\usepackage{amsmath}
\usepackage{amssymb}

\usepackage{graphicx}
\usepackage{xcolor}
\usepackage{multirow}

\def\be{\begin{equation}}
\def\ee{\end{equation}}
\def\bea{\begin{eqnarray}}
\def\eea{\end{eqnarray}}

\newcommand{\com}{centre-of-mass }
\newcommand{\Process}{e^+e^-\rightarrow\chi\chi\gamma}

\title{WIMP Dark Matter at the International Linear Collider}

\date{\today}

\addauthor{Moritz Habermehl}{\institute{1}}
\addauthor{Mikael Berggren}{\institute{1}}
\addauthor{Jenny List}{\institute{1}}

\addinstitute{1}{Deutsches Elektronen-Synchrotron DESY}

\abstract{

In this study, the sensitivity of future lepton colliders to WIMP dark matter is evaluated 
assuming WIMP pair production 
accompanied by a photon from initial state radiation, through which the process can be identified. 
A full detector simulation for the International Large Detector (ILD) concept at the International Linear Collider (ILC) is performed for a \com energy of 500\,GeV. Energy scales of up to $3$\,TeV can be tested for different effective operators for WIMP masses almost up to half the \com energy.
The sensitivity benefits from the polarised beams, which can reduce the main SM background from neutrino pair production substantially. In addition, systematic uncertainties are shown to be significantly reduced by combining data with several different polarisation configurations.
In comparison to a previous study, the reconstruction of the forward detectors has been improved and the systematic uncertainties are fully treated.
The results are also extrapolated to other \com energies, luminosities and beam polarisations. This allows to provide results for the full ILC programme, i.e.\ from 250\,GeV to 1\,TeV, as well as to give approximate results for other planned lepton colliders.}

\titlecomment{This work was carried out in the framework of the ILD detector
concept group}

\graphicspath{ {./logos/}{./figures/}{.} }

\begin{document}

\titlepage

\section{Introduction}
\label{sec:introduction}
\begin{figure}[htb]
  \begin{center}
    \includegraphics[width=0.4\textwidth]{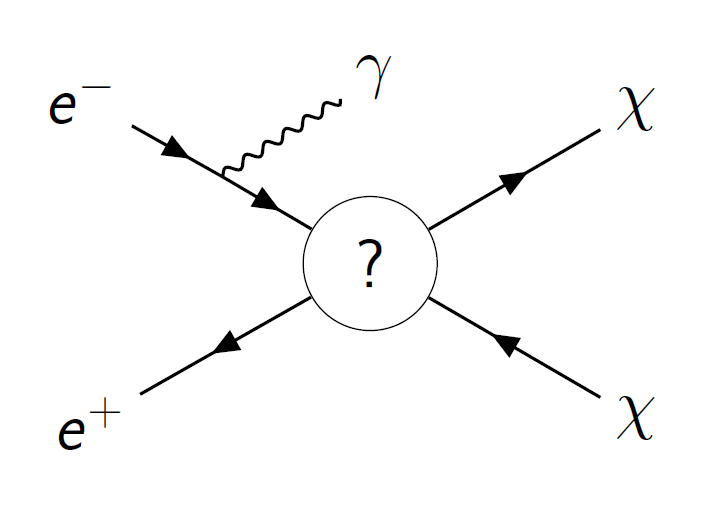}
    \caption{Visualisation of the mono-photon process $\Process$ as a pseudo-Feynman diagram.}
    \label{fig:feynman_sig}
   \end{center}  
\end{figure}

The nature of dark matter, which today is known to account for about $27\%$ of the total energy density in the universe~\cite{Aghanim:2018eyx} and thus contributes more than 5 times as much as ordinary matter, is one of the most important questions the Standard Model of particle physics (SM) fails to answer in a satisfactory way. 
Weakly Interacting Massive Particles (WIMPs) are among the primary candidates for dark matter and are being searched for via many different experimental approaches. These comprise searches for astrophysical WIMPs through direct and indirect detection as well as collider searches which offer the possibility of producing WIMPs -- either directly or in the decay of other, more heavy exotic particles. Furthermore, lepton colliders offer unique capabilities to probe WIMPs via energy scans (to determine their mass) and beam polarisations (to determine their couplings).
At collider searches, the tested scenarios range from simplified signatures to complete models~\cite{kahlhoefer2017review}. For complete models the full particle content, their mass spectra, and interactions among themselves and with SM particles are available for experimental tests.
In a simplified setup, 
the generic signature to search for is WIMP pair production via an effective coupling between WIMPs and SM particles. The effective coupling can be modeled using an intermediate mediator particle of a certain mass.
While dark matter particles do not interact with the detector material and escape detection, 
visible particles recoiling against WIMPs can be used to identify this signature.
This process is being searched for at the LHC considering all kinds of visible particles~\cite{Aaboud:2019yqu,Sirunyan:2017jix, Sirunyan:2018dsf}. In case of lepton colliders, WIMP searches are performed using a photon from initial state radiation (ISR), as illustrated in Fig.~\ref{fig:feynman_sig}.  To date the only lepton collider bounds on WIMPs via the mono-photon signature have been derived from LEP results~\cite{Fox:2011fx,Abdallah:2003np,Abdallah:2008aa}.

While direct detection experiments and WIMP production at hadron colliders always require non-vanishing coupling to quarks, searches at lepton colliders depend only on the couplings to leptons, specifically the electron. Therefore, results from lepton colliders cannot
be compared to those from hadron collider or direct detection without making model assumptions on the relative strengths of the coupling to leptons and quarks.

A connection between the couplings to leptons and to quarks can be made by requiring that the relic abundance of a thermally produced WIMP should not be higher than the relic density of dark matter observed today, which constrains the couplings of the WIMP and the mediator to the various SM particles from below. Based on this constraint, the interplay between direct and indirect detection as well as collider searches has been studied in a global likelihood analysis based on an effective field theory (EFT) ansatz for the example of a singlet-like Majorana fermion WIMP~\cite{Matsumoto:2016hbs}. This example shows that there are significant regions in parameter space not probed by the LHC that can be covered by future high-energy lepton colliders. 
The coverage by a future $e^+e^-$ collider is based on projections obtained in a rather old study~\cite{CBDA1}.

In this paper, we re-analyse the mono-photon signature using a full simulation of the ILD detector concept for the ILC at $\sqrt{s}=500$\,GeV, with a detailed treatment of the machine environment, including the luminosity spectrum and high cross-section $\gamma\gamma$-induced processes, and a careful consideration of systematic uncertainties. 
The results of this paper supersede those of~\cite{CBDA1}.
The presented study is based on~\cite{thesis} which can be consulted for additional material. Sec.~\ref{sec:setup} is dedicated to the ILC and the simulation of the accelerator environment as well as the detector simulation and the event reconstruction.

The sensitivity to a WIMP signal will be calculated using the photon energy distribution, which looks different for signal and background.
The dominant irreducible SM background to the signature shown in Fig.~\ref{fig:feynman_sig} is neutrino pair production with an associated photon (radiative neutrino pair production).  
Since neutrino events are indistinguishable from WIMP events on an event-by-event basis, the event selection is designed so that the majority of these events survive. 
Because of the high polarisation dependence of the most dominant background, different polarisation data sets offer different discovery prospects for dark matter.
Beyond the irreducible neutrino background, any process with a photon in the final state can contribute to the total background provided that all other particles escape detection. SM processes which contain either jets or charged particles are comparably easy to distinguish from a WIMP event, leading to only negligible contribution to the total background~\cite{CBDA1}. This, however, does not apply to radiative Bhabha scattering, i.e.\ electron-positron pair production with an associated photon from initial or final state radiation, which has a huge cross-section and can mimic the signal if both leptons escape undetected, for example through the beam pipes. The modeling of these two most important background
processes (neutrino pair production and Bhabha scattering) has been improved considerably with respect to earlier studies and will be discussed in detail in Sec.~\ref{sec:analysis}, followed by a description of the event selection and of the relevant systematic uncertainties.

In order to cover a large range of potential signatures, the general approach of effective operators~\cite{perelstein,dreiner2013illuminating,effoplep,beltran2010maverick,bai2010tevatron,goodman2010constraints,zheng2012constraining,yu2012constraining} is chosen for the presentation of the results in Sec.~\ref{sec:results500}. In the EFT framework, the sensitivity depends on the type of operator describing the WIMP production, the mass and spin of the WIMP, and on the parameter $\Lambda$, which defines the energy scale at which the new physics becomes important. In this study, three different operators with vector, axial-vector and scalar tensor structure as presented in Table~\ref{tab:operators} are used. The energy scale $\Lambda$ is related to the cross-section as $\sigma\propto 1 / \Lambda^4$.
At lepton colliders, the probed energy scales are typically much higher than
the centre-of-mass energy, so that the validity of the EFT is ensured.

The polarised double-differential cross-section formulas for the WIMP pair production with one ISR photon $\frac{d^2\sigma}{dE_\gamma d\cos{\theta_\gamma}}$ are taken from~\cite{perelstein}. For the considered operators, WIMP pair production is only possible for either opposite helicity of the two colliding particles (vector operator) or same sign helicity (axial-vector and scalar operators).

\begin{table}[h!]
 \begin{center}
  \begin{tabular}[h]{ l c c c }
\hline
   & four-fermion operator & $\sigma(e^-_L,e^+_R)=\sigma(e^-_R,e^+_L)$ & $\sigma(e^-_L,e^+_L)=\sigma(e^-_R,e^+_R) $ \\ \hline
  vector & $(\overline{f}\gamma^\mu f)(\overline{\chi}\gamma_\mu\chi)$ & $\sigma\propto 1 / \Lambda^4$ & 0 \\
  axial-vector & $(\overline{f}\gamma^\mu\gamma^5f)(\overline{\chi}\gamma_\mu\gamma_5\chi)$ & 0 & $\sigma\propto 1 / \Lambda^4$ \\
  scalar & $(\overline{\chi}\chi)(\overline{f}f)$ & 0 & $\sigma\propto 1 / \Lambda^4$ \\
\hline
  \end{tabular}
  \caption[Effective operators used in this analysis.]{Effective operators used in this analysis and their chiral properties.}
  \label{tab:operators}
 \end{center}
\end{table}

As opposed to the lepton collider case, mono-X (where X stands for any SM particle) searches at the LHC cannot 
probe new physics scales above the centre-of-mass energy, and thus EFT is not 
a suitable approach for their interpretation~\cite{busoni2014validity,malik2015interplay}.
Instead, simplified models have to be used~\cite{alwall2009simplified}, in which $\Lambda$ relates to the mediator mass and its coupling to the SM fermions $g_{SM}$ and its coupling to the WIMPs $g_\chi$ in the following way: $ \Lambda = M_{\textrm{mediator}} / \sqrt{g_{\textrm{SM}}g_\chi}$. 
So, results from lepton and hadron colliders cannot be compared directly, but the sensitivity presented in terms of the single parameter $\Lambda$ has to be expressed using at least three free parameters ($M_{\textrm{mediator}}$, $g_{\textrm{SM}}$ and $g_\chi$).
Furthermore, lepton colliders probe the coupling of the mediator to leptons
$g_{\textrm{SM}}^{\textrm{ee}}$, while hadron colliders probe its coupling 
$g_{\textrm{SM}}^{\textrm{qq}}$ to quarks. So, 
results from lepton and hadron colliders cannot be compared without assuming 
a specific model, which specifies not only $M_{\textrm{mediator}}$ and $g_\chi$, but also 
$g_{\textrm{SM}}^{\textrm{ee}}$ and $g_{\textrm{SM}}^{\textrm{qq}}$.

In the final step, we extrapolate the obtained results to other centre-of-mass energies, integrated luminosities and beam polarisations, as described in Sec.~\ref{sec:otherCOM}. Though still based on the ILD simulation, these extrapolations can be regarded as good approximations for the capabilities of other future $e^+e^-$ colliders than the ILC. 
We summarise the conclusions in Sec.~\ref{sec:conclusion}.

\section{Experimental considerations} 
\label{sec:setup}
In this section the experimental environment is discussed, with a focus on those aspects of the ILC and the ILD detector which are of particular 
relevance for the WIMP search.

\subsection{The International Linear Collider} \label{sec:ilc}

The ILC is a planned electron-positron collider based on a superconducting acceleration technology~\cite{TDRvol1,TDRvol2,TDRvol3I,TDRvol3II,TDRvol4}. It is a lepton collider at the energy frontier with centre-of-mass energies in the range of 250-500 GeV and 1 TeV after an upgrade, and unprecedented luminosities. Both beams are foreseen to be polarised, the electrons with 80\% and the positrons with 30\% polarisation. The particles are brought to collisions with a crossing angle of 14\,mrad. 

For the requirement of high luminosity, small beam sizes are a prerequisite.
The resulting strong fields lead to the emission of beamstrahlung photons~\cite{Yokoya:1985xx}. Due to this energy loss, the beam energy spectra have a characteristic distribution. The beamstrahlung photons can produce background processes, like electron-positron pairs and hadronic events, both characterised by small transverse momenta. 
As these processes happen simultaneously and independently of the electron-positron process, they form pile-up polluting SM background events as well as the signal-like events.
Hence, in order to efficiently detect signal events, one needs to also include events having a certain detector activity in addition to the signal photon, rather than only select events with a signal photon and nothing else. See Sec.~\ref{sec:event_selection} on the event selection.
The effect of these pile-up processes are simulated by overlaying them on top of the $e^+e^-$ process and hence this type of background will be referred to as \textit{overlay}.

\begin{figure}[htb]
 \begin{center}
  \includegraphics[width=0.5\textwidth]{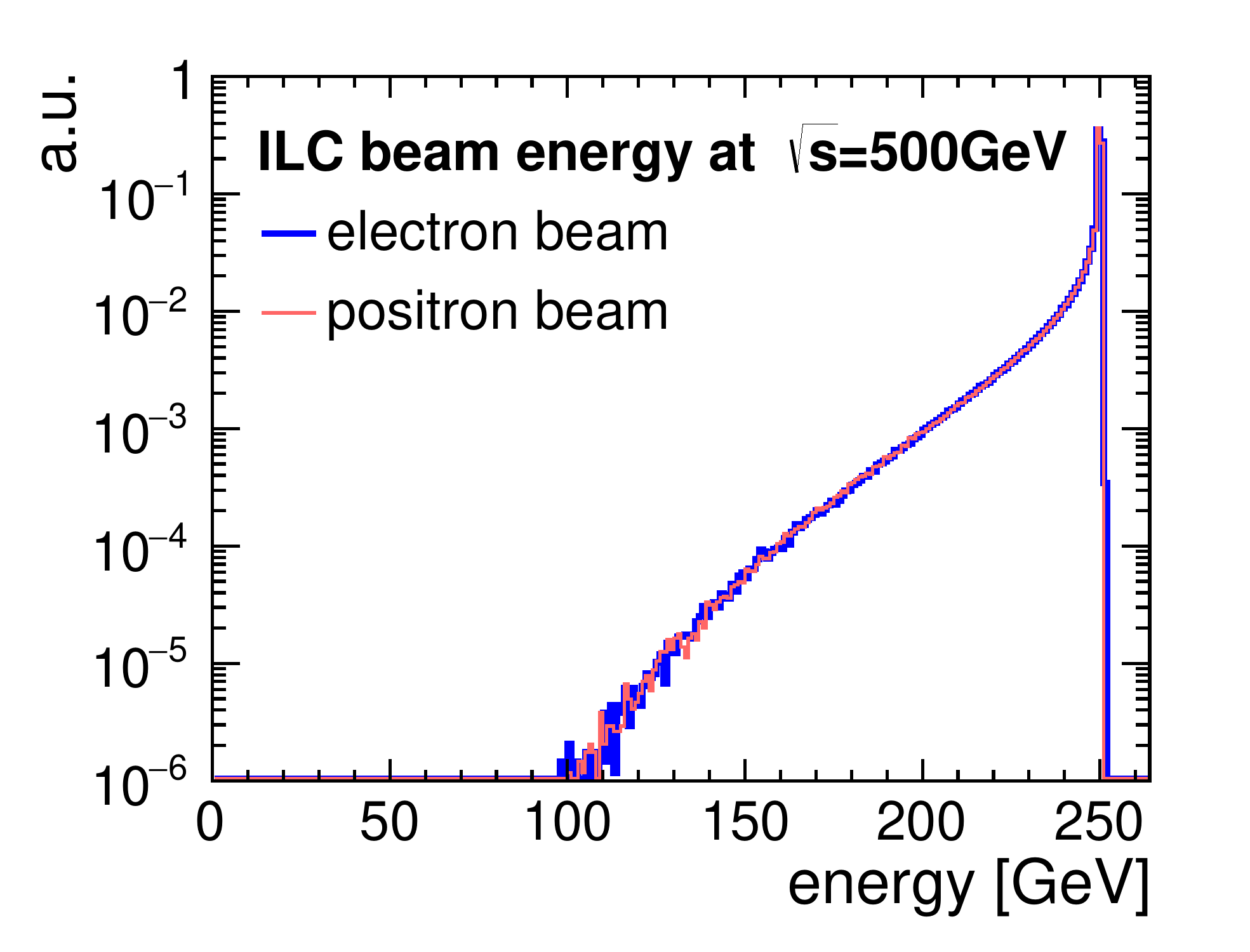} 
 \end{center}
 \caption{Beam energy spectra of the electron (blue, thick) and positron (red, thin) beams at the interaction point for baseline parameters and a \com energy of 500\,GeV, generated with \textsc{GuineaPig}~\cite{schulte}.}
 \label{fig:bs_500}
 \end{figure}

\subsubsection{The simulation of the accelerator environment} \label{sec:overlay}

This analysis is based on a Monte Carlo simulation with a realistic description of both the detector and accelerator environment. 
The particle beam propagation and beam-beam interactions are simulated using \textsc{GuineaPig}~\cite{schulte}, which provides the overlay events from coherent and incoherent $e^+e^-$ pair production as well as the expected beam energy spectra, which are shown in Fig.~\ref{fig:bs_500}. The latter are required for a realistic description of the centre-of-mass energy distribution, i.e.\ the luminosity spectrum, in the event generation.
In the second preparational step for the event generation, the luminosity spectrum is parametrised with the programme \textsc{Circe2}~\cite{circe2}.

Modelling the processes involving the beamstrahlung photons is crucial to provide a realistic description of the experimental environment at ILC energies. 
This is especially important for the WIMP study, because this overlay is the main detector activity besides the photon in signal events.
In the simulation, an \textit{anti-DID}~\cite{antiDID} magnetic field is included which directs the majority of pairs into the holes of outgoing beam pipes. In Fig.~\ref{fig:bcal_layer}, this increase of the energy of electron-positron pairs towards the centre of the forward calorimeter BeamCal can be seen.

\subsubsection{Operation scenario} \label{sec:H20}

The main results of this WIMP search are the sensitivities expected for the data taken assuming a programme for 20 years of operation (called \textit{H20} scenario in~\cite{Barklow:2015tja}). This scenario foresees in particular a data set of $4$\,ab$^{-1}$ to be collected at a centre-of-mass energy of $\sqrt{s}=500$\,GeV. More recently, a staged version of the ILC has been proposed which would start by collecting $2$\,ab$^{-1}$ at $\sqrt{s}=250$\,GeV before extending the linacs to reach $\sqrt{s}=500$\,GeV~\cite{Fujii:2017vwa}. The ILC has been designed for up to $\sqrt{s}=1$\,TeV. While the full detector simulation study presented in this paper has been performed for the $\sqrt{s}=500$\,GeV stage, extrapolations of the results to lower and higher energies will be provided. 

At the ILC, both beams are foreseen to be polarised. Thus, the particles can be brought to collision with four different polarisation configurations: sgn$\left(P(e^-),P(e^+)\right)=(-,-)$, $(-,+)$, $(+,-)$ and $(+,-)$, where ``$-$'' denotes left-handed and ``$+$'' denotes right-handed helicity and $|P(e^-)|=80$\% and $|P(e^+)|=30$\% are the nominal absolute values at the ILC. At $\sqrt{s}=500$\,GeV, the standard sharing between the different polarisation sign configurations according to the H20 scenario is 40\% for $(-,+)$ and $(+,-)$, each, and 10\% for each of the equal-sign configurations.

\subsection{The modeling of the International Large Detector concept}\label{sec:simulation}

The production of the Monte Carlo events for the neutrino and Bhabha scattering background processes comprises three steps: the event generation using \textsc{Whizard}, the simulation of the interaction of the generated particles with the detector material using the \textsc{Geant4}-based~\cite{geant4_1,geant4_2,geant4_3} detector simulation \textsc{Mokka}~\cite{Mokka} (version~08-00-03 in \textsc{ILCSoft} version~01-16-02), and the event reconstruction in the \textsc{Marlin} framework~\cite{Marlin} (\textsc{ILCSoft} version 01-17-11). The setup of the event generation will be discussed in more detail in Sec.~\ref{sec:samples}, as the correct and complete modeling of events with one or several photons from initial or final state radiation is a crucial ingredient to this analysis.
The detector simulation and reconstruction includes all overlay-type of backgrounds from beamstrahlung and $\gamma \gamma \to $ low-$p_t$ hadron processes.

The detector model \texttt{ILD\_o1\_v05}, an implementation of the International Large Detector (ILD) concept~\cite{TDRvol4}, is used for the simulation. Together with SiD~\cite{TDRvol4}, ILD is one of the two concepts for a multi-purpose detector at the ILC, with a TPC-based tracking system in the inner part, followed by electromagnetic and hadronic calorimeters, embedded in a magnetic field. 

In the mono-photon analysis, the tracking system is needed to discriminate photons from electrons, i.e.\ neutral from charged electromagnetic showers. The electromagnetic calorimeter (ECAL) is essential for this analysis as it measures the energy and angle of the photon. The hadronic calorimeter is required to veto hadronic events. The muon system, the outermost part of the detector, is used to reject muon events. 

An excellent hermeticity in the forward region, down to polar angles of about $6$\,mrad, is achieved by three additional calorimeters on both sides of the detector, of which BeamCal is especially important to suppress background from radiative Bhabha scattering. The impact of a larger uninstrumented region around the beam pipes, which could be required e.g.\ in cases where the final focus quadrupoles of the accelerator need to be
much closer to the experiment, will be discussed in Sec.~\ref{sec:detector_effects}.
With respect to~\cite{CBDA1}, the modeling of the background from $e^+e^-$ pairs from beamstrahlung as well as the reconstruction and identification of the clusters from isolated high-energy $e^{\pm}$ in BeamCal~\cite{BCal_arxiv} have been improved both in terms of efficiency and realism.

\subsubsection{Photon reconstruction} \label{sec:photon_reco}

The photons deposit most of their energy in ECAL, which consists of 30 readout layers interleaved with tungsten absorbers. The pixel size of $5\times5$\,mm$^2$ provides a high granularity which allows for good pattern recognition and good separation of showers. The energy resolution, demonstrated with a prototype detector under test beams, is given by $\sigma_{E}/E = 16.53\%/\sqrt{E([\textrm{GeV}])}\oplus1.07\%$~\cite{ecalres}.

The reconstruction of photon candidates has been improved with respect to~\cite{CBDA1} by employing a  dedicated photon reconstruction algorithm~\cite{BorouXu}, which is part of the Pandora Particle Flow Algorithm software package~\cite{Pandora}. The potential of the highly granular ECAL is exploited, so that nearby photons can also be separated. At the same time, the photon splitting occurs at a maximum as low as 1.02 reconstructed photons per generated photon~\cite{thesis} as opposed to 3.5 in earlier studies~\cite{CBDA1}.

\subsubsection{BeamCal reconstruction} \label{sec:beamcalreco}

\begin{figure}[htb]
 \begin{center}
 \includegraphics[width=0.5\textwidth]{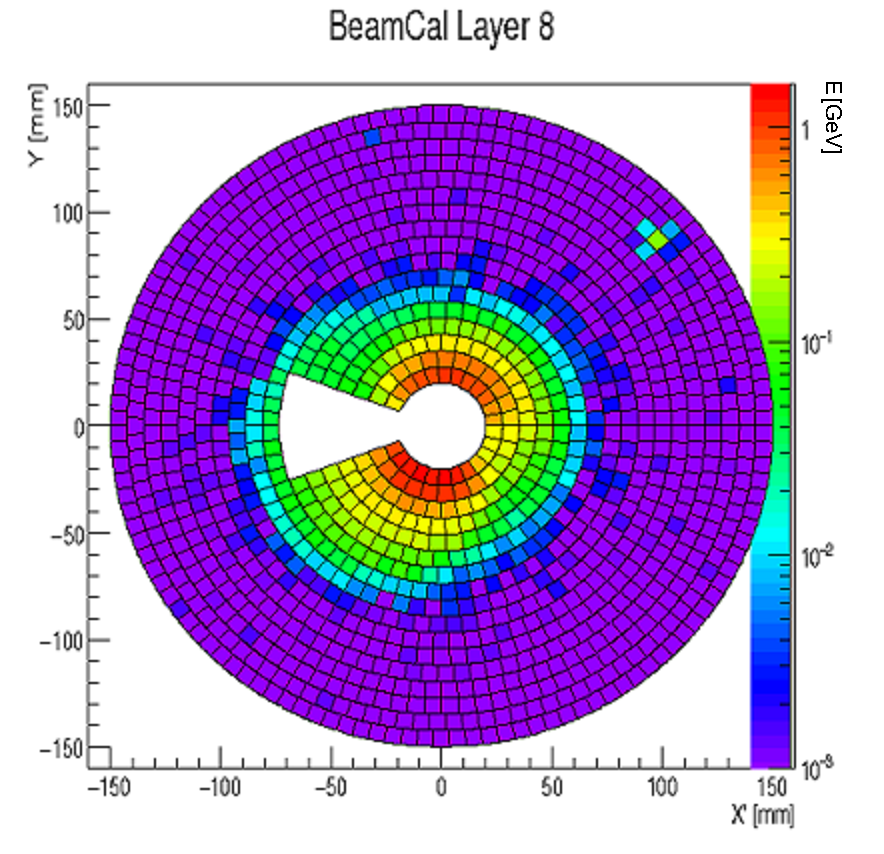}
 \end{center}
 \caption[BeamCal layer.]{Energy deposition in a layer of the forward calorimeter BeamCal integrated over one bunch crossing. The energy deposition stemming from overlay increases exponentially towards the centre.}
 \label{fig:bcal_layer}
\end{figure}

The electromagnetic calorimeters in the very forward region (BeamCal)~\cite{OldBeamCalReco} are crucial for several aspects of the WIMP search. The BeamCal is placed on each side of the detector, centred around the outgoing beam pipes covering a polar angle range from  $5.6$ to $42.9$\,mrad.\footnote{Due to a redesign of the forward region~\cite{newL*}, these values have slightly increased in more recent ILD models.} They are thus the subdetectors closest to the beam pipe with an inner opening of only $\sim$20\,mm. Each BeamCal consists of 40 layers of 0.3\,mm diamond sensors interleaved with 3.5\,mm thick tungsten absorbers. In Fig.~\ref{fig:bcal_layer}, the energy depositions in one of the BeamCal layers is shown. The round opening in the centre of BeamCal accommodates the outgoing beam pipe, while the left part of the keyhole-shaped opening hosts the incoming beam pipe. While the isolated cluster near $x=y=10$\,cm corresponds to a high-energy electron, as e.g.\ from Bhabha scattering, the large energy deposition around the beam pipes originates from the overlay of $e^+e^-$ pairs. 
With these energy depositions, the beam parameters can be determined~\cite{grah}. The obtained precision can be used to calculate the systematic uncertainties on the luminosity spectrum, as we will discuss in more detail in Sec.~\ref{sec:systematics}. On the other hand, the pair background makes the identification of particles from hard-scattering events, e.g.\ the outgoing leptons from low-angle Bhabha scattering, challenging.

For the BeamCal reconstruction, the \textsc{Marlin} processor \textsc{BeamCalClusterReco}~\cite{BCal_arxiv} is used. In order to prevent reconstructing hits which are likely to stem from overlay, any energy deposition in a pad is ignored if it is less than two standard deviations above the average expected from pair background or if the energy is below a threshold of 0.01\,GeV. The first layer is not considered in the reconstruction, because the overlay is dominated by low-energy particles which deposit their energy mainly in the first layers. As the next step, cluster candidates are formed from \textit{towers}, which comprise a minimum of six pads in consecutive layers, which passed the criteria above. If the cluster energy is higher than 36\,GeV, it is considered as a reconstructed particle. The cuts have been optimised to reconstruct electrons above 50\,GeV, while minimising the number of fakes, \emph{i.e.} clusters formed from hits from overlay. In this way, reconstructed objects are likely to be high-energetic particles from the hard interaction, e.g.\ Bhabha scattering leptons and therefore events with reconstructed BeamCal object are vetoed.

\begin{figure}[htb]
  \begin{center}
   \includegraphics[width=0.55\textwidth]{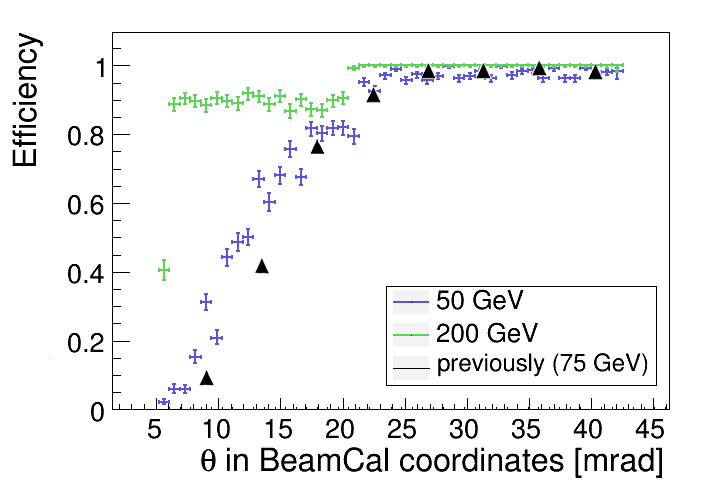}
   \end{center}
   \caption{Efficiency to reconstruct electrons in BeamCal. The green (blue) crosses are obtained with the \textsc{BeamCalClusterReco} processor~\cite{BCal_arxiv} for 200\,GeV (50\,GeV) electrons. The previous reconstruction algorithm (black triangles)~\cite{OldBeamCalReco} shows a worse performance for 75\,GeV electrons.}
\label{fig:bcaleff}
\end{figure}

In Fig.~\ref{fig:bcaleff}, the electron identification efficiency in BeamCal averaged over the azimuthal angle is shown as a function of the polar angle. For large angles, corresponding to the outer part of the calorimeter, the identification is perfect for 200\,GeV electrons and above 95\% for 30\,GeV electrons, but decreases for smaller angles. At about 20\,mrad, there is a step-like drop since the averaging over the azimuthal angle includes the keyhole-shaped uninstrumented region for the incoming beam pipe. Closer to the centre, the efficiency decreases further because of the increasing overlay (visible in Fig.~\ref{fig:bcal_layer}). The decrease occurs at  lower angles for higher electron energies. The performance of \textsc{BeamCalClusterReco} for electrons with an energy of $50$\,GeV is better than with the previous BeamCal reconstruction for electrons with an energy of $75$\,GeV (black triangles)~\cite{OldBeamCalReco}. In addition, the new approach is deemed more realistic because it also models the possibility of additional fake electron candidates.

\section{Analysis} 
\label{sec:analysis}
In this section, we describe the modeling of signal and background events, in particular with respect to the photon radiation, as well as the event selection and the evaluation of systematic uncertainties.

\subsection{The data samples} \label{sec:samples}

The Monte Carlo simulation is based on data samples for the background processes of neutrino pair production and Bhabha scattering at a \com energy of 500\,GeV. For the neutrino pair production, all three generations ($\nu_e$, $\nu_\mu$, $\nu_\tau$) are considered and in total about 16,000,000 events are produced for the two possible configurations of opposite beam polarisation. For Bhabha scattering, all four polarisation combinations are generated with a total of about 12,000,000 events. The events are generated with 100\% polarisation and are weighted to the studied polarisation configurations and the integrated luminosity of the running scenario.

The events are generated using \textsc{Whizard}~2.4.4~\cite{Kilian:2007gr}, which provides a realistic treatment of the ILC environment by taking into account the beam energy spectrum, the polarisation of the beams, and initial state radiation (ISR).

\subsubsection{Modelling of ISR photons} \label{sec:isr}

The ISR photon is the only detectable particle in the signal process and in the neutrino background. The photon distributions have to be carefully modelled because both the energy distribution rises towards soft photons and the angular distribution rises towards collinearity. In addition, more than one ISR photon can be emitted with polar angles high enough to interact with the detector.

For ISR, the standard routine within \textsc{Whizard} is used. The dedicated parametrisation comprises all orders of soft and soft collinear photons and the first three orders of hard collinear photons. With this routine, the most accurate total cross-section is obtained. This description, however, does not provide the correct distribution of the number of photons and their angular and energy distributions.
Thus, these photons should not be taken as the signal photon. To avoid a reconstruction in the detector, their polar angle is restricted to zero.\footnote{In later \textsc{Whizard} versions the ISR routine has been updated to give a better description of the recoil, however not of the photons.}

In order to obtain the correct differential distributions of detectable photons, the photons are included in the matrix elements for neutrino and Bhabha scattering processes. In this way, the number of photons can be controlled and the energy and angle distributions are modelled correctly. All photon multiplicities with a cross-section not less than four orders of magnitude lower than the leading order, i.e.\ with one ISR photon, are taken into account by separate data sets with different numbers of photons in the matrix element (1-4 photons in the case of neutrino pair production and 1-3 for Bhabha scattering).

\textsc{Whizard} allows to set cuts on the kinematics of \textit{all} particles, including the ISR photons, which is crucial to reduce computational time. The exact definitions of the cuts can be found in Sec.~5.3.2 of~\cite{thesis}. The phase space is adjusted to the requirements on the photon of the signal definition, see Sec.~\ref{sec:photon_selection}. At least one of the generated matrix element photons has to fulfill $4^\circ<\theta<176^\circ$ and $p_T>1$\,GeV.

\subsubsection{Modelling of Bhabha scattering}\label{sec:bhabha_modelling}

For the generation of Bhabha scattering events, an inclusive Bhabha
generator like \textsc{Bhwide}~\cite{bhwide} could not be used because it does
not allow to place cuts on the energy and angle of the photons at event 
generation time, which is required in order to obtain a sufficiently large 
sample of events in the signal definition in a practical way. Instead, 
\textsc{Whizard} is used which allows to apply generator-level cuts directly on the 
photons.
The target is to cover at least the full detector angular range, so that the effect of the acceptance loss can be studied. 

Compared to neutrino pair production, additional cuts have to be applied in the case of Bhabha scattering, because the cross-section diverges for certain combinations of the particles' momenta and hence these regions in parameter space have to be avoided.

One of the cases is already avoided by the signal definition. 
By requiring a minimum transverse momentum for the photon, which is balanced by the transverse momentum of the outgoing leptons, the probability that at least one lepton hits the detector is increased, which allows an identification of this background process.
For the full set of cuts in the event selection see Sec.~\ref{sec:event_selection}.



\begin{figure}[htb]
 \begin{center}
\includegraphics[width=0.5\textwidth]{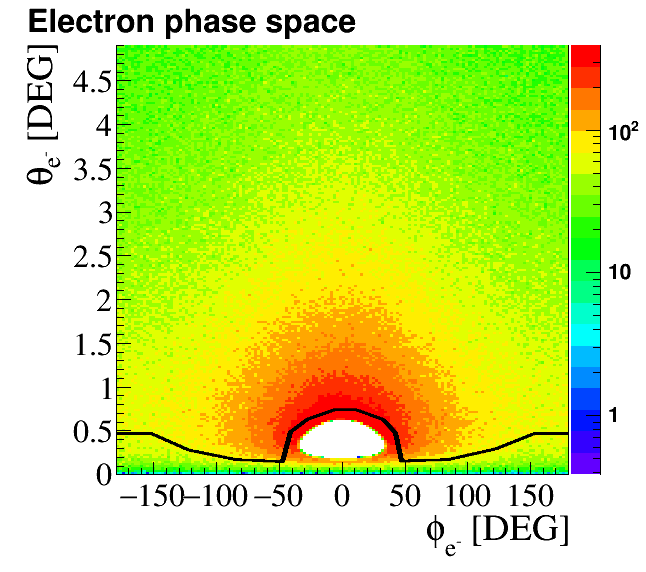}
 \end{center}
 \caption{Phase space of the generated electrons in Bhabha scattering events in the forward region of the detector.}
 \label{fig:thetaphiM4_M1}
 \end{figure}

Some of the remaining divergencies have to be avoided by additional cuts at the generator level. Either the invariant mass or the four-momentum transfer of the problematic pairs of particles is constrained (see~\cite{thesis} Sec.~5.3.2.2 for details). 
As a consequence, not the full phase space is generated, i.e.\ there are no leptons in the very forward region.
Even though the distribution of one of the leptons is shifted away from zero (because of the transverse momentum cut in the signal definition), the tails of the distribution should also extend to lower angles and hence the cross-section might be underestimated. In Fig.~\ref{fig:thetaphiM4_M1}, the empty region around $\phi=0^\circ$ and $\theta<0.5^\circ$ visualises the missing phase space.
Without this unavoidable cut the cross-section might be higher. The moderate effect of a different level of the Bhabha scattering background is studied in Sec.~\ref{sec:detector_effects}.

Nevertheless, the angular range lies outside of the coverage of ILD which extends to $\theta\approx 0.5^\circ$. The $\phi$-dependent shape of the inner rim of the most forward detector BeamCal is indicated by the black line in Fig.~\ref{fig:thetaphiM4_M1}. As the complete instrumented region is modelled, the detector activity, and hence also the event selection, are described in a realistic way.

A part of the missing phase space is modelled using an angular cut instead of the four momentum transfer. The resulting complementary data set with opposite cuts than the standard set is added to the background. An illustration of the impact of these samples will be given later in Fig.~\ref{fig:Egamma} in Sec.~\ref{sec:event_selection} when the event selection is discussed. As this data set is available only at the generator level, the event selection is simplified (see~\cite{thesis} Sec.~6.1.6.3 for details).


\subsection{Strategy to obtain the signal events}
	   \label{sec:sigrew}

As the mono-photon analysis is designed to be sensitive to WIMPs with a large variety of properties, a flexible approach to model the distribution of the WIMP observables is chosen. Since mono-photon WIMP production is, on an event-by-event basis, indistinguishable from mono-photon neutrino production, the radiative neutrino events are used to obtain the WIMP distributions by a reweighting procedure. Each event receives a weight according to the ratio of the polarised differential cross-sections for pair production of WIMPs (with a certain set of properties like mass, operator, etc.) and for SM neutrino pair production:
\begin{equation}\label{eq:signal_weight}
w_{\textrm{signal,pol}}=\frac{d\sigma_{\chi\chi\gamma}}{dE'_\gamma}(\sqrt{s'},P_{e^-},P_{e^+};E'_\gamma,\theta'_\gamma, \phi'_\gamma) \quad / \quad \frac{d\sigma_{\nu\bar{\nu}\gamma}}{dE'_\gamma }(\sqrt{s'},P_{e^-},P_{e^+};E'_\gamma,\theta'_\gamma, \phi'_\gamma),
\end{equation}
where the $'$-quantities are calculated at the generator level in the reference frame of the $\chi\chi\gamma$ and the $\nu\bar{\nu}\gamma$ system, respectively, with $\gamma$ being the photon with the highest transverse momentum.

It is sufficient to consider the single differential cross-sections $d\sigma/dE_\gamma$, since the similarity of the dependence on the polar angle allows to integrate over the polar angle. 
Two different sets of weights are applied, with integration boundaries of the polar angle adjusted to the two different ($\phi$-dependent) minimum transverse momentum cuts.
Only for very forward angles a third set of integration boundaries is required. 

As described in Sec.~\ref{sec:samples}, the radiative neutrino events are generated with the ILC luminosity spectrum, soft collinear ISR photons and up to four matrix element photons. 
Since the reweighting addresses only the hard subprocess, the effects of the luminosity spectrum and those of the additional ISR photons are transferred to the WIMP events, so that the effects of a realistic description of the ILC environment are also taken into account in the signal events.

The WIMP cross-section is evaluated for different WIMP masses $M_\chi$ and the three operators using the formulae derived in~\cite{perelstein}. The SM cross-sections are obtained using \textsc{Whizard} in bin sizes for $E_\gamma'$ and $\sqrt{s'}$ of up to 10\,GeV. More details can be found in~\cite{thesis} Sec.~7.1.1. 

\begin{figure}[htb]
\begin{center}
     \includegraphics[width=0.6\textwidth]{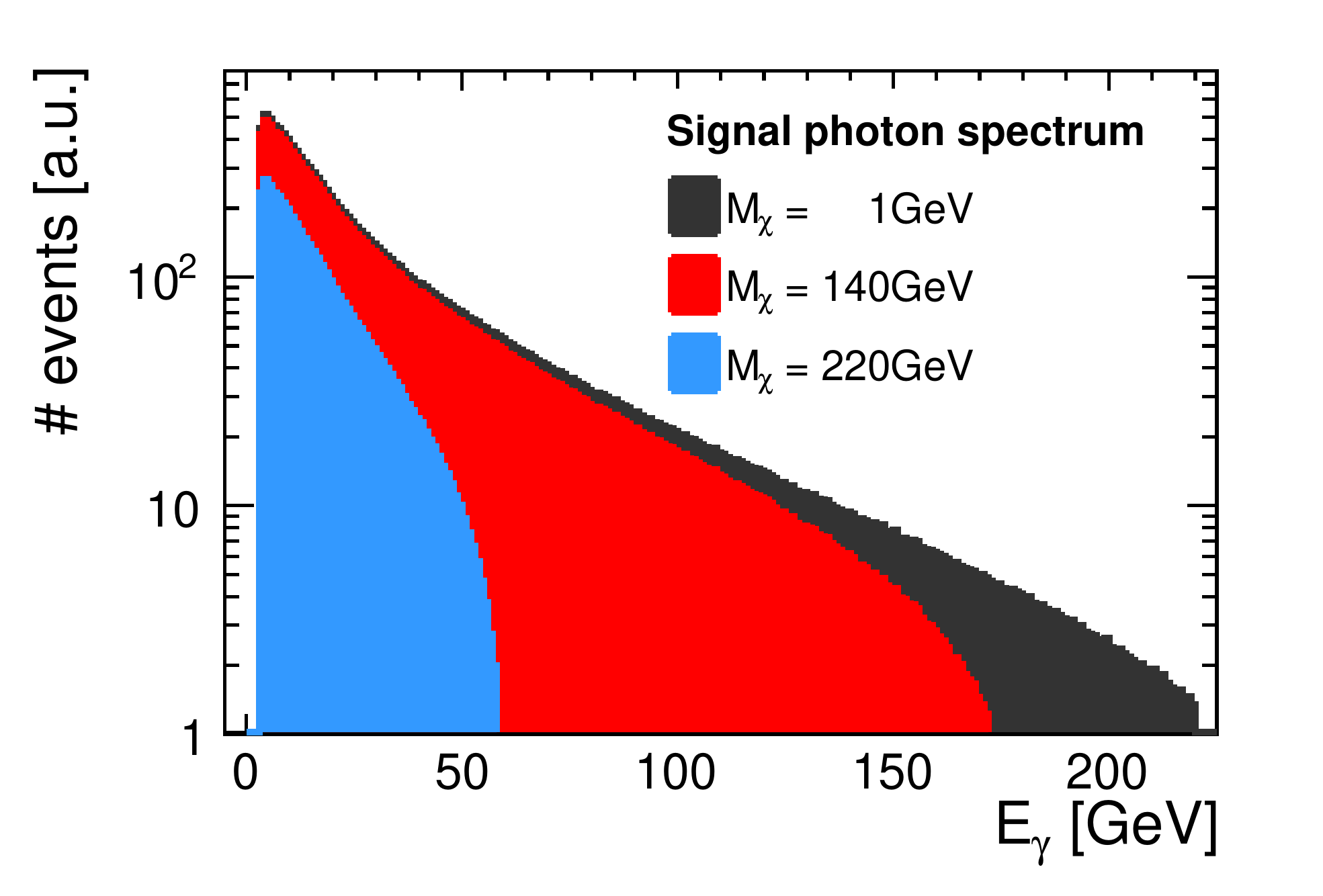} 
\end{center}
      \caption[Photon spectrum for different assumed WIMP masses.]{Photon energy spectra for different WIMP masses, assuming that the WIMP production is mediated by a vector operator. The endpoint moves to smaller energies with increasing WIMP mass.} 
      \label{fig:egammaWIMPmass}
 \end{figure}

In Fig.~\ref{fig:egammaWIMPmass}, three examples of photon energy distributions for different WIMP masses are shown, assuming that the WIMP production is mediated by a vector operator, obtained by the reweighting procedure. As expected, with increasing WIMP mass the photon spectrum is more and more truncated towards smaller energies.\footnote{Note that the endpoint at $E_\gamma=220$\,GeV for the lowest WIMP mass is given by one of the cuts, presented in the next section.} 
The height of the peak region is only changing moderately for a large range of WIMP masses. Only for the highest mass, the total number of signal events starts to decrease more significantly.

\subsection{Event selection} \label{sec:event_selection}

In accordance with the characteristics of the mono-photon signature, the event selection requires the presence of an isolated high-$p_t$ photon and the absence of any other significant detector activity.
Both requirements are explained in detail in the following subsections. An overview of all cuts can be found in Table~\ref{tab:sigdef}. Since the WIMP signal is on an event-by-event basis indistinguishable from the SM neutrino events, the event selection has been optimised to retain as many $\nu\bar{\nu} + 1\gamma$ events as possible while rejecting events with several photons as well as Bhabha scattering events.

\begin{savenotes}
\begin{table}[h!]
\begin{center}
 \begin{tabularx}{\textwidth}{  l  l  }
\hline
\textbf{selection criteria} & \textbf{explanation}\\
\hline
photon cuts (signal definition)  &  \\
\hline
\hspace{0.8cm}$7^\circ<\theta_\gamma<173^\circ$			& distinguish from charged electromagnetic particles\\ 
\hline
\hspace{0.8cm}$p_{T,\gamma}>1.92$\,GeV\hphantom{71}for $|\phi_\gamma|>35^\circ$ 	&ensure identification of Bhabha scattering events\\
\hspace{0.8cm}$p_{T,\gamma}>5.65$\,GeV\hphantom{97}for $|\phi_\gamma|\leq 35^\circ$ 	&\\
\hspace{0.8cm}(in BeamCal coordinates)						&   \\
\hline
\hspace{0.8cm}E$_\gamma>2$\,GeV 							& distinguish from noise\\
\hline
\hspace{0.8cm}E$_\gamma<$\,220\,GeV 							& avoid $Z$ return
\\
\hline\hline
veto conditions &\\
\hline
\hspace{0.8cm}charged particle with $p_T>3$\,GeV 	& suppress background, allow overlay \\
\hspace{0.8cm}electron with $p_T>0.5$\,GeV		& (mainly hadrons)\\
\hline
\hspace{0.8cm}visible energy above 10\,GeV		& suppress background, allow overlay  \\
\hspace{0.8cm}or 30\,GeV if rest is hadrons		& (mainly hadrons) \\
\hline
\hspace{0.8cm}BeamCal cluster				& suppress Bhabha scattering in forward region \\\hline
\end{tabularx}
\caption{Event selection: criteria one photon has to fulfill for the signal definition and event vetos to reject further detector activity.}
\label{tab:sigdef}
\end{center}
\end{table}%
\end{savenotes}

\subsubsection{Photon selection}\label{sec:photon_selection}
In each event, the reconstructed photon with the largest transverse momentum is considered as the ``signal photon'', which has to fulfill the set of cuts summarised in the upper half of Table~\ref{tab:sigdef}, referred to as {\em signal definition} later. By requiring the polar angle of the signal photon to be in the range of $7^\circ<\theta_\gamma<173^\circ$, the forward region is ignored, which is outside of the tracking acceptance and hence a photon cannot be distinguished from charged electromagnetic particles. A minimum transverse momentum of the photon is required to suppress Bhabha scattering events using BeamCal, as explained in Sec.~\ref{sec:bhabha_modelling}. In order to follow the inner rim of BeamCal, the cuts are $\phi$-dependent ($p_{T,\gamma}>1.92$\,GeV for $|\phi_\gamma|> 35^\circ$ and $p_{T,\gamma}>5.65$\,GeV for $|\phi_\gamma|\leq 35^\circ$, where $\phi$ is defined in the range $-180^\circ < \phi < 180^\circ$) and are expressed in the coordinate system of BeamCal, which is centred around the outgoing beam pipe and is thus tilted by half the crossing angle of 7\,mrad with respect to the main detector axis. 
As the cross-section steeply rises with $p_{t,\gamma}$, the cuts are chosen as loosely as possible and the precise values correspond exactly to one BeamCal pad as safety margin.
The $p_T$ cuts are complemented by an energy threshold with a similar value ($2$\,GeV), but expressed in the main coordinate system, to exclude soft reconstructed objects which might be noise in the calorimeter. A maximum energy cut of $220$\,GeV is applied to avoid the large background rates around the photon energy corresponding to the resonance of the radiative return to the $Z$ boson at $E_\gamma \simeq 242$\,GeV for $\sqrt{s}=500$\,GeV.

\subsubsection{Veto on further detector activity}
Events with only little detector activity besides the photon are selected by three criteria based on (i) charged particles in the event, (ii) the visible energy in the detector and (iii) the activity in the forward region of the detector. 
In order to keep as many signal-like events in the presence of overlay, the cuts are designed exploiting the properties of the overlay. 
The fact that the overlay contains typically particles with low energies and transverse momenta can be used to keep them. As the overlay is hadronic the cuts on non-hadronic particles are tighter.

\renewcommand{\labelenumi}{(\roman{enumi})}
\begin{enumerate}
\item The $p_t$ distribution of the reconstructed charged particles in $\nu\bar{\nu} + N\gamma$ and $e^+e^- + N\gamma$ events is shown in Fig.~\ref{fig:pt_charged}, categorised in terms of the true particle type. The categories are formed by testing consecutively if any fraction of the hits contributing to the reconstructed track are created from overlay or photons or, in the case of Bhabha scattering, electrons.
Tracks identified as electrons (or positrons), shown in Fig.~\ref{fig:pt_charged:elec} and~\ref{fig:pt_charged:elec_zoom}, are dominated by the $e^{\pm}$ from Bhabha scattering, but also receive significant contributions from photon conversions. The contribution from overlay backgrounds, which are present at the same level in the signal events (not shown), dominate at the lowest $p_t$ bin. In case of the tracks identified as other charged particles, i.e.\ hadrons or muons, shown in Fig.~\ref{fig:pt_charged:other} and~\ref{fig:pt_charged:other_zoom}, the contribution from overlay is much more prominent and extends to higher $p_t$ values. Therefore, only charged particles with a transverse momentum above 3\,GeV lead to a rejection of the event, while the veto condition is tightened to 0.5\,GeV in the case of electrons.  

\begin{figure}[htb]
\begin{subfigure}{0.495\textwidth}
\includegraphics[width=\textwidth]{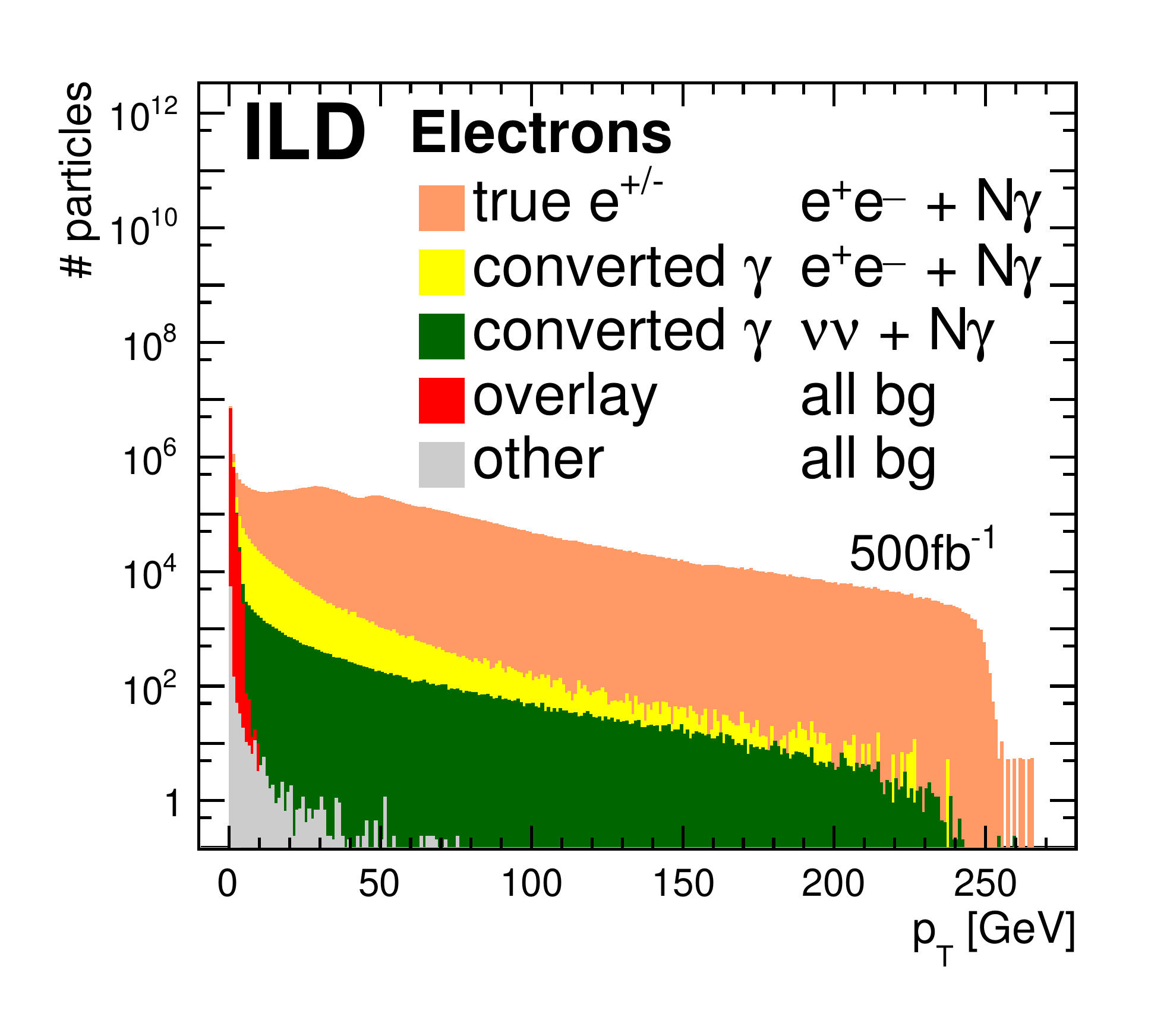} 
\caption{\label{fig:pt_charged:elec}}
\end{subfigure}
\begin{subfigure}{0.495\textwidth}
\includegraphics[width=\textwidth]{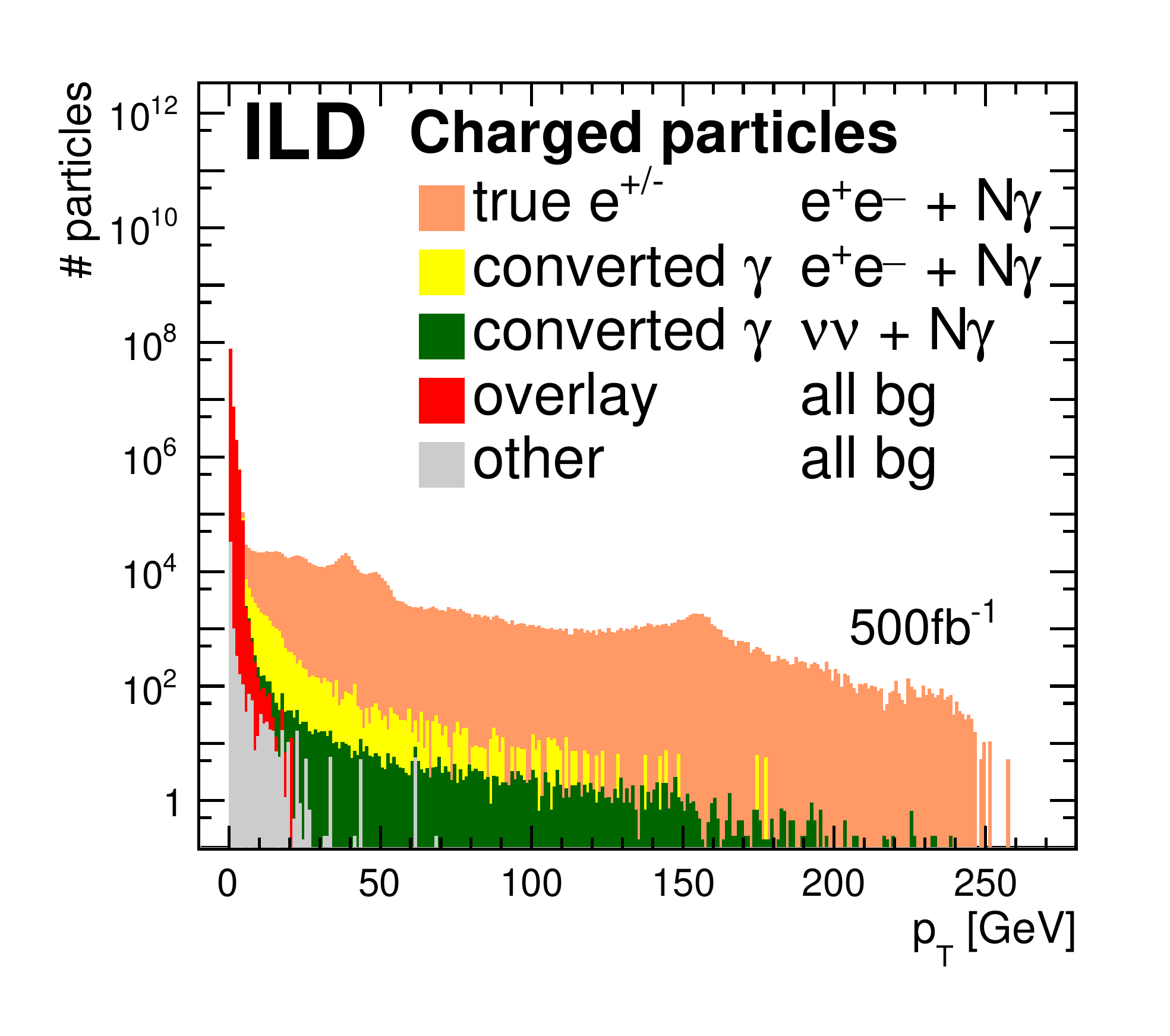} 
\caption{\label{fig:pt_charged:other}}
\end{subfigure}
\begin{subfigure}{0.495\textwidth}
\includegraphics[width=\textwidth]{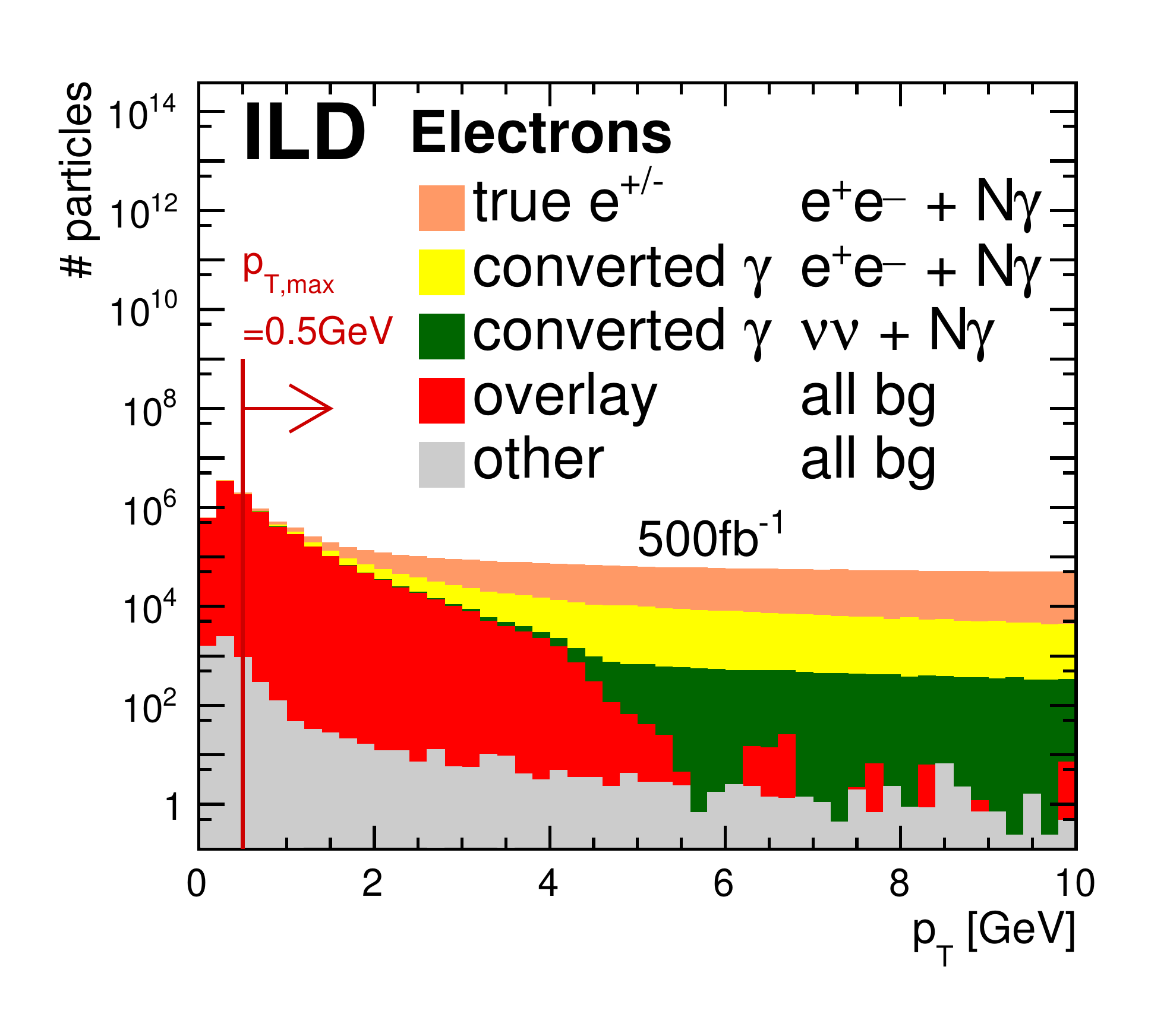} 
\caption{\label{fig:pt_charged:elec_zoom} zoom of low $p_t$ part of (a)}
\end{subfigure}
\begin{subfigure}{0.495\textwidth}
\includegraphics[width=\textwidth]{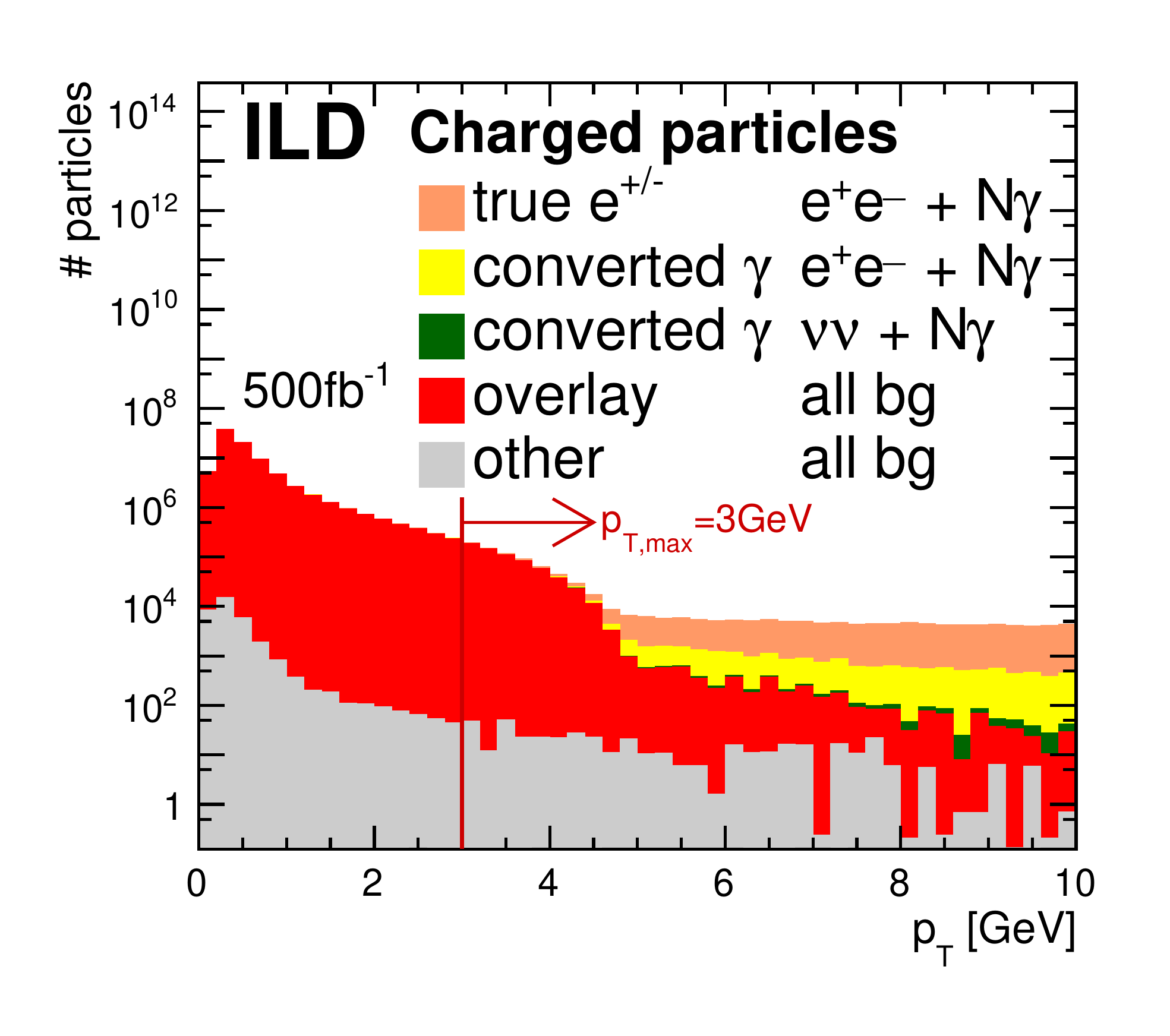} 
\caption{\label{fig:pt_charged:other_zoom} zoom of low $p_t$ part of  (b)}
\end{subfigure}
\caption{Transverse momentum distribution of (a,c) particles reconstructed as electrons or positrons, and (b,d) other charged particles reconstructed as hadrons or muons, categorised in terms of their original particle type. }\label{fig:pt_charged}
\end{figure}

\item In Fig.~\ref{fig:evis:all}, the energy sum $E_{\mathrm{all}}$ of all reconstructed particles with individual energies above $5$\,GeV for $\nu\bar{\nu} + N\gamma$ and $e^+e^- + N\gamma$ events are shown, not including the signal photon and energy depositions in BeamCal. In Fig.~\ref{fig:evis:nohad}, the analogous sum $E_{\mathrm{nohad}}$ when omitting hadrons (i.e.\ reconstructed charged pions and neutrons) is shown. Events with  $E_{\mathrm{all}}>30$\,GeV or $E_{\mathrm{nohad}}>10$\,GeV are rejected.

\begin{figure}[htb]
\begin{subfigure}{0.495\textwidth}
\includegraphics[width=\textwidth]{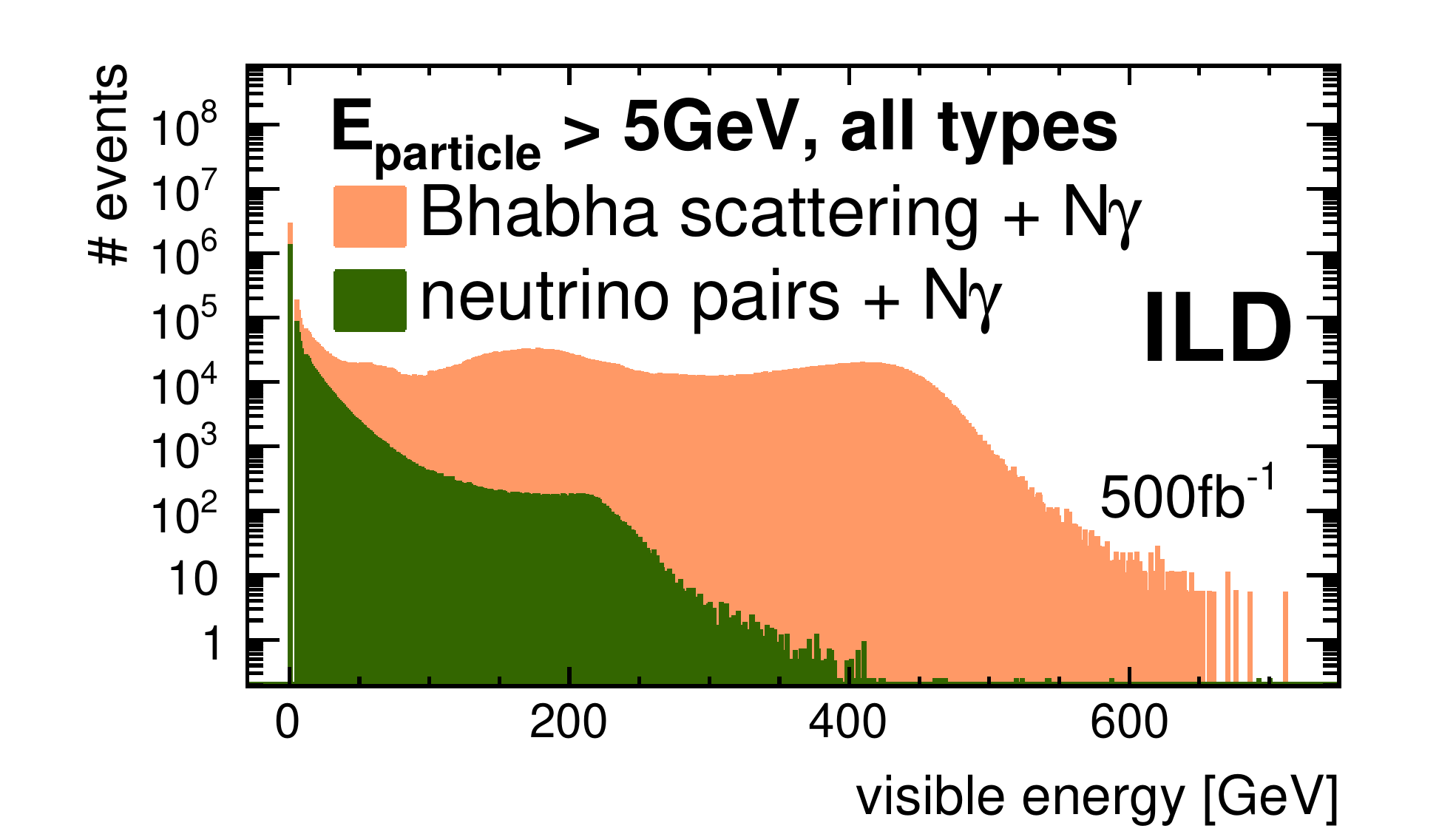} 
\caption{\label{fig:evis:all}}
\end{subfigure}
\begin{subfigure}{0.495\textwidth}
\includegraphics[width=\textwidth]{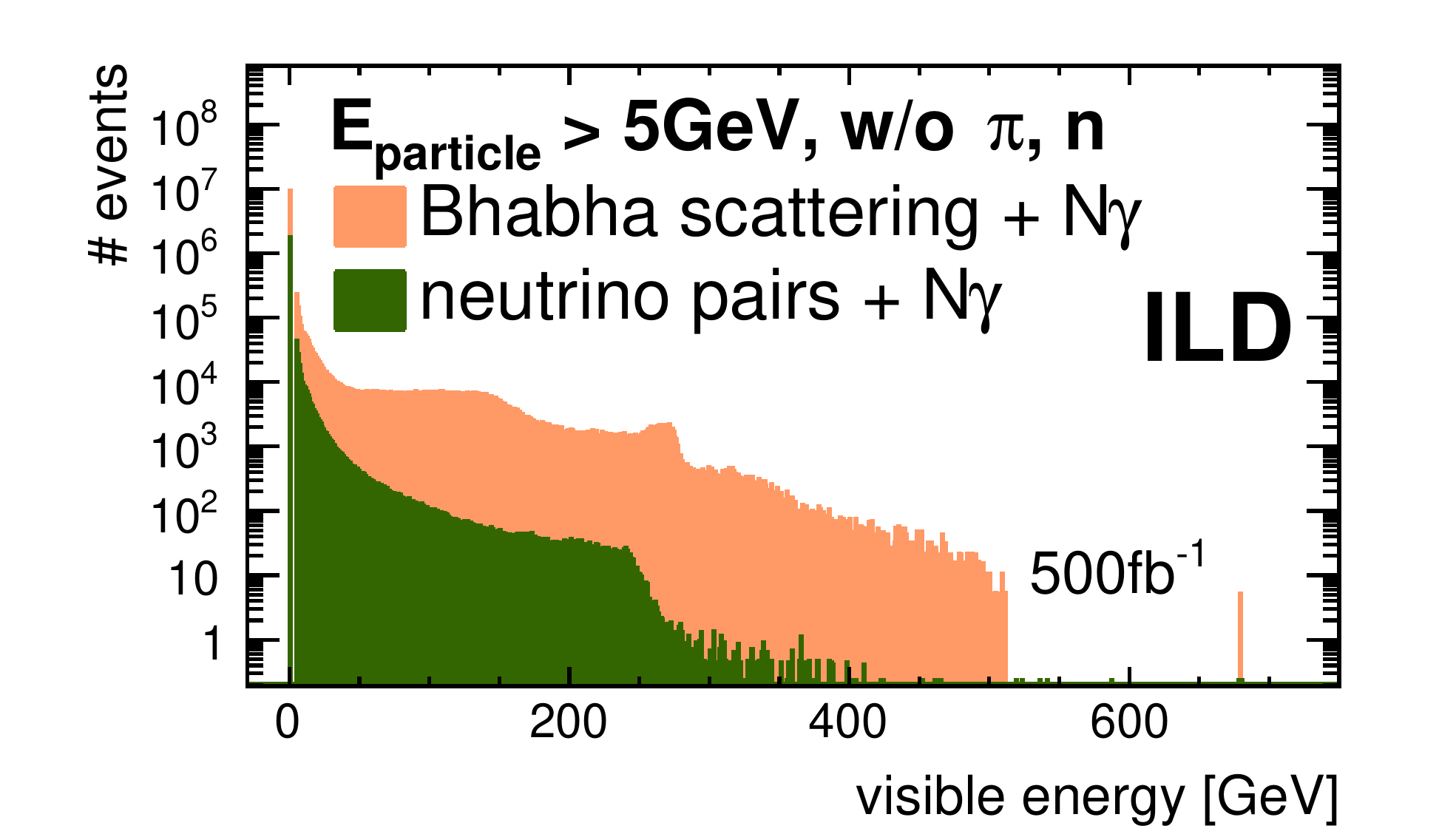} 
\caption{\label{fig:evis:nohad}}
\end{subfigure}
\caption{Energy sum of all reconstructed particles above $5$\,GeV, outside of BeamCal and apart from the signal photon, (a) summing all particle types and (b) without reconstructed pions and neutrons.}
\label{fig:evis}
\end{figure}

\item The last criterion is tat no clusters may be reconstructed in BeamCal.\footnote{Due to the different approach in the reconstruction of the BeamCal response (as described in Sec.~\ref{sec:simulation}), BeamCal objects are not considered in the previous cuts.} As shown in Fig.~\ref{fig:bcal_clusters}, Bhabha scattering is effectively suppressed, whereas only 2\% of neutrino events contain a BeamCal cluster, which is mainly due to additional ISR photons. In the $\nu\bar{\nu} + 1\gamma$ sample, which by construction does not contain photons in BeamCal, fake clusters from fluctuations in the energy depositions from pair background occurs in about 0.5\% of the events.  

\begin{figure}[h]
\begin{center}
 \includegraphics[width=0.5\textwidth]{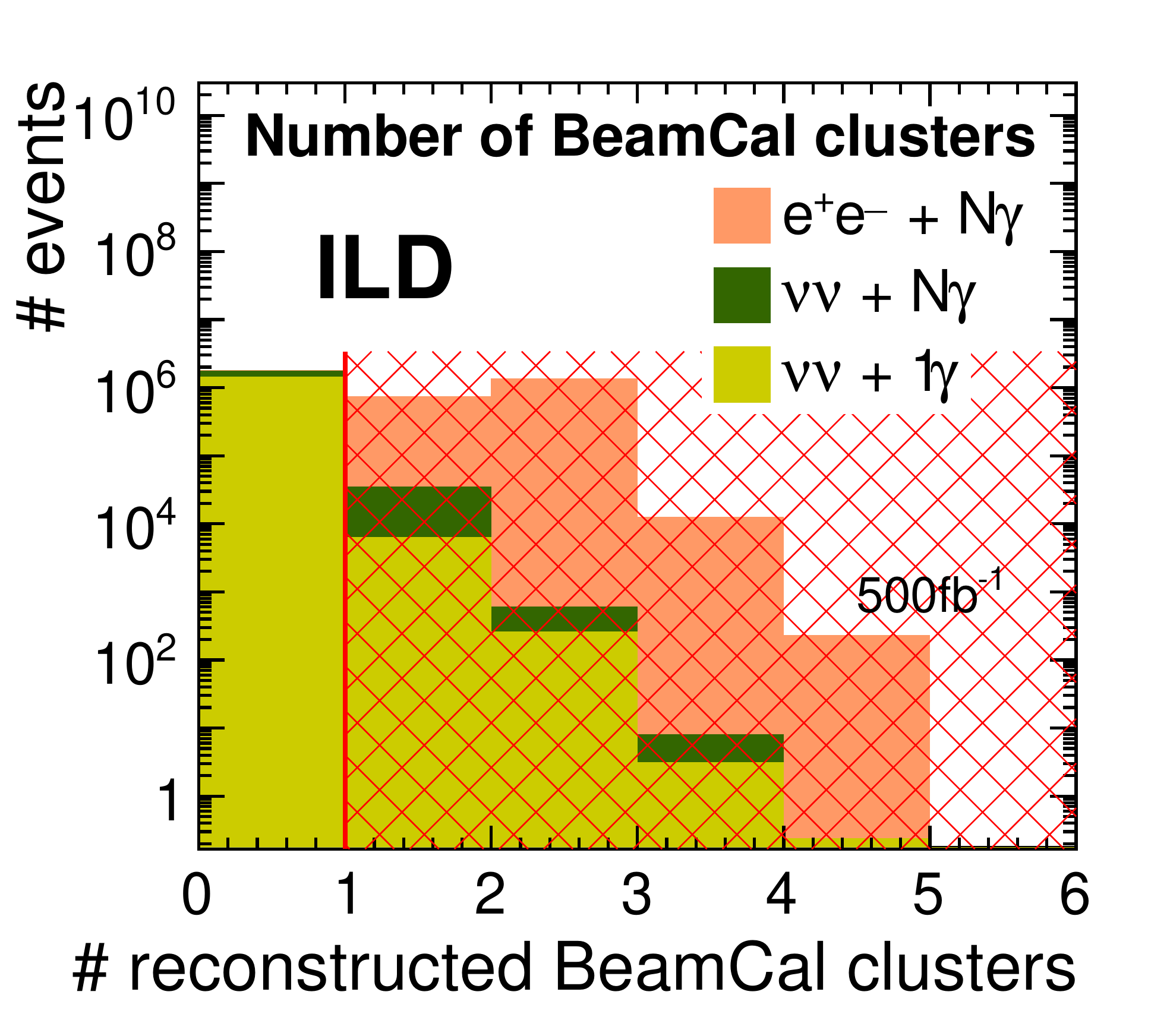} 
\end{center}
 \caption{Number of clusters reconstructed in BeamCal in $\nu\bar{\nu} + 1\gamma$, $\nu\bar{\nu} + N\gamma$ and $e^+e^- + N\gamma$ events. At least one BeamCal cluster is found in about 0.5\% of the $\nu\bar{\nu} + 1\gamma$ events. In these cases, fluctuations in the energy depositions from pair background are falsely reconstructed as cluster. }
\label{fig:bcal_clusters}
\end{figure}

\end{enumerate}

The effect of these cuts on the radiative neutrino and Bhabha scattering events is summarised in Tables~\ref{tab:cut_flow} and~\ref{tab:efficiencies}, normalised to $1$\,fb$^{-1}$ of data. In Table~\ref{tab:cut_flow}, the efficiency of the veto cuts is given with respect to the number of events fulfilling the signal definition. The efficiency to select signal-like events, i.e.\ neutrino pair production events with a single ISR photon, is 83\% for unpolarised beams.

\begin{table}[h!t]
 \begin{center}
  \begin{tabular}[h]{ l l r r r r }
  \cline{3-6}
\multicolumn{2}{l}{$\mathbf{P_{e^-}=0}$\textbf{\%,} $\mathbf{P_{e^+}=0}$\textbf{\%}} & signal & low $p_T$ & low visible & BeamCal  \\
 & & definition & of charged & energy & veto \\\hline
\boldmath$\nu\bar{\nu}+1\gamma$ & number of events & 3608 & 3189 & 2999 & 2986 \\
& selection efficiency & & 88.4\% & 83.1\% & 82.8\% \\\hline
\boldmath$\nu\bar{\nu}+$N$\gamma$ & number of events & 4534 & 3988 & 3565 & 3495 \\
& selection efficiency & & 88.0\% & 78.6\% & 77.1\% \\\hline
\boldmath$e^+e^-+$N$\gamma$& number of events & 50508 & 19647 & 4261 & 113 \\
& selection efficiency & & 38.9\% & 8.4\% & 0.2\% \\\hline
\end{tabular}
\caption{Cut flow assuming unpolarised beams. Event numbers for 1\,fb$^{-1}$ and selection efficiencies with respect to the signal definition are given.}\label{tab:cut_flow}
\end{center}
\end{table}

In Table~\ref{tab:efficiencies} the number of selected events for various polarisation choices is shown. While the number of Bhabha scattering events is practically the same for all polarisation combinations, the number of selected neutrino events is highly polarisation-dependent and varies by one order of magnitude between $P(e^-,e^+)=(-80\%,+30\%)$ and $(+80\%,-30\%)$.

\begin{figure}[h]
\begin{subfigure}{0.495\textwidth}
\includegraphics[width=\textwidth]{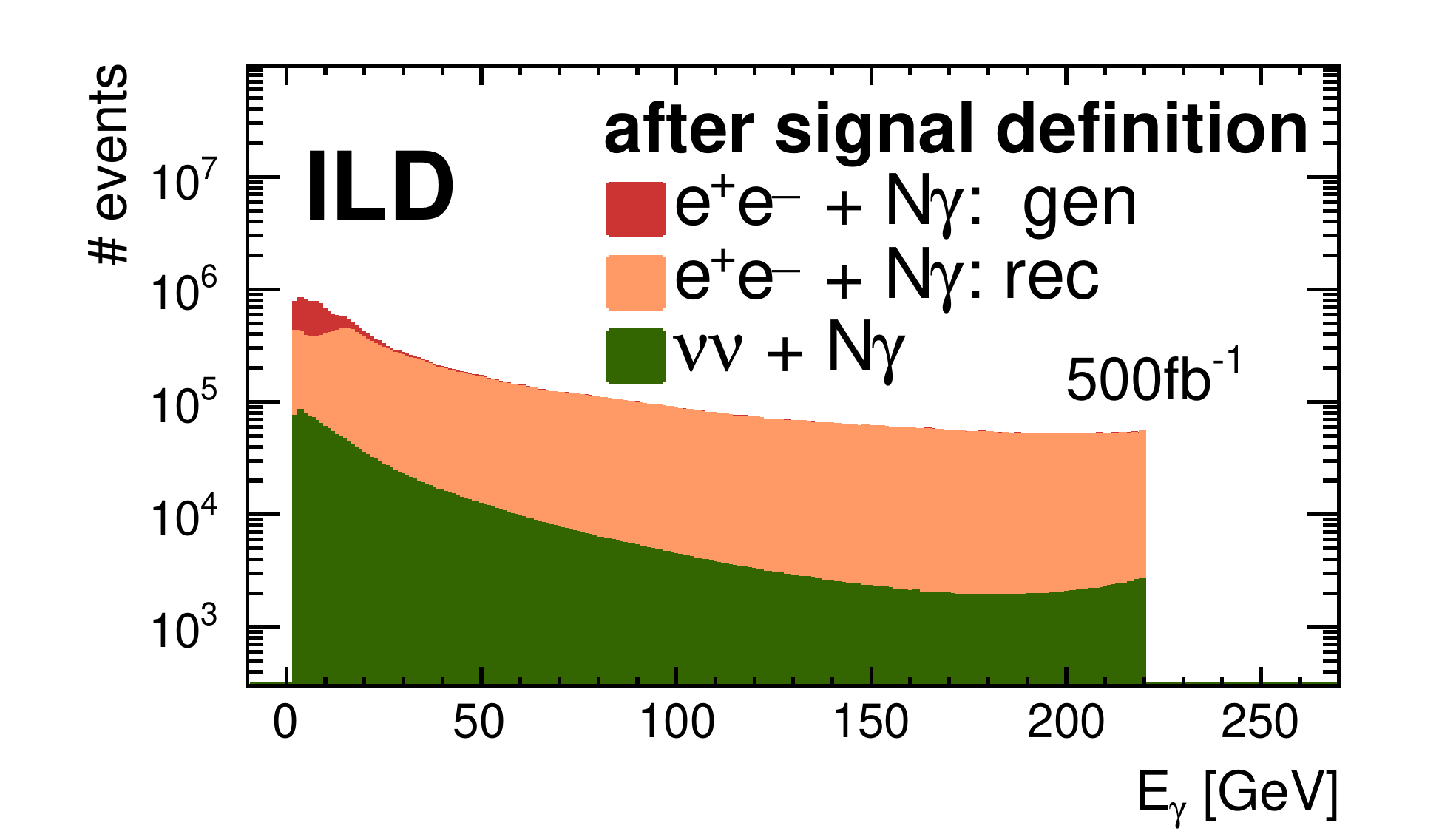} 
\caption{\label{fig:Egamma:photon}}
\end{subfigure}
\begin{subfigure}{0.495\textwidth}
\includegraphics[width=\textwidth]{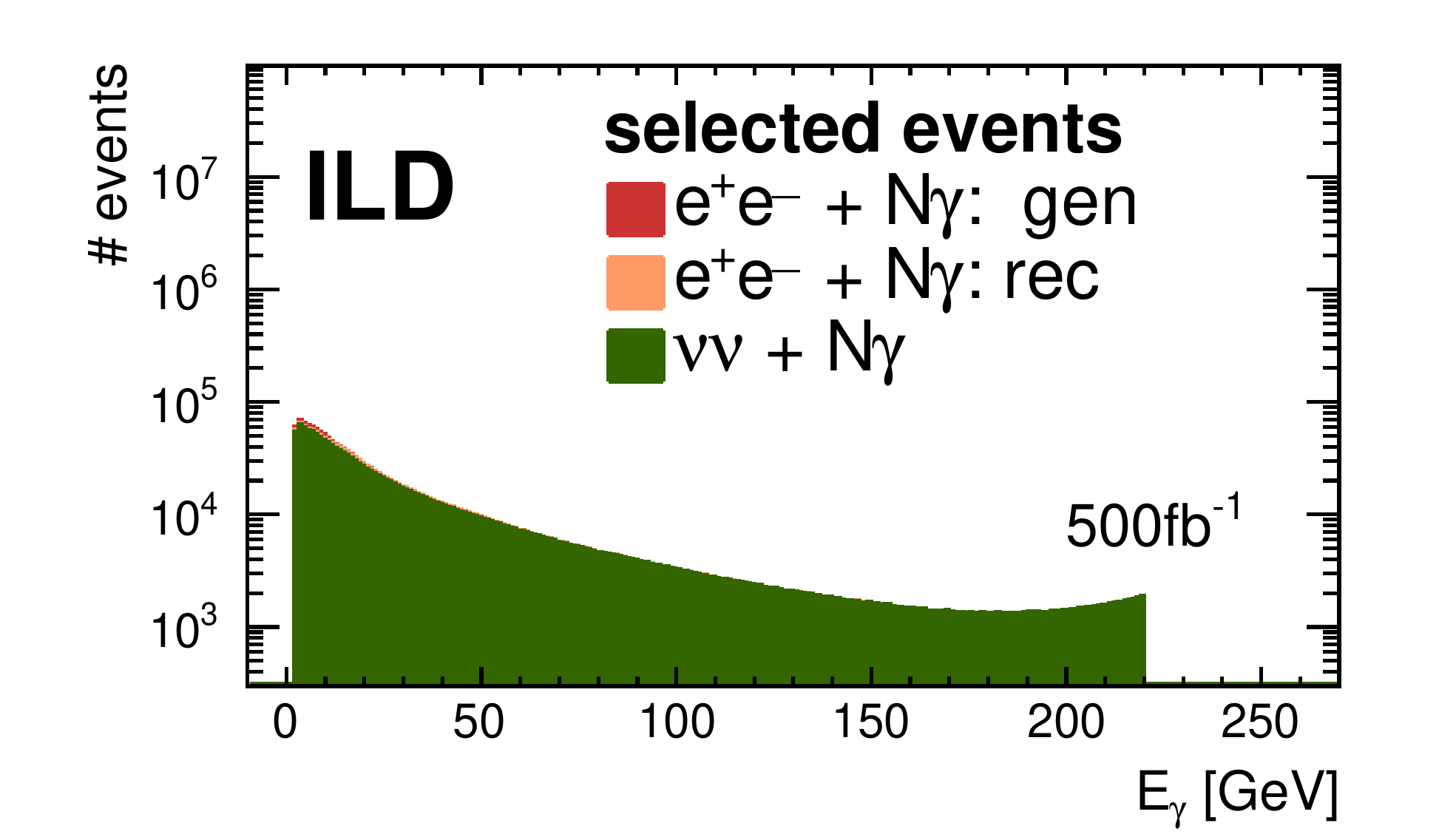} 
\caption{\label{fig:Egamma:final}}
\end{subfigure}
\caption[$E_\gamma$ with the additional Bhabha scattering events.]{Photon energy distribution for unpolarised beams, including the additional Bhabha scattering events, (a) after the signal definition and (b) after the event selection.}
\label{fig:Egamma}
\end{figure}

The photon energy distribution as obtained after applying the signal definition is shown in Fig.~\ref{fig:Egamma:photon}, normalised to $500$\,fb$^{-1}$ of unpolarised data. The photons from the additional Bhabha scattering data set, discussed in Sec.~\ref{sec:bhabha_modelling}, are shown in dark red and smoothly fill up the Bhabha scattering distribution at low photon energies. In Fig.~\ref{fig:Egamma:final}, the same distribution after the application of all cuts is shown. Whereas Bhabha scattering is the dominant background after the signal definition cuts, it is suppressed to the per mille level, in particular by the BeamCal veto. On the other hand, $77\%$ of the signal-like neutrino events are kept, making this the dominant background to the WIMP signal. The efficiency for the radiative neutrinos depends slightly on the choice of beam polarisation, varying between $71$\% and $78$\%. This occurs due to the different relative contribution of the $s$-channel $Z$  exchange diagram, for which the radiative return to the $Z$ pole increases the probability to have additional ISR photons with substantial energy. These ISR photons (in addition to the signal photon) can thus lead to a higher visible energy and are more likely to be reconstructed if they hit BeamCal and hence a larger fraction of those events is rejected.

\begin{table}[h!t]
 \begin{center}
  \begin{tabular}[h]{ l r r r r r }
  \hline
$\mathbf{P_{e^-}}$			& 0	& $-$80\%	& $-$80\%	& $+$80\%& $+$80\%\\
$\mathbf{P_{e^+}}$			& 0	& $-$30\%	& $+$30\%	& $-$30\%& $+$30\%\\\hline
\boldmath$\nu\bar{\nu}+$N$\gamma$ 	& 3495	& 4280		& 7906	& 762	& 1033\\  
\boldmath$e^+e^-+$N$\gamma$ 		&  113	&  114		&  113		& 113	& 113\\\hline
  \end{tabular}
  \caption{Number of selected events for different polarisation combinations per fb$^{-1}$.}
  \label{tab:efficiencies}
\end{center}
\end{table}

\subsection{Sensitivity calculation}\label{sec:tsyslimit}

The energy spectra of the selected photon candidates for signal and background events are shown in Fig.~\ref{fig:egammaWIMPmass} and Fig.~\ref{fig:Egamma:final}, respectively. These spectra, together with the systematic uncertainties, are input to the sensitivity calculation (discussed in Sec.~\ref{sec:systematics}), which follows a frequentist approach based on fractional event counting~\cite{Bock:2004xz}.
For each of the three operators defined in Table~\ref{tab:operators}, WIMP mass hypotheses between 1 and 250\,GeV have been tested in steps of 1\,GeV.

The sensitivity will be presented in terms of the expected exclusion reach at 95\% confidence level and the expected  $5\sigma$ discovery reach.
The confidence levels of the expected exclusion limits are calculated according to the ``$CL_s$ method'', i.e.~$CL_{exclusion} = 1 - CL_{s+b}/CL_b$~\cite{read2000modified,junk1999confidence} and the confidence level of a discovery is calculated according to $CL_{discovery}= 1 - CL_b$.
The photon energy is taken as a discriminating variable.
The package \textsc{TSysLimit} is used, which was originally written for a leptoquark analysis at HERA~\cite{tsyslimit_hera}. The approach is optimised for systematic uncertainties by downgrading the weights entering the test statistic of bins with large systematic uncertainties. This is important due to the substantial amount of remaining background. 

The sign of the beam polarisations is foreseen to be flipped continuously on the time scale of tenths of seconds, namely from bunch train to bunch train, i.e.\ much faster than the typical time scales of experimental systematic effects, which depend e.g.\ on alignment, calibration and configuration of the detectors and the accelerator. Due to this fast helicity reversal, the experimental uncertainties are expected to be correlated to a high degree between the data sets collected at the same energy, but with different polarisation sign configurations. Note that this correlation does not apply between data sets of different centre-of-mass energies, which need to be taken consecutively, typically in different years, and thus cannot be assumed to receive the same size and magnitude of systematic effects. More information on the interplay of polarised beams with fast helicity reversal and systematic uncertainties can be found in~\cite{Karl:2019hes}.

\subsection{Systematic uncertainties}\label{sec:systematics}

The following sources of systematic uncertainties are taken into account in the sensitivity calculation:
\renewcommand{\labelenumi}{(\roman{enumi})}
\begin{enumerate}
\item {\bfseries Shape of luminosity spectrum:} The effect of the \textit{shape} of the luminosity spectrum has been specifically studied for this paper. It is the dominant source of systematic uncertainties, with values between a few per mille and a few percent, depending on the physics process, the WIMP mass, beam polarisation and photon energy.
\item {\bfseries Luminosity:} The uncertainties in the luminosity measurement by counting Bhabha scattering events in the forward detectors LumiCal add up to 2.6 per mille~\cite{jelisavvcic2013luminosity}. 
\item {\bfseries Beam polarisation:} The precision of 0.2-2.5 per mille is obtained by combining the polarimeter measurements with collision data~\cite{karl2017polarimetry}.
\item {\bfseries Event selection:} The uncertainty on the event selection is estimated from a fit to the peak of the radiative return to the $Z$ boson~\cite{CBDA1}. With the higher integrated luminosity of this study the uncertainty is scaled to 2 per mille (c.f.\ Sec.\ 7.4.3 in~\cite{thesis}).
\item {\bfseries Theory:} The uncertainty on the cross-section is given by 1.3 per mille, a typical value for the uncertainty obtained by the matrix element calculation of \textsc{O'Mega}~\cite{Moretti:2001zz} within \textsc{Whizard}.
\end{enumerate}
(ii)-(v) are considered as normalisation uncertainties only. Additional shape-dependent components are assumed to be covered by the rather conservative treatment of (i), which will be discussed in the following.
The effect of these uncertainties on the sensitivity will be discussed in Sec.~\ref{sec:systematics_effect}.

The shape of the luminosity spectrum can be determined with two complementary techniques: BeamCal has been designed to provide online monitoring of the energy depositions from beamstrahlung. From the shape of these energy depositions, optionally assisted by the measurement of GamCal, an even more forward calorimeter, the beam parameters can be extracted for every couple of bunches~\cite{grah}. This time-resolved method is complemented by a long-term average determination of the luminosity spectrum from low-angle Bhabha scattering~\cite{poss2014luminosity}. The following analysis is based on the online method only and can thus be considered as a conservative estimate of the effect of the luminosity spectrum -- or it can be considered as a proxy for other, additional shape uncertainties, e.g.\ from selection efficiencies.

\begin{figure}[htb]
\begin{subfigure}{0.495\textwidth}
\includegraphics[width=\textwidth]{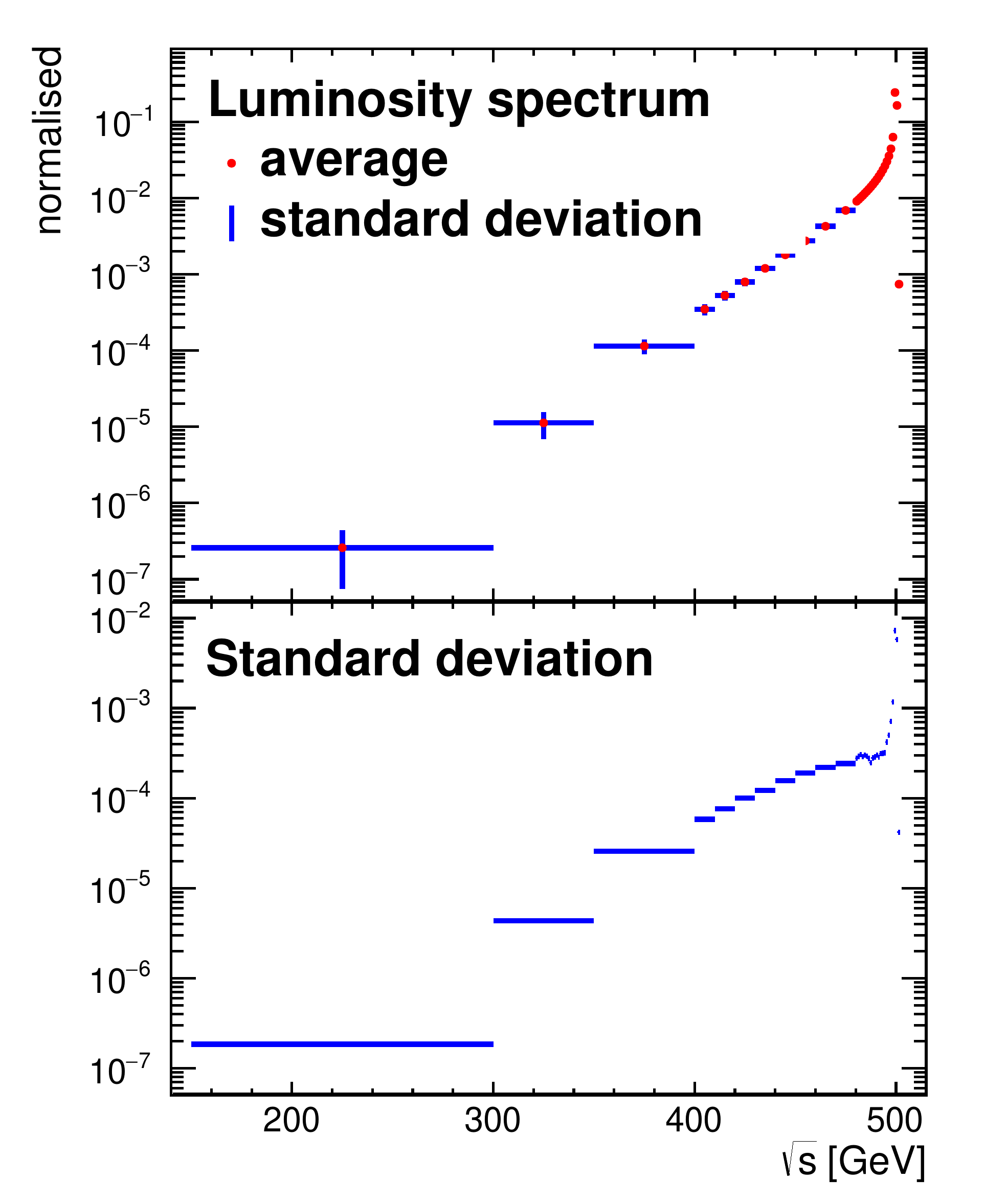} 
\caption{\label{fig:ls_sigmas_envelopes:sigmas}}
\end{subfigure}
\begin{subfigure}{0.495\textwidth}
\includegraphics[width=\textwidth]{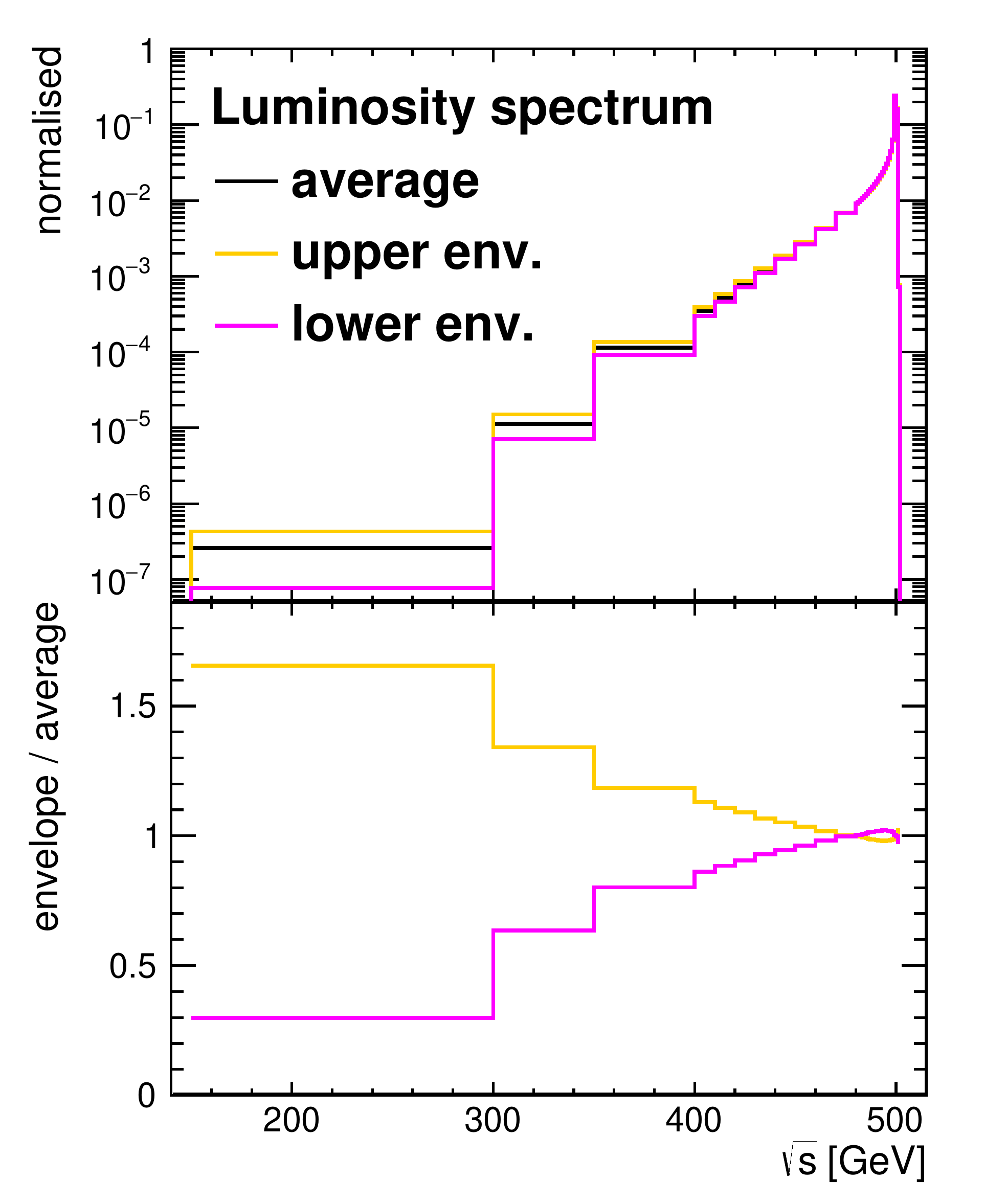} 
\caption{\label{fig:ls_sigmas_envelopes:envelopes}}
\end{subfigure}
\caption[Average luminosity spectrum shape and upper and lower envelopes.]{(a) Average luminosity spectrum shape obtained with 200 simulations of the beam-beam interactions. (b) Nominal luminosity spectrum and the normalised upper and lower envelopes of the $1\sigma$ uncertainties shown in (a).}
\label{fig:ls_sigmas_envelopes}
\end{figure}

In the first step, the shape of the luminosity spectrum is obtained by simulating the beam-beam interactions using \textsc{GuineaPig}~\cite{schulte}. The simulation is repeated 200 times with random variations of number of particles per bunch, the horizontal emittance and the horizontal $\beta$ function, which have been identified to be the parameters which influence the shape of the spectrum the most. The size of the variations is chosen according to the 1$\sigma$ uncertainties obtained in~\cite{grah}. In Fig.~\ref{fig:ls_sigmas_envelopes:sigmas}, the bin-wise average values and standard deviations of the 200 luminosity spectra are shown. 

In order to study the effect of this variation, two spectra with maximally different shapes within 1$\sigma$ uncertainties are constructed, referred to as \textit{upper} and \textit{lower envelope} in the following and displayed in Fig.~\ref{fig:ls_sigmas_envelopes:envelopes}. The upper envelope is obtained by taking the central value plus the $1\sigma$ uncertainty in each bin and the lower envelope with the $1\sigma$ uncertainty subtracted. As the integral of the curve, i.e.~the total luminosity, is measured independently, the two envelopes are normalised such that they differ only in shape. 

\begin{figure}[htb]
\begin{center}
\begin{subfigure}{0.495\textwidth}
\includegraphics[width=\textwidth]{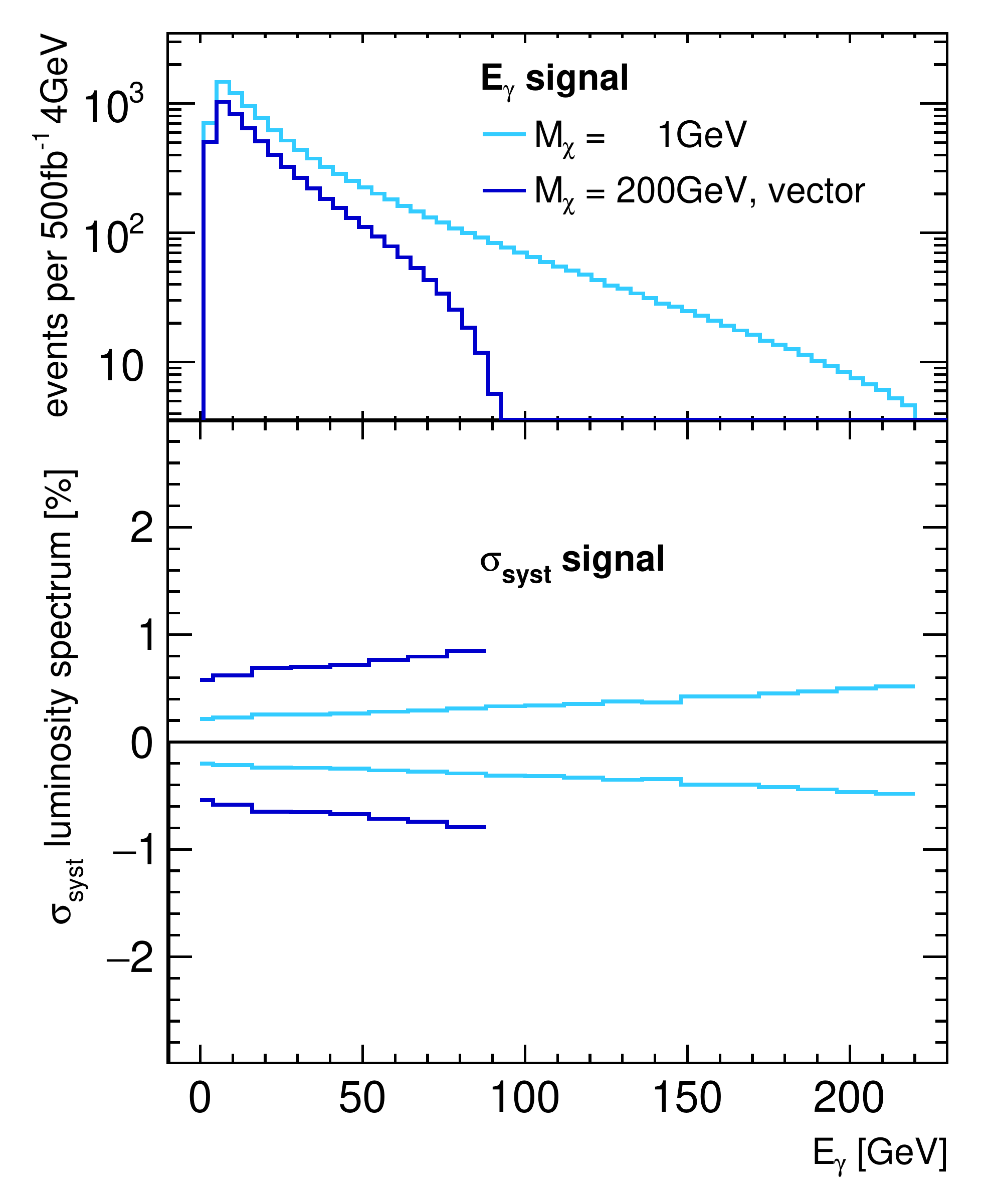} 
\caption{\label{fig:ls_sys_final:sig}}
\end{subfigure}
\begin{subfigure}{0.495\textwidth}
\includegraphics[width=\textwidth]{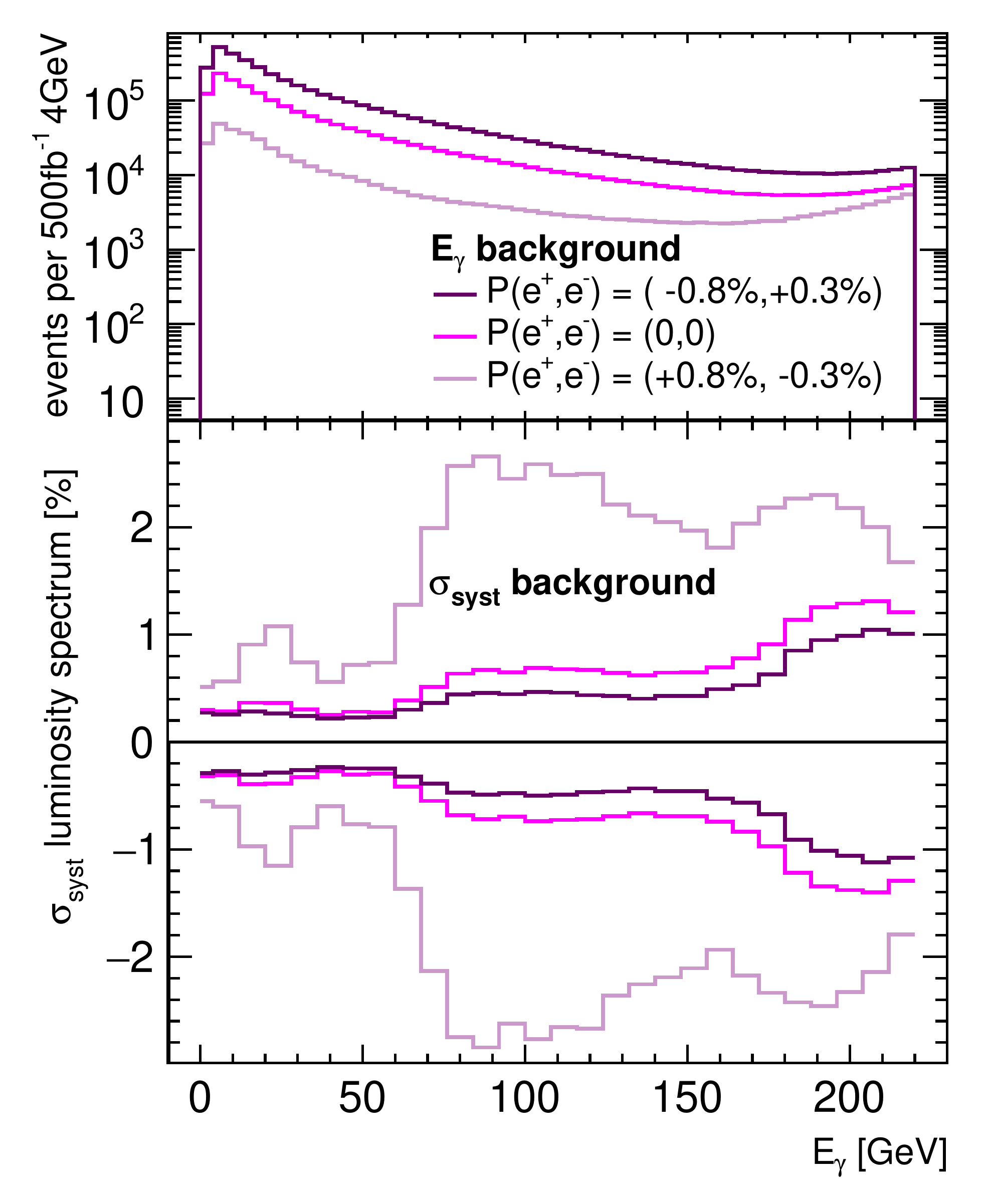} 
\caption{\label{fig:ls_sys_final:bkg}}
\end{subfigure}
\caption[Systematic uncertainties induced by the luminosity spectrum.]{Systematic uncertainties induced by the luminosity spectrum as a function of the photon energy, as they will be used in the sensitivity calculation. (a) Effect on the signal for two example values of $M_\chi$. (b) Effect on the background for different beam polarisation configurations. The background photon spectrum for the unpolarised case corresponds to Fig.~\ref{fig:Egamma:final}.}
\label{fig:ls_sys_final}
\end{center}
\end{figure}

The uncertainty on the \textit{photon} spectrum is obtained by weighting the individual events with the ratio ``envelope'' to ``nominal'' for the \com energy of the event, obtained using generator-level information. The resulting uncertainties on the signal and background photon spectra, corresponding to the two envelopes, are shown in Fig.~\ref{fig:ls_sys_final}. In case of the background, the resulting uncertainty on the photon energy spectrum exhibits a similar shape for all polarisation settings, but differs in magnitude. The larger the relative contribution of the $s$-channel $Z$ exchange, the larger are the uncertainties, reaching nearly $3\%$ in the worst case. In the case of the signal, the values are independent of the polarisation, but depend on the WIMP mass and increase with increasing WIMP mass to a maximum value of about 1\%. For more details such as the small dependency on the type of operator, see Sec.~7.5. in~\cite{thesis}.

\section{Results for 500\,GeV} 
\label{sec:results500}
 
In this section, we present the sensitivity of the ILC to WIMP production solely based on the full detector simulation study at a \com energy of 500\,GeV. The sensitivity will be presented in the plane of the new physics scale $\Lambda$ versus the WIMP mass $M_{\chi}$. All $\Lambda$ values below the curve can be discovered or excluded, depending on the tested hypothesis. The testable energy scales are always well above the \com energy and hence effective operators are a suitable approach. A grey area indicates parameter space which cannot be tested using effective operators, taken here as~$\Lambda\leq\sqrt{s}$. The normalisation and shape-dependent systematic uncertainties presented in Sec.~\ref{sec:systematics} are taken into account in the results.

\begin{figure}[htb]
\begin{subfigure}{0.495\textwidth}
  \includegraphics[width=\textwidth]{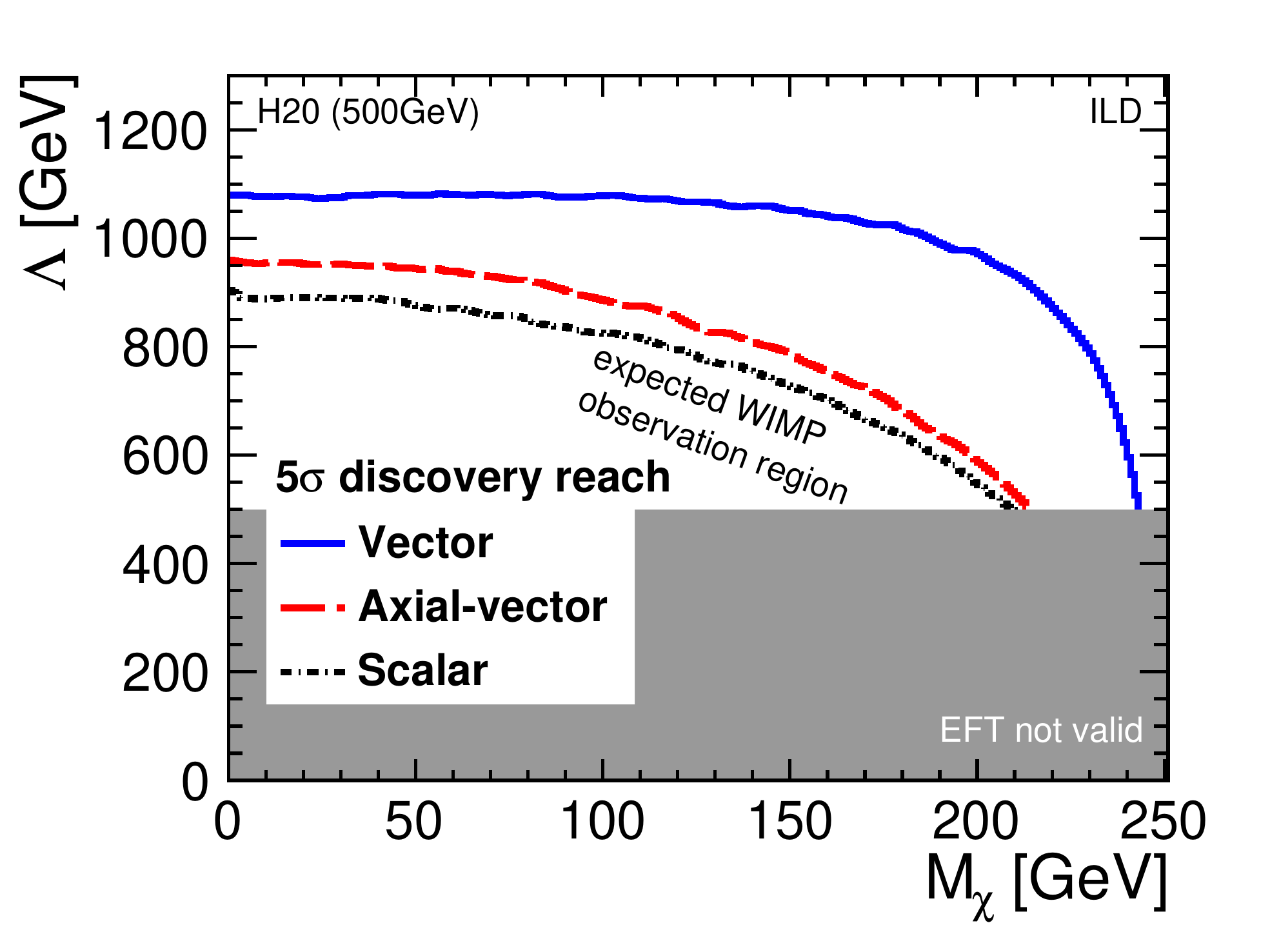} 
   \caption{$5\sigma$ discovery reach \label{fig:H20_500:discovery}}
\end{subfigure}
\begin{subfigure}{0.495\textwidth}
  \includegraphics[width=\textwidth]{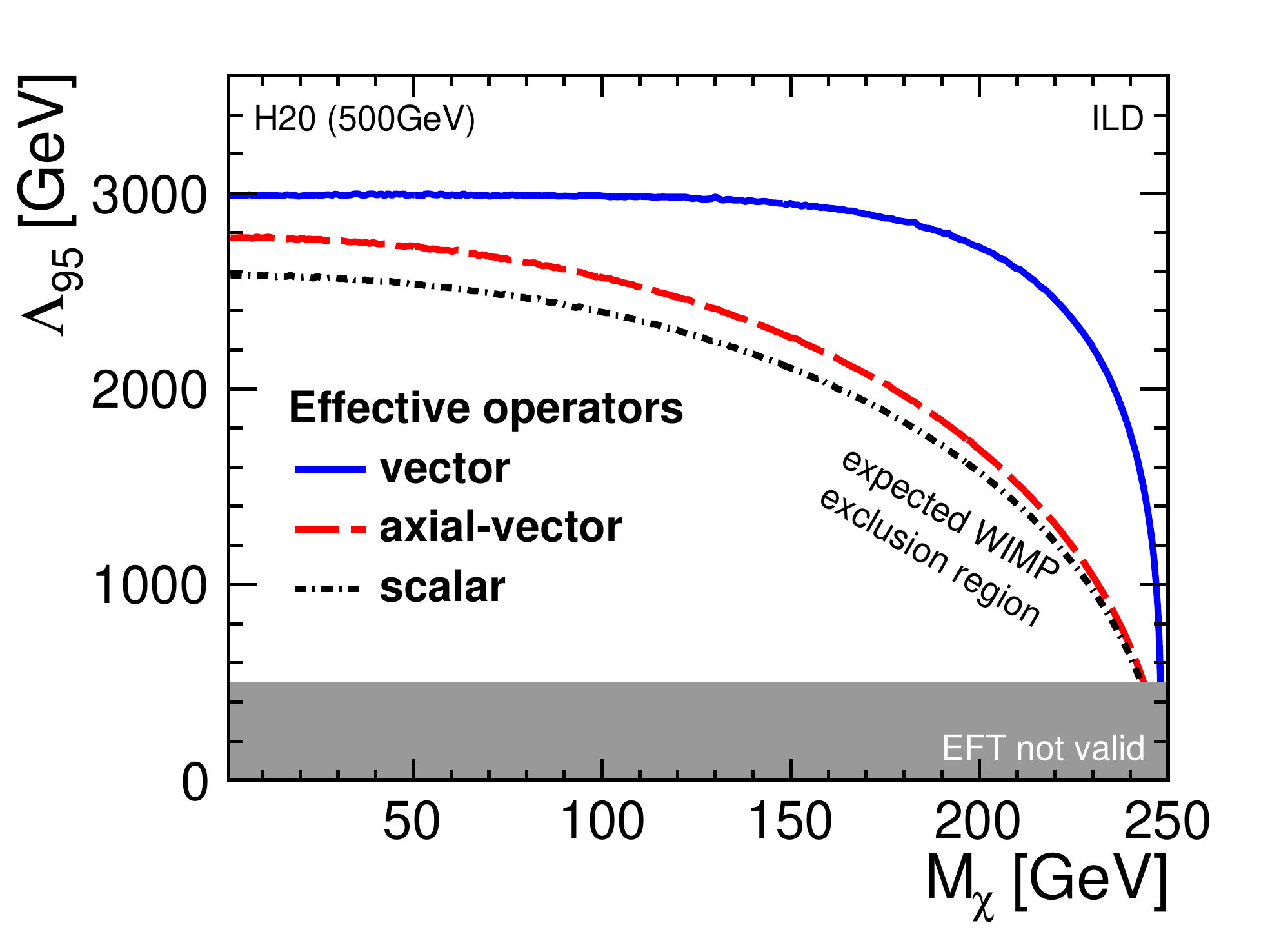} 
   \caption{95\% exclusion reach\label{fig:H20_500:exclusion}}
\end{subfigure}
  \caption{The expected sensitivity for the different effective operators assuming $4$\,ab$^{-1}$ at 500\,GeV.   }
  \label{fig:H20_500}
 \end{figure}

At a \com energy of 500\,GeV, the baseline running scenario for the ILC comprises 4\,ab$^{-1}$ of data shared between the different polarisation sign configurations of 40\% for each of $(-,+)$ and $(+,-)$, and 10\% for each of the equal-sign configurations, as introduced in Sec.~\ref{sec:H20}.
Fig.~\ref{fig:H20_500} shows the discovery and exclusion reach for this data set, based on confidence levels of $5\sigma$ (i.e.~$99.99994$\%) and $95$\%, respectively. WIMP masses up to almost half the \com energy can be tested. The sensitivity decreases for higher WIMP masses. In the case of the vector operator, the plateau with constant testable energy scales continues to significantly higher WIMP masses than for the other operators. 
The independence of the WIMP mass for a large range of masses can be explained by the small difference of the dominating peak regions of the corresponding photon spectra (see Fig.~\ref{fig:egammaWIMPmass}).
WIMP production could be discovered for new physics energy scales $\Lambda$ up to about $1$\,TeV, with the best sensitivity to the vector operator, followed by the axial-vector and scalar operators. At 95\% confidence level, energy scales up to $3$, $2.8$ and $2.6$\,TeV can be probed for the vector, axial-vector and scalar operators, respectively.

\subsection{Impact of beam polarisation} \label{sec:pol}

The beam polarisation plays a very important role for the mono-photon signature as it strongly suppresses 
the $\nu\bar{\nu}\gamma$ background in the case of a mainly right-handed electron beam. The interplay of beam polarisation and systematic uncertainties will be discussed in the next section.

\begin{figure}[htb]
  \begin{subfigure}{0.495\textwidth}
    \includegraphics[width=\textwidth]{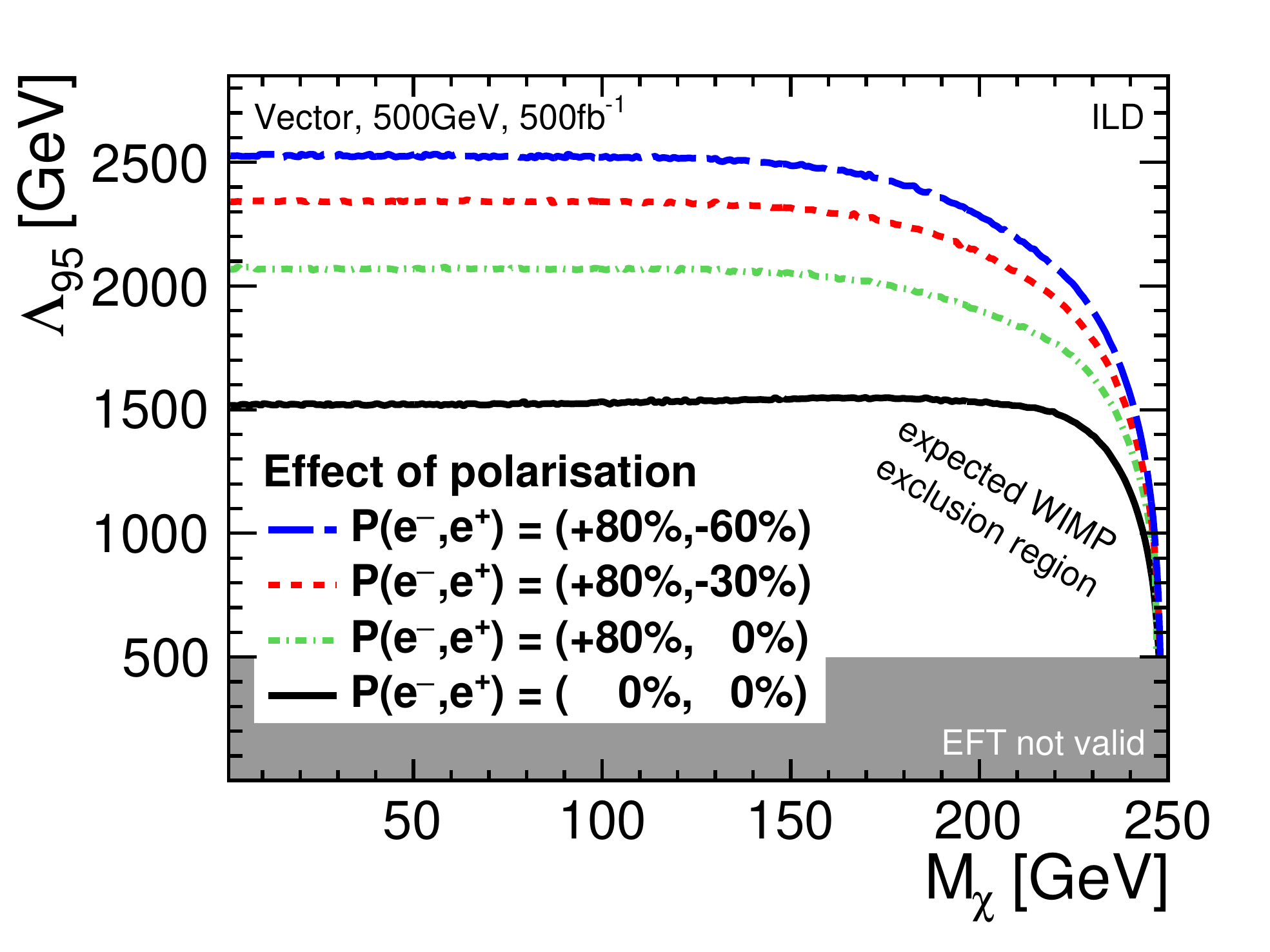} 
    \caption{\label{fig:bg_sig_pol:sens}} 
  \end{subfigure}
  \begin{subfigure}{0.495\textwidth}
    \includegraphics[width=\textwidth]{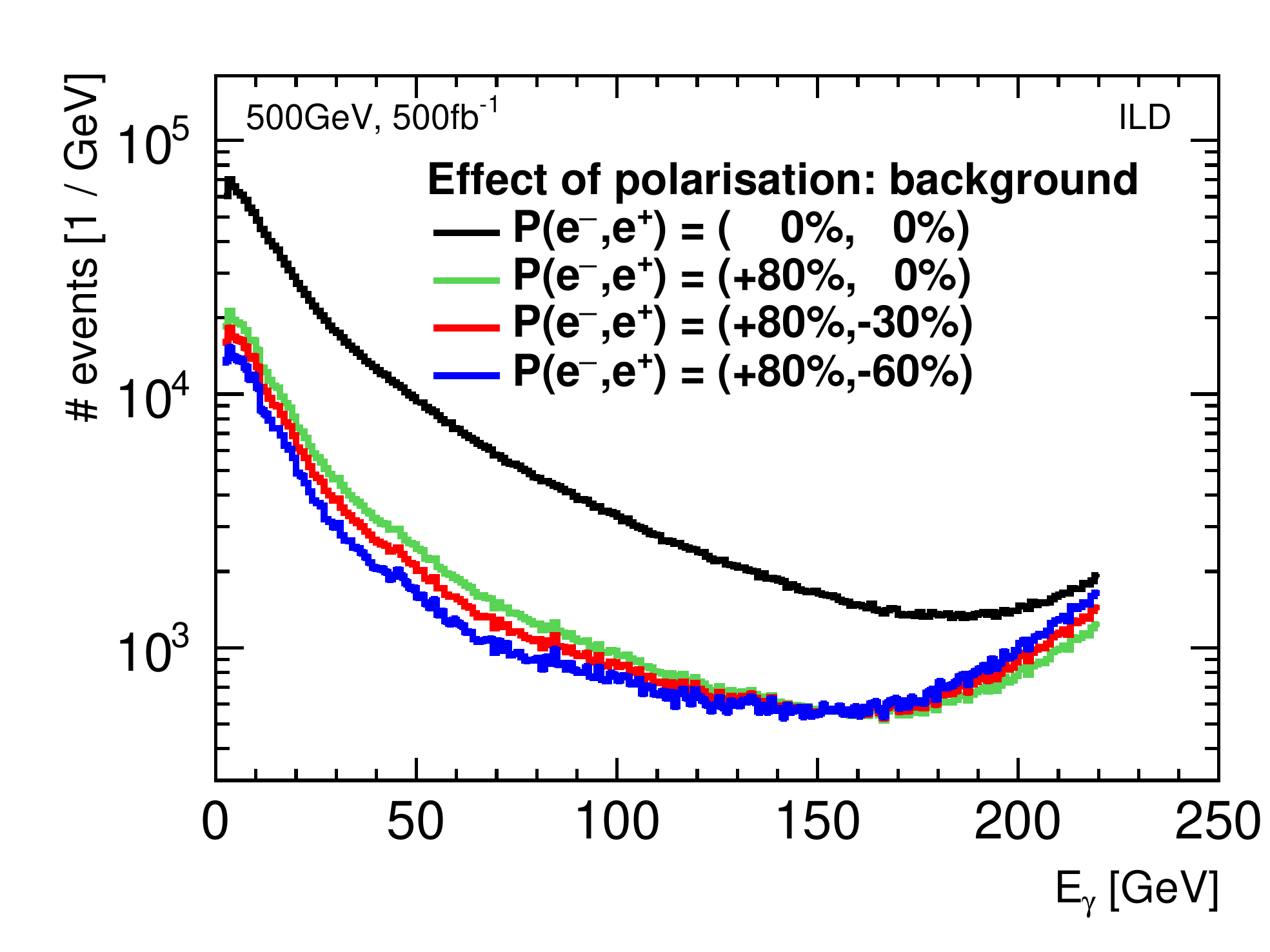} 
    \caption{\label{fig:bg_sig_pol:bg}} 
  \end{subfigure}
  \caption{(a) Effect of the beam polarisation on the sensitivity for the example of the vector operator at 500\,fb$^{-1}$. (b) The effect is by far dominated by the suppression of the $\nu\bar{\nu}\gamma$ background  in the case of a mainly right-handed electron beam.}
  \label{fig:bg_sig_pol}
\end{figure}

In Fig.~\ref{fig:bg_sig_pol:sens}, the expected sensitivity at 95\% confidence level for the example of the vector operator for four cases are compared: (1) both beams unpolarised, (2) only polarised electrons with $+$80\%, and additional positron polarisation with (3) $-30$\% and (4) $-60$\%. The integrated luminosity is 500\,fb$^{-1}$ in all cases. It shows that the reach in $\Lambda$ is increased by up to $60\%$ by choosing the optimal polarisation configuration. This is mostly due to the background suppression in the case of a mainly right-handed electron beam, as illustrated in Fig.~\ref{fig:bg_sig_pol:bg}. With left-handed positrons, the total number of background events can be further reduced, especially at lower energies. The effect of the beam polarisation on the signal cross-section is smaller and depends on the type of operator. Thus the polarised cross-sections can be used to characterise the WIMP-SM interaction in case a signal is observed, as has been studied e.g.\ in~\cite{CBDA1}.

\subsection{Effect of the systematic uncertainties} \label{sec:systematics_effect}

\begin{figure}[h]
 \begin{subfigure}{0.495\textwidth}
  \includegraphics[width=\textwidth]{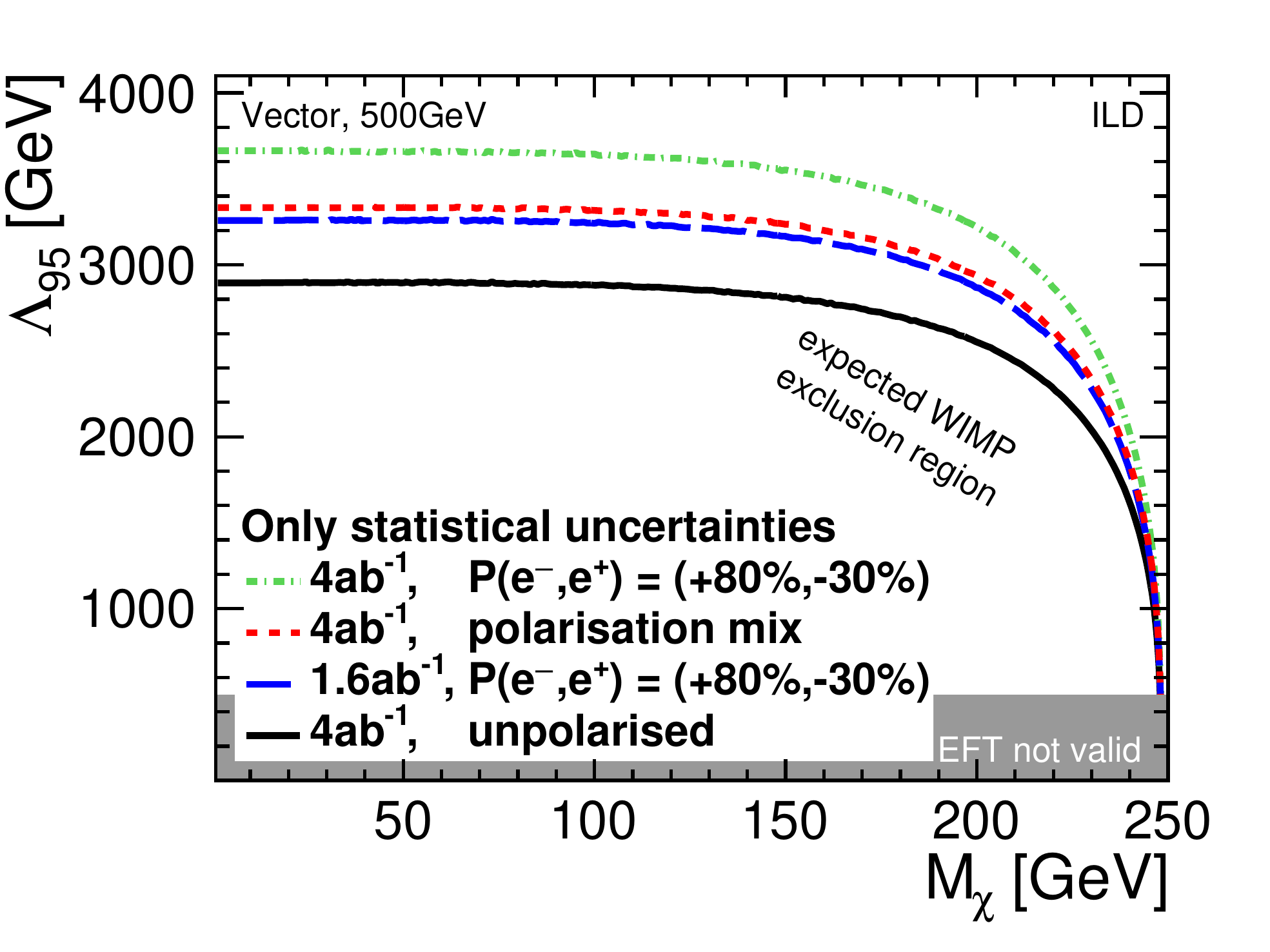} 
 \caption{\label{fig:systematics:without}}
 \end{subfigure}
 \begin{subfigure}{0.495\textwidth}
  \includegraphics[width=\textwidth]{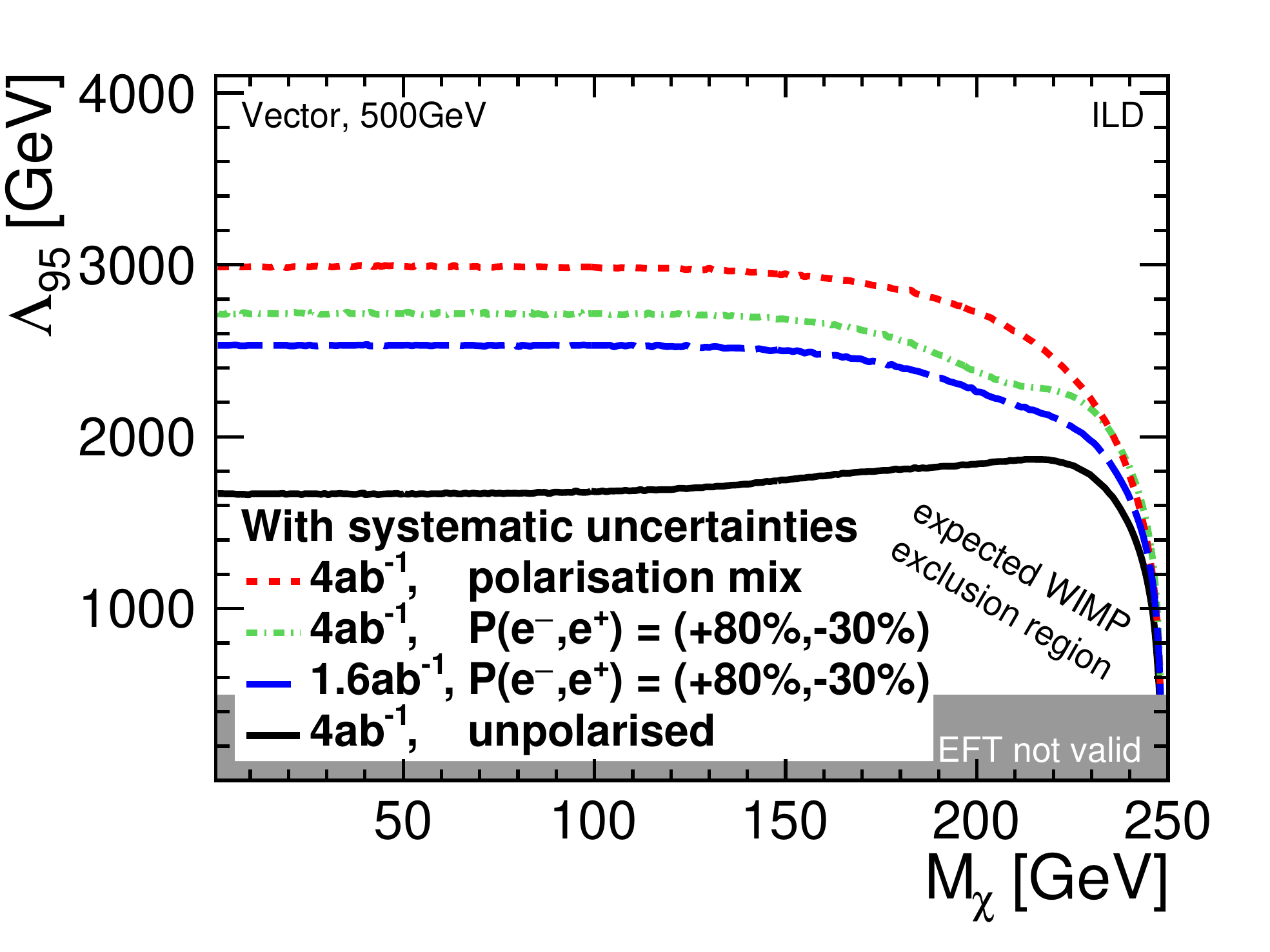} 
 \caption{\label{fig:systematics:with}}
 \end{subfigure}
 \caption{Expected exclusion limits (left) with only statistical uncertainties and (right) with systematic uncertainties taken into account.}
 \label{fig:systematics}
\end{figure}

In order to illustrate the impact of the systematic uncertainties, and the role of their correlation across data sets with different beam polarisations taken ``quasi-concurrently'' due to the fast helicity reversal, the expected sensitivity at 95\% confidence level for the vector operator has been calculated for different assumptions on the polarisation: the unpolarised case, the optimal beam polarisation alone, and the canonical polarisation mix of the H20 scenario (c.f.\ Sec.~\ref{sec:H20}). In Fig.~\ref{fig:systematics:without}, the results are compared when considering statistical uncertainties only, whereas, in Fig.~\ref{fig:systematics:with}, the same scenarios receive the full treatment of the systematic uncertainties.

For the case of statistical uncertainties only, the limits mainly depend on the number of signal and background events and hence the largest sensitivity is obtained when investing the full $4$\,ab$^{-1}$ into the configuration which suppresses the background most, namely $P(e^-,e^+)=(+80\%,-30\%)$, which is of course not a realistic scenario. But even for the $1.6$\,fb$^{-1}$ data set with $P(e^-,e^+)=(+80\%,-30\%)$ contained in the H20 scenario alone the performance is significantly better than for $4$\,ab$^{-1}$ of unpolarised data. The sensitivity of the H20 mix is dominated completely by its share with $P(e^-,e^+)=(+80\%,-30\%)$.
 
When the systematic uncertainties are considered, the picture changes completely. The smaller the signal-to-background ratio and the larger the data set (thus the smaller the statistical uncertainties are), the bigger the loss in sensitivity due to the systematic uncertainties. This means that the unpolarised data set suffers most and, especially for small WIMP masses, the probed reach in $\Lambda$ shrinks by nearly a factor of two. For the pure $P(e^-,e^+)=(+80\%,-30\%)$ data sets, the reach is reduced by $20-25\%$. The largest reach now is provided by the data set with the polarisation mix, which loses only $10\%$ in reach and outperforms now even a data set with the same integrated luminosity and the optimal polarisation combination alone. It can thus be concluded that beam polarisation is important to provide separate data sets with different helicity combinations, resulting in different signal-to-noise ratios, where data sets with a suppressed signal production act as a kind of control sample which separates the effect of systematic uncertainties from the presence of a signal. 

The mass dependence of the limit curve also differs from case to case. For higher masses, there is a signal-free control region at high photon energies. In the unpolarised case, the sensitivity to higher WIMP masses is therefore better than to lower masses. Since the background at highest photon energies rises with polarisation, as can be seen in Fig.~\ref{fig:bg_sig_pol:bg}, the effect of the control region is diminished. 

\begin{figure}[htb]
  \begin{center}
  \includegraphics[width=0.5\textwidth]{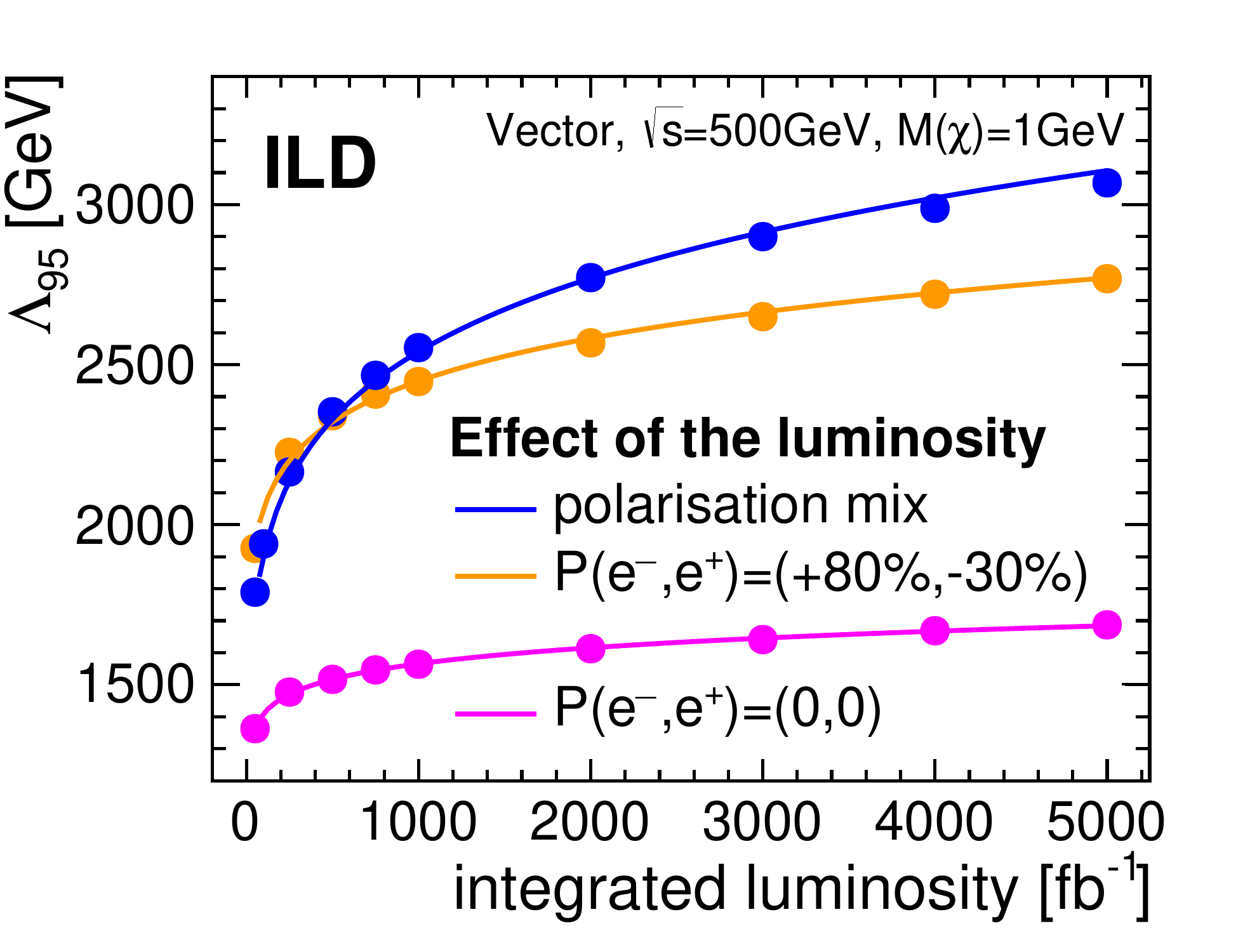}  
  \end{center}
  \caption{Scan of the expected exclusion limits for the vector operator and a WIMP mass of 1\,GeV over different values of the integrated luminosity for different WIMP masses and polarisation configurations. For the corresponding results for the axial-vector and scalar operator see~\cite{thesis} Fig.~7.10.}
  \label{fig:lumi_scan}
\end{figure}

In order to understand better how the reachable energy scale $\Lambda$ evolves with the integrated luminosity  depending on the beam polarisation configuration, the different scenarios are tested with integrated luminosities between 50\,fb$^{-1}$ to 5\,ab$^{-1}$. The evolution of the sensitivities for the vector operator and WIMP masses of 1\,GeV is shown in Fig.~\ref{fig:lumi_scan}. In Table~\ref{tab:lumi_fits} the fitted proportionalities of the sensitivities to the integrated luminosities are shown for all three operators and WIMP masses of $1$ and $200$\,GeV. By far the worst results would be obtained for the pessimistic assumptions of having no beam polarisation. At very low integrated luminosities, the polarisation mix is worse than the unrealistic assumption to operate at the optimal polarisation configuration alone. But already at about 500\,fb$^{-1}$, the lines intersect and, for larger data sets, the polarisation mix is clearly better. So, if the limits are based on several input parameters, the rise is significantly steeper and almost independent of WIMP mass and operator. Here the effect of the systematic uncertainties is significantly smaller and these proportionalities are close to the expectations for statistical uncertainties only ($\Lambda\propto\mathcal{L}^{1/8}$).

\begin{table}[htb]
\begin{center}
\begin{tabular}[h]{l l l l l l l}
\hline
& \multicolumn{2}{c}{vector} & \multicolumn{2}{c}{axial-vector} & \multicolumn{2}{c}{scalar} \\
\hline
$P(e^-,e^+)$ & $M_\chi=1$\,GeV & 200\,GeV & 1\,GeV & 200\,GeV & 1\,GeV & 200\,GeV \\\hline
$(0$\%$,0$\%$)$ & $\Lambda\propto \mathcal{L}^{1/22}$ & $\mathcal{L}^{1/12}$ & $\mathcal{L}^{1/22}$ & $\mathcal{L}^{1/10}$ & $\mathcal{L}^{1/23}$ & $\mathcal{L}^{1/10}$ \\
$(+80$\%$,-30$\%$)$ & $\Lambda\propto \mathcal{L}^{1/13}$ & $\mathcal{L}^{1/16}$ & $\mathcal{L}^{1/13}$ & $\mathcal{L}^{1/15}$ & $\mathcal{L}^{1/13}$ & $\mathcal{L}^{1/14}$ \\
H20 & $\Lambda\propto \mathcal{L}^{1/8}$ & $\mathcal{L}^{1/9}$ & $\mathcal{L}^{1/9}$ & $\mathcal{L}^{1/9}$ & $\mathcal{L}^{1/9}$ & $\mathcal{L}^{1/9}$ \\
\hline
\end{tabular}
\caption{Proportionalities of the expected exclusion limits $\Lambda$ to the integrated luminosity $\mathcal{L}$ for the different operators, WIMP masses and polarisation configurations.}
\label{tab:lumi_fits}
\end{center}
\end{table}

\subsection{Detector effects} \label{sec:detector_effects}

Two aspects of the detector performance which could influence the mono-photon WIMP search are the hermeticity of the detector and the photon reconstruction in the electromagnetic calorimeter. 

The impact of the ECAL resolution is tested by smearing the generated energy with the unrealistically optimal energy resolution of $\sigma_E/E = 1$\%$/\sqrt{E}$. The result is not modified in any significant way with respect to the full simulation case, so it can be concluded that the design of ECAL is sufficient for the WIMP study.

\begin{figure}[ht]
	    \begin{subfigure}{0.495\textwidth}
	      \includegraphics[width=\textwidth]{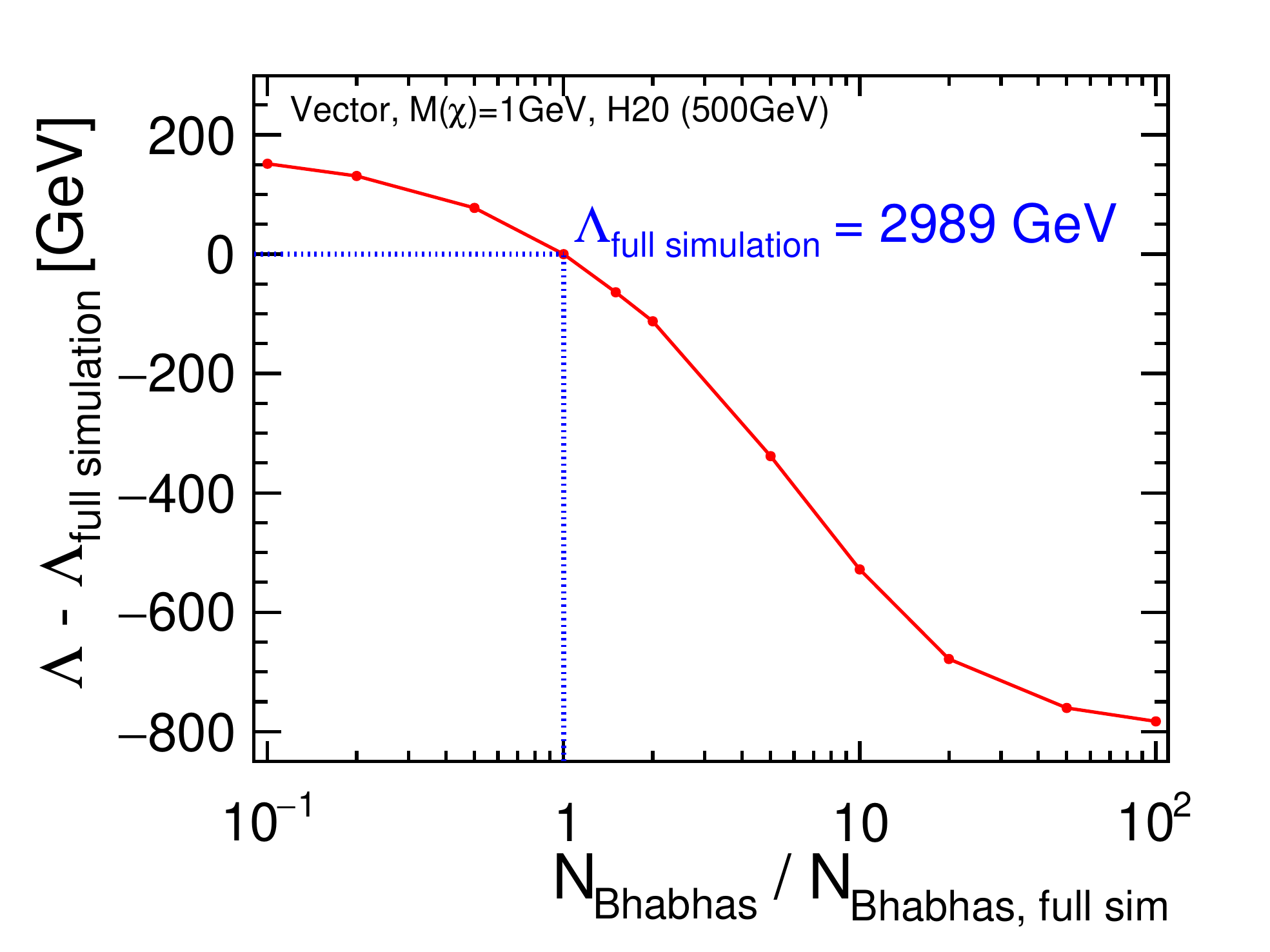} 
               \caption{\label{fig:lambda_herm:Nbh}}
	    \end{subfigure}
	    \begin{subfigure}{0.495\textwidth}
	      \includegraphics[width=\textwidth]{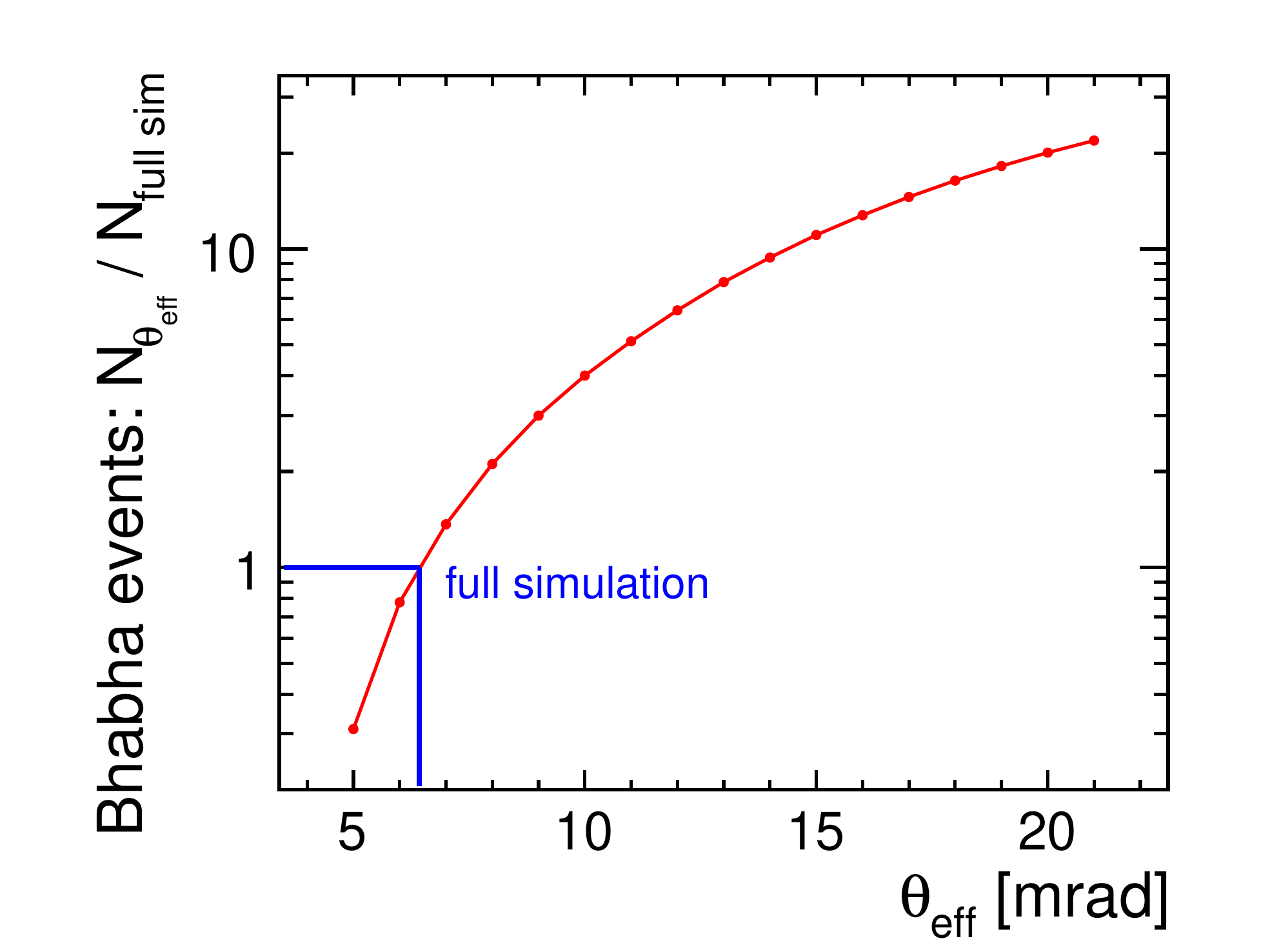} 
               \caption{\label{fig:lambda_herm:theta_eff} }
	    \end{subfigure}
	      \caption{(a) $\Lambda_{95}$ as a function of the amount of remaining Bhabha scattering background, for the example of a vector operator and a WIMP mass of $1$\,GeV with the expected exclusion limit from Fig.~\ref{fig:H20_500} as reference point. (b) Expected $\Lambda_{95}$ as a function of the lowest polar angle under which high energetic electrons from Bhabha scattering can be efficiently detected, $\theta_{\mathrm{eff}}$.}   
              \label{fig:lambda_herm}
	   \end{figure}

The situation is very different in case of the hermeticity, where we find a strong dependence on the lowest polar angle under which high energetic electrons from Bhabha scattering can be efficiently detected. 
In order to give a quantitative estimate on how the Bhabha scattering background influences the reachable sensitivity, the number of background events is scaled up or down in the sensitivity calculation. In Fig.~\ref{fig:lambda_herm:Nbh}, it is shown how the expected exclusion limit changes with respect to the limit for the full simulation as a function of this modified background level. With a better background rejection, the improvement would be moderate, i.e.~less than 10\% for 10 times less Bhabha background. If, on the other hand, more background events survived the event selection, the testable energy scale would drop by about $600$\,GeV for a 10 times higher remaining background. 

Based on the predicted angular spectrum of radiative Bhabha scattering, the number of remaining events can be translated into the lowest polar angle $\theta_{\mathrm{eff}}$ under which the detector has to be able to tag high energetic electrons in order to reach a certain background level. The result is shown in Fig.~\ref{fig:lambda_herm:theta_eff}. For the ILD detector and in the presence of the $e^+e^-$ pair background expected from beamstrahlung, this angle is about $\theta_{\mathrm{eff}} = 7$\,mrad.\footnote{Despite the pair background, this is remarkably close to the inner rim of the BeamCal at $5.6$\,mrad, c.f.\ Sec.~\ref{sec:beamcalreco}.}

The detector concepts proposed for future circular colliders do not foresee instrumentation so close to the beam pipe, since their final focus quadrupoles have to be inside the main detector in order to achieve the high luminosities. The acceptance typically starts with the luminosity calorimeters at $\theta = 20$\,mrad (CEPC~\cite{CEPCStudyGroup:2018ghi}) or even $\theta = 30$\,mrad (FCC-ee~\cite{Abada:2019zxq}). For $\theta_{\mathrm{eff}} = 20$\,mrad, the expected $\Lambda_{95}$ would be reduced by about $700$\,GeV.

\section{Effect of the centre-of-mass energy} 
\label{sec:otherCOM}
Even though the full simulation is performed at a \com energy of $500$\,GeV, approximate results can be given for other energies. In Sec.~\ref{sec:otherCOM1} and~\ref{sec:otherCOM2} two different approaches are presented: In the first approach the signal and background photon spectra are modified to correspond to other ILC \com energies, and then the full sensitivity calculation procedure is applied. Since this approach is CPU-time consuming and limited to \com energies lower than the $500$\, GeV of the full simulation analysis, the second approach is based on an extrapolation directly in terms of $\Lambda$. In Sec.~\ref{sec:other_colliders}, approximate results for other planned lepton colliders are compared to the sensitivity of the ILC.

\subsection{Adaptation of photon spectra to other \com energies}\label{sec:otherCOM1}

With a few fairly simple modifications, approximate signal and background spectra for other \com energies can be obtained from the photon energy distribution of the full simulation study at $500$\,GeV (i.e.~Fig.~\ref{fig:Egamma:final}). Based on these, the sensitivity can be computed for different WIMP masses including the full treatment of the systematic uncertainties. As this requires to calculate the CPU-intensive reweighting of neutrino to WIMP events, only a limited number of configurations can be tested. This approach can be applied to \com energies smaller than the $500$\,GeV of the full simulation in order to have data available over the full range of accessible photon energies.

For the {\em signal} photon energy spectrum, the parameter $\sqrt{s'}$ in the procedure to reweight neutrino to WIMP events  (c.f.\ Sec.~\ref{sec:sigrew}) is adjusted with the ratio of the tested \com energy and the \com energy of the full simulation ($500$\,GeV). Due to the energy-dependence of the cross-section, the signal height decreases noticeably with the centre-of-mass energy.
The shape of the {\em background} photon energy spectrum is approximated by rescaling the energy axis with the ratio of the \com energies. The overall height is modified by scaling the distribution with the cross-sections at the tested \com energy and the one at $500$\,GeV, which are calculated with Whizard using a simplified signal definition at the generator level. 
For simplicity, the effect of the systematic uncertainties is assumed to be the same as for the full statistics at $500$\,GeV. Even though the effect of the luminosity spectrum is expected to be smaller for lower energies, the effect of this change on the final result is expected to be negligible within the accuracy of this setup.

\begin{figure}[htb]
\begin{center}	     \includegraphics[width=0.6\textwidth]{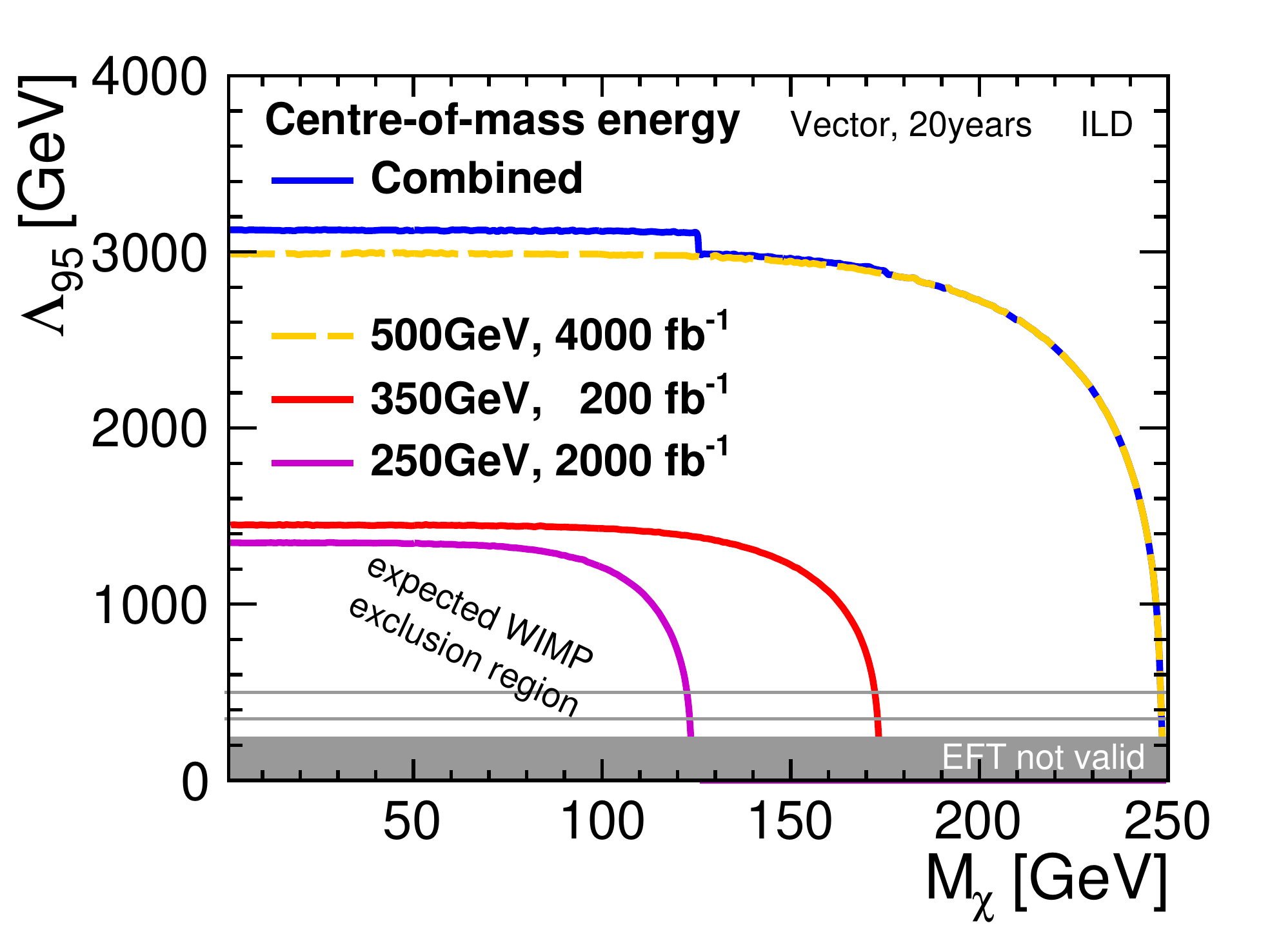} 
\end{center}
\caption{Expected sensitivity at 95\% confidence level for the vector operator at \com energies of $250$, $350$ and $500$\,GeV, assuming  the integrated luminosities and polarisation configurations of the H20 running scenario. The corresponding results for axial-vector and scalar operators can be found in~\cite{thesis}, Fig.~8.6 and~8.7.}
\label{fig:otherCOM_mass_vector}
\end{figure}

In Fig.~\ref{fig:otherCOM_mass_vector}, the results for $250$ and $350$\,GeV, i.e.~the other two energy stages in the H20 running scenario (c.f.~Sec.~\ref{sec:H20}), are shown together with the expected exclusion limit at $500$\,GeV from the full simulation study. In general, WIMP masses up to approximately half the \com energy can be tested. The probed energy scales $\Lambda$ also depend strongly on the \com energy: The small integrated luminosity of $200$\,fb$^{-1}$ at $350$\,GeV is still sufficient to surpass the reach of a ten times larger data set  ($2$\,ab$^{-1}$) at 250\,GeV. This shows that the rise of $\Lambda$ with the \com energy is much faster than with the integrated luminosity. For lower WIMP masses, the combination of the three limits shows a small improvement of about $5\%$ over the result at $500$\,GeV.

\subsection{Extrapolation of the sensitivity to other \com energies}\label{sec:otherCOM2}

As a faster alternative an approximate formula of the energy scale $\Lambda$ which can be probed for a fixed WIMP mass as a function of the \com energy and the integrated luminosity has been derived in Sec.~8.4 of~\cite{thesis}. This approach can also be used to extrapolate to higher \com energies.
This method is based on an approximate energy-dependence of the fiducial background cross-sections calculated with  \textsc{Whizard}~\cite{Kilian:2007gr}, and the assumption that the {\em energy}-dependence of $\Lambda_{95}$ is dominated by the change in signal and background statistics. It has been verified explicitly in~\cite{thesis} that for constant luminosity, this assumption holds to a very good precision. As shown in Sec.~\ref{sec:systematics_effect}, however, the systematic uncertainties strongly affect the dependence of $\Lambda_{95}$ on the integrated luminosity. Therefore, the {\em luminosity}-dependence is taken as determined in the full simulation analysis, given in Table~\ref{tab:lumi_fits}.
 
\begin{figure}[ht]
\begin{subfigure}{0.495\textwidth}  
  \includegraphics[width=\textwidth]{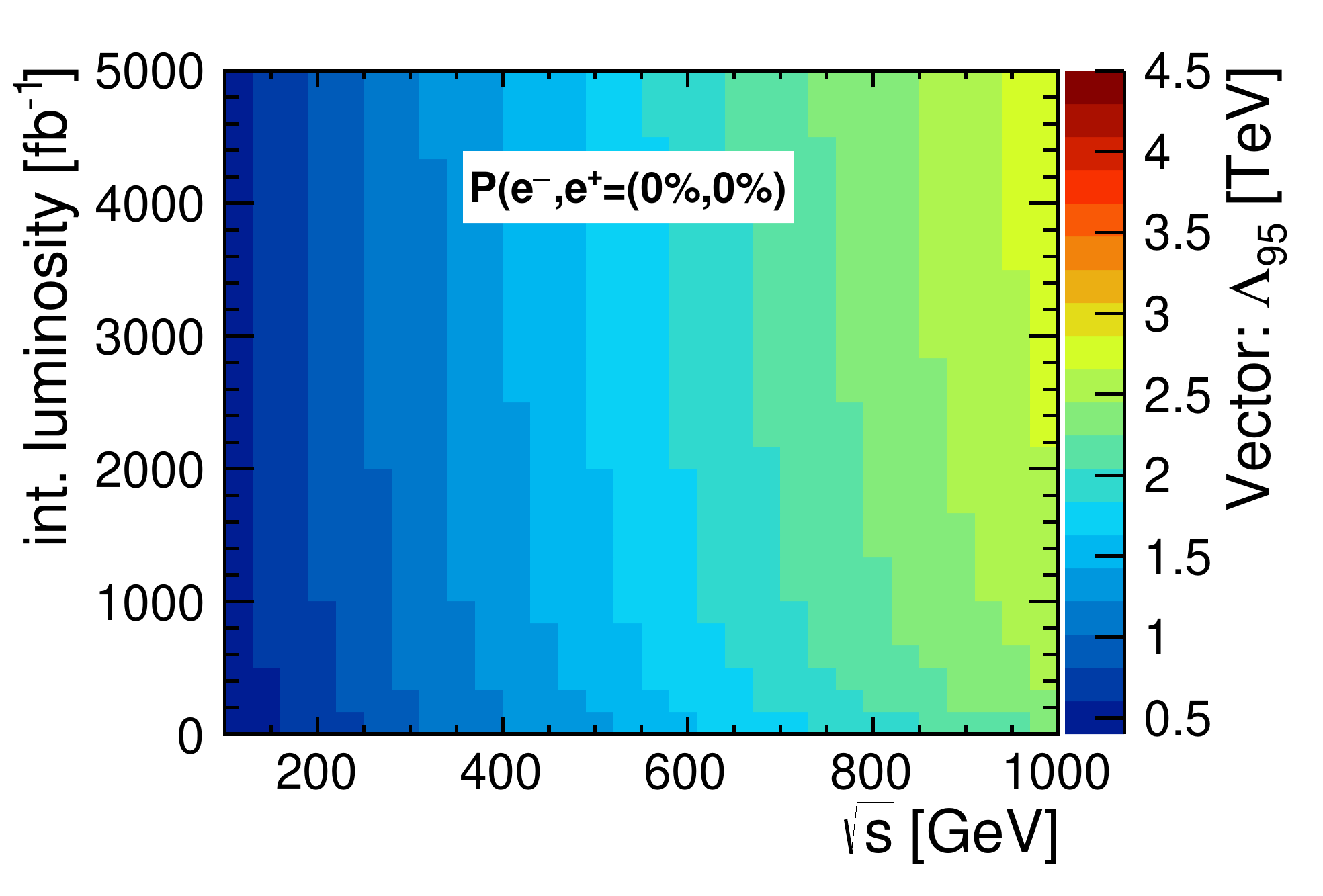} 
   \caption{\label{fig:vector_extrapolation:unpol} Vector, $P(e^-,e^+)=(0,0)$ }
\end{subfigure}
\begin{subfigure}{0.495\textwidth}
  \includegraphics[width=\textwidth]{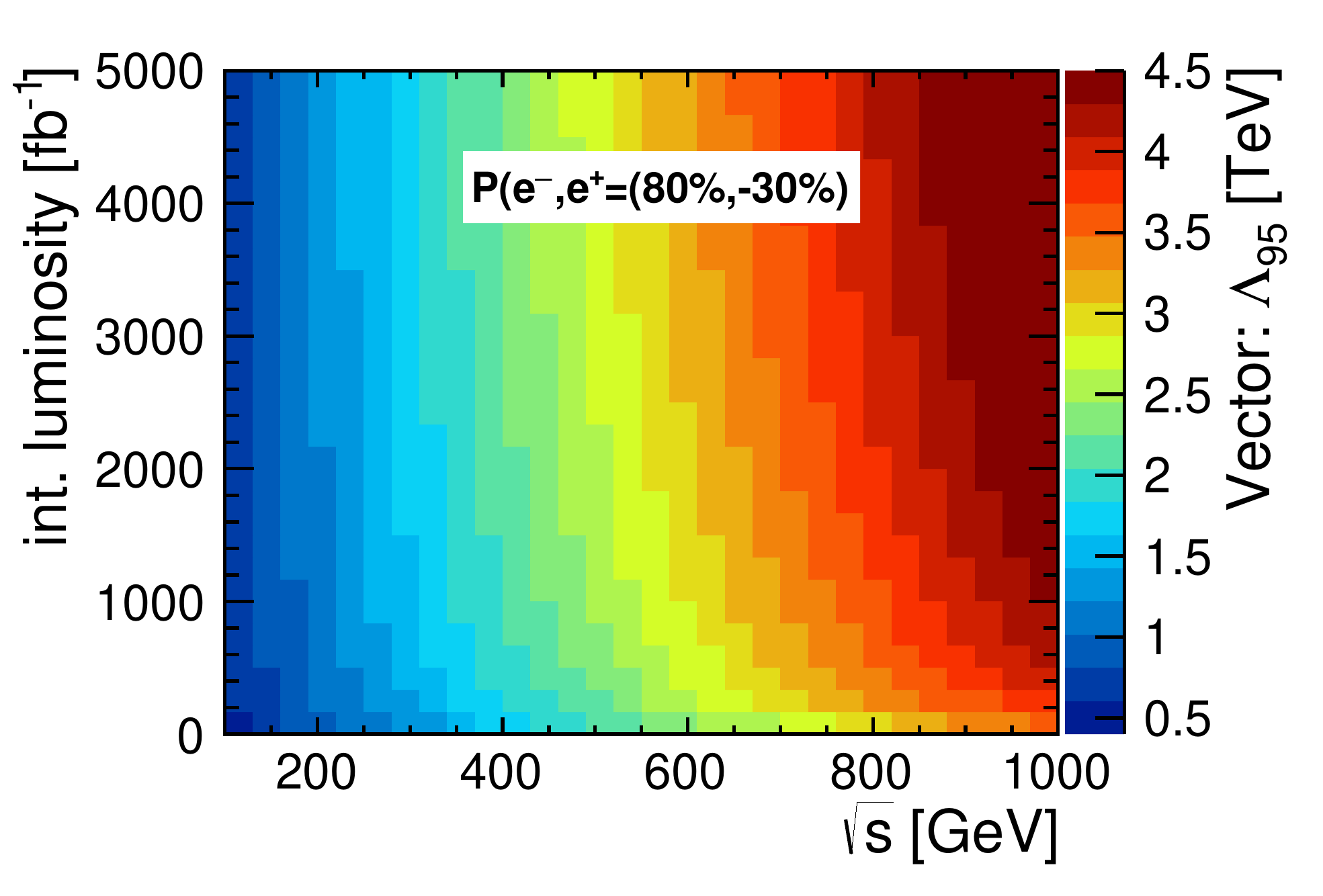} 
   \caption{\label{fig:vector_extrapolation:pol} Vector, $P(e^-,e^+)=(+80\%,-30\%)$ }
\end{subfigure}
\begin{subfigure}{0.495\textwidth}  
  \includegraphics[width=\textwidth]{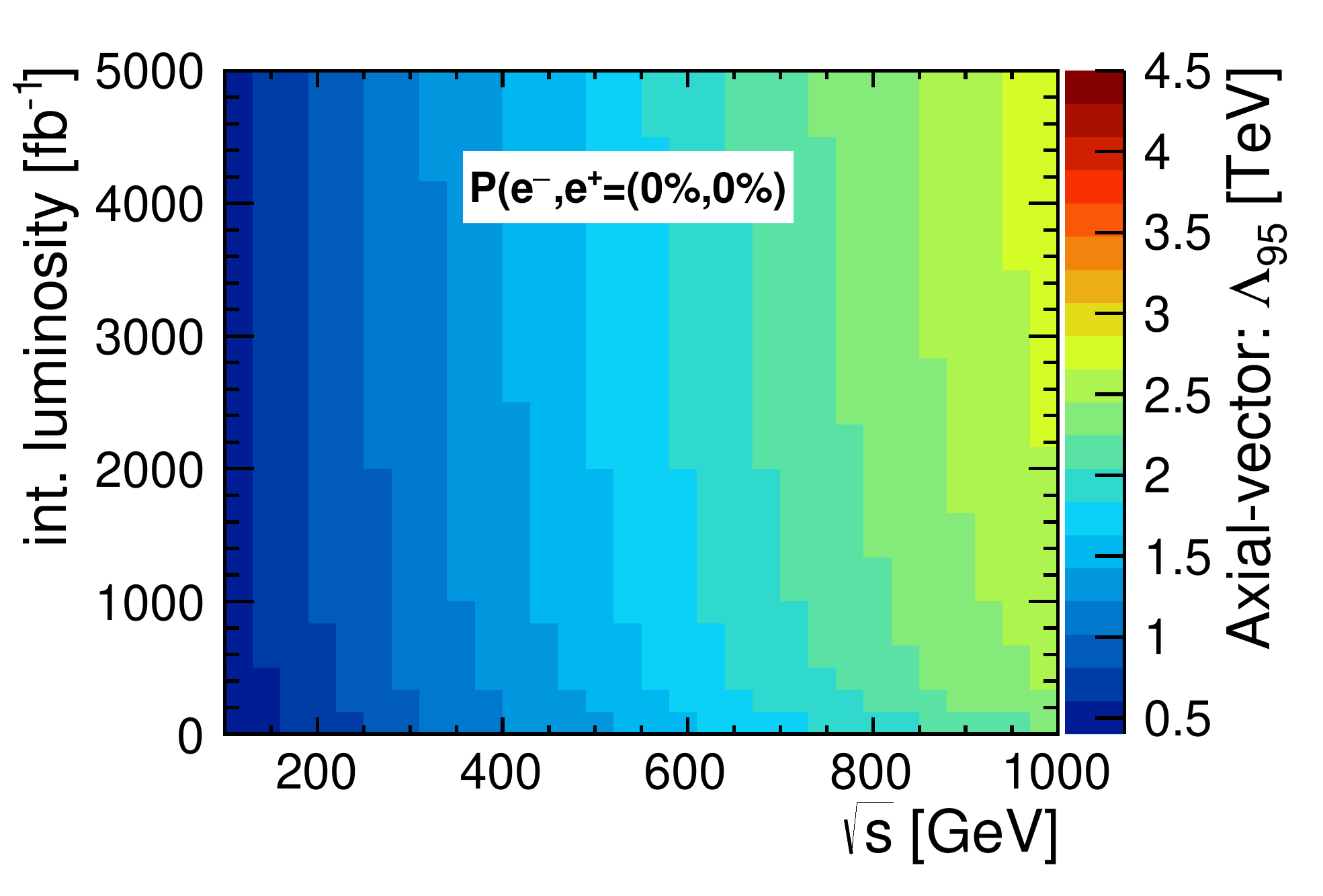} 
   \caption{\label{fig:axial_vector_extrapolation:unpol} Axial-vector, $P(e^-,e^+)=(0,0)$ }
\end{subfigure}
\begin{subfigure}{0.495\textwidth}
  \includegraphics[width=\textwidth]{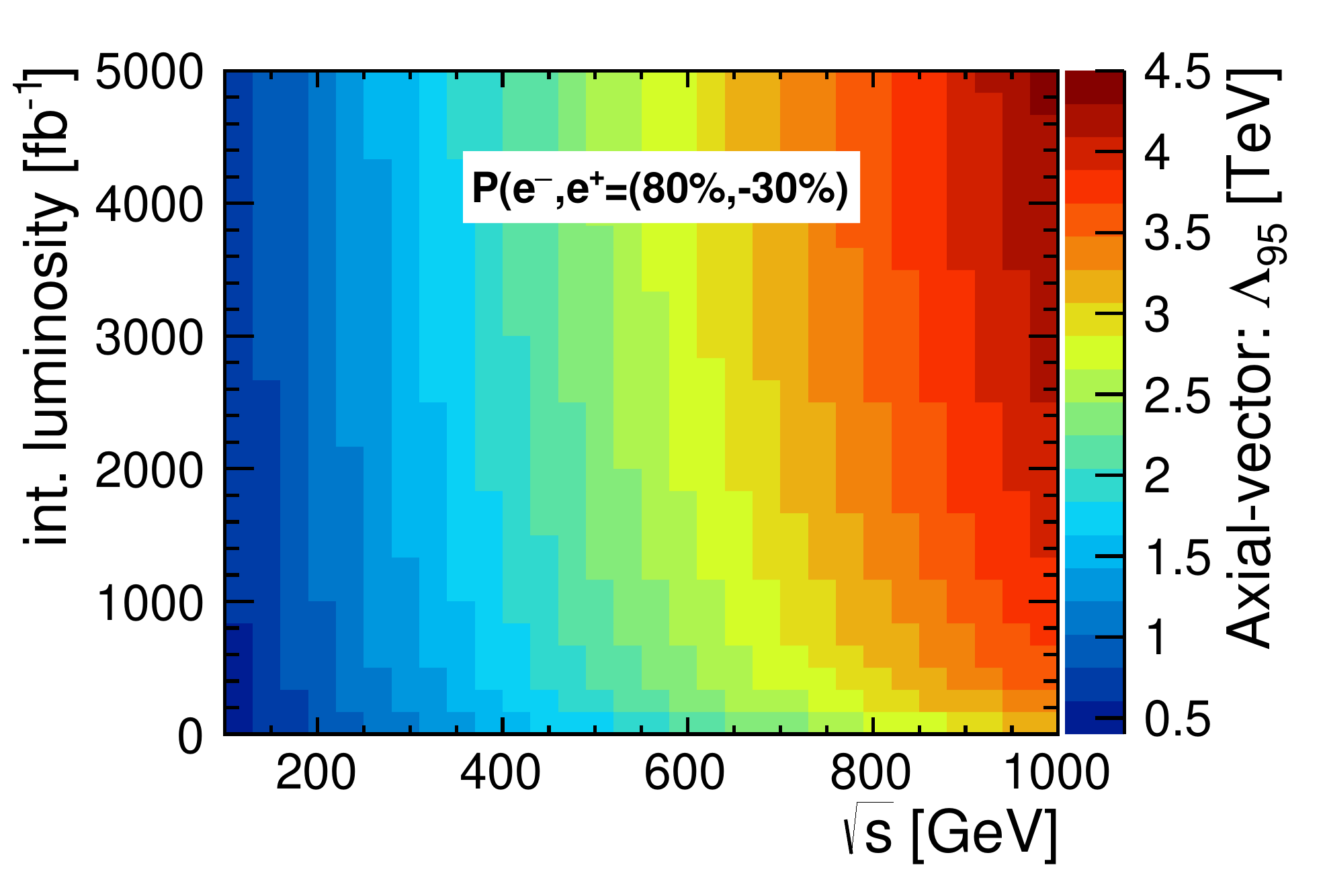} 
   \caption{\label{fig:axial_vector_extrapolation:pol} Axial-vector, $P(e^-,e^+)=(+80\%,-30\%)$ }
\end{subfigure}
\caption{Expected exclusion limits at 95\% confidence level obtained with the extrapolation formalism (a,b) for the vector operator and (c,d) for the axial-vector operator for two different beam polarisations.}
\label{fig:vector_extrapolation}
\end{figure}

While being very fast, this procedure is limited to one data set at a time, i.e.~with one polarisation combination and cross-section, and to low WIMP masses where the limits show no variation with mass. In Fig.~\ref{fig:vector_extrapolation}, such a sensitivity scan is shown for the pessimistic assumption of unpolarised beams and for the optimal polarisation configuration $P(e^-,e^+)=(+80\%,-30\%)$, for a WIMP mass of $1$\,GeV. This two-dimensional presentation clearly confirms that a higher \com energy is more beneficial than collecting large amounts of data at a lower \com energy.

With this extrapolation formalism, approximate results for a \com of $1$\,TeV, i.e.~for an upgrade of the ILC, can be given for the first time. The canonical integrated luminosity for $1$\,TeV is $8$\,ab$^{-1}$, with the same sharing between polarisation configurations as for $500$\,GeV~\cite{Barklow:2015tja}, thus $3.2$\,ab$^{-1}$ will be collected with the most favourable sign configuration of $P(e^-,e^+)=(+80\%,-20\%)$\footnote{The effect of the slightly lower positron polarisation of $20\%$ can safely be neglected here, see e.g.\ Fig.~\ref{fig:bg_sig_pol:bg}.}. Considering only this most powerful part of the data, the expected sensitivity at 95\% confidence level for production of low mass WIMPs mediated by a vector (axial-vector) operator is $\Lambda_{95}=4760$\,GeV ($\Lambda_{95}=4220$\,GeV). Comparing this e.g.\ to the case of collecting the full $8$\,ab$^{-1}$ without any beam polarisation, which would yield  $\Lambda_{95}=2850$\,GeV ($\Lambda_{95}=2850$\,GeV) for the vector (axial-vector) case, this shows again the potential of polarised beams.

\subsection{Comparison to other proposed lepton colliders} \label{sec:other_colliders}

The extrapolation techniques introduced in the previous subsections can also be used to address the question of how the reach of the mono-photon search compares for the different luminosities, energies, and polarisation settings offered by the currently proposed lepton colliders. We neglect the impact of differences in the detector acceptance discussed in Sec.~\ref{sec:detector_effects}, as well as differences in the luminosity spectrum (c.f.~Sec.~\ref{sec:systematics}). The systematic uncertainty associated with the ILC beam energy spectrum, which would not apply to circular colliders, reduces the expected sensitivity at 95\% confidence level by about $260$\,GeV in the full simulation study~\cite{thesis}. On the other hand, the effect of a worse detector acceptance due to final focus magnets inside the main detector volume as required for circular colliders amounts to about $700$\,GeV. Ignoring both effects in our extrapolations is thus a generous approach towards circular colliders.

\begin{figure}[htb]
\begin{center}
  \includegraphics[width=0.5\textwidth]{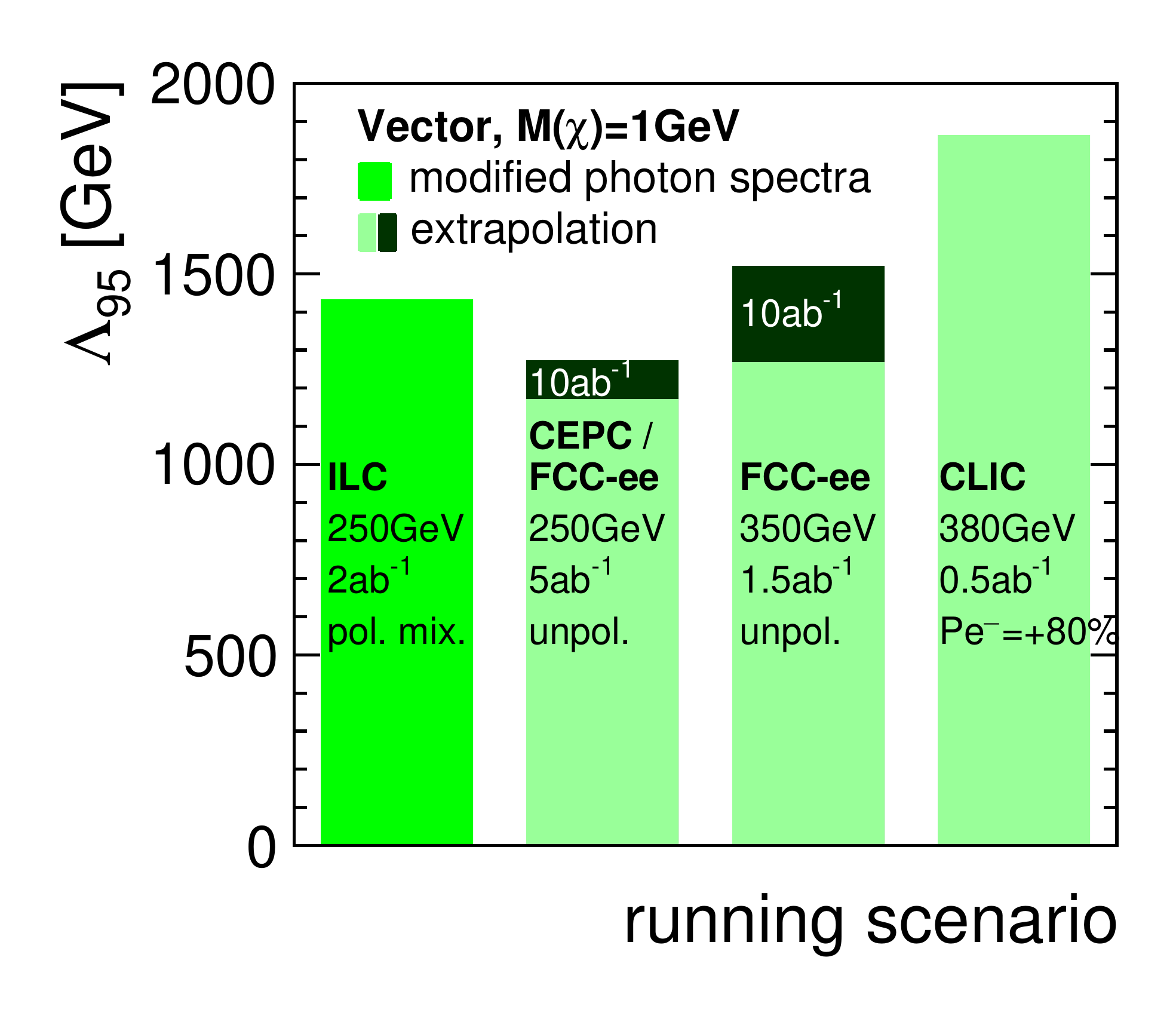} 
\end{center} 
 \caption{Expected sensitivities at 95\% confidence level for $M_\chi = 1$\,GeV, the vector operator and different settings for \com energy, integrated luminosity and polarisation combination.}
 \label{fig:manhattan_first_stages}
\end{figure}

\begin{figure}[htb]
\begin{center}
  \includegraphics[width=0.8\textwidth]{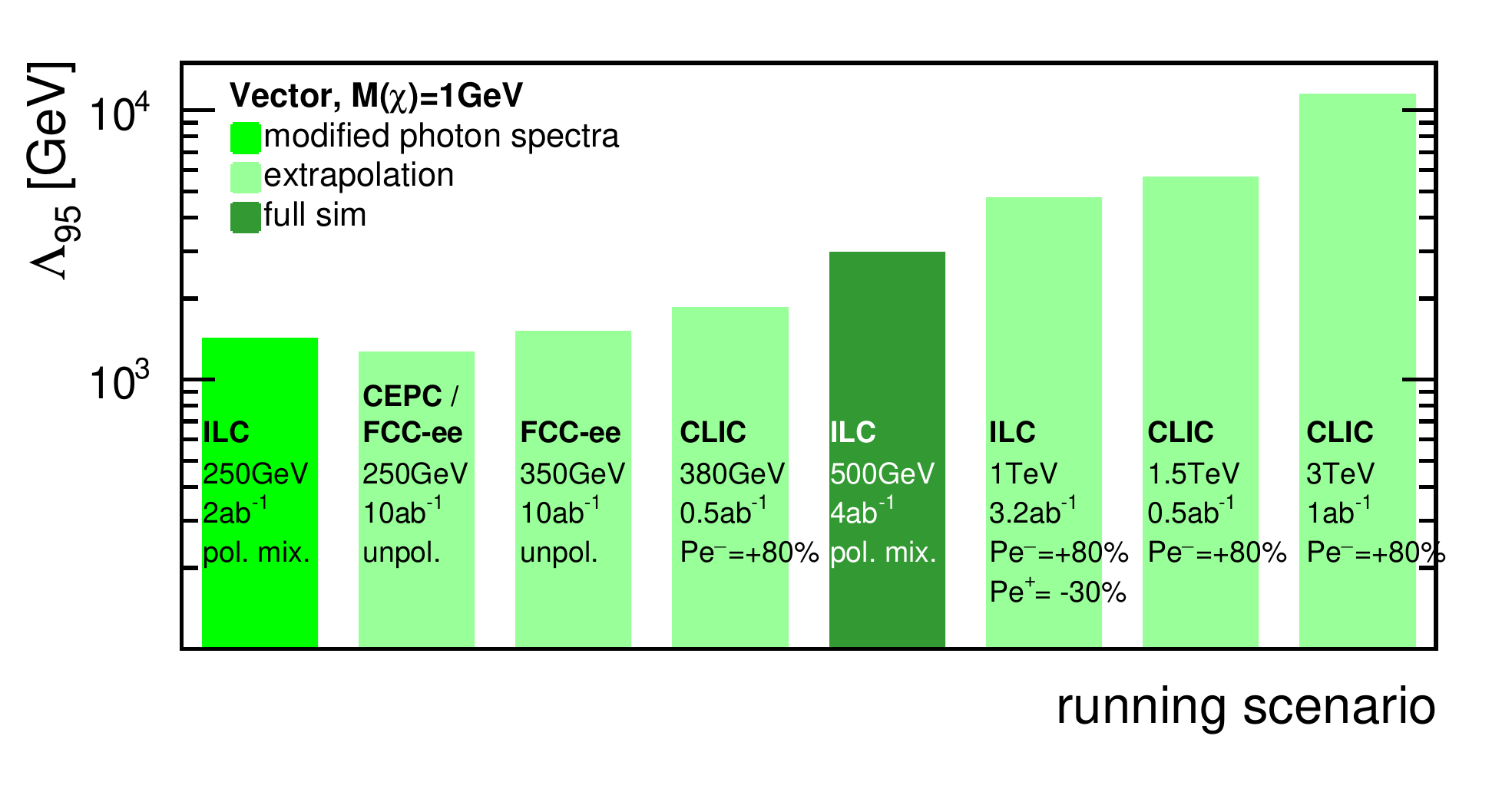} 
\end{center} 
 \caption{Expected sensitivities at 95\% confidence level for $M_\chi = 1$\,GeV, the vector operator and different settings for \com energy, integrated luminosity and polarisation combination. }
 \label{fig:manhattan}
\end{figure}

In Fig.~\ref{fig:manhattan_first_stages} and~\ref{fig:manhattan}, the expected sensitivities at 95\% confidence level for the vector operator and small WIMP masses are compared for different running scenarios. In Fig.~\ref{fig:manhattan_first_stages}, the sensitivities of the anticipated first stages of the proposed experiments are shown (in linear scale) which are complemented by approximate results for higher \com energies in Fig.~\ref{fig:manhattan} (here shown in logarithmic scale).

The ILC result for the 250\,GeV stage corresponds to the lowest line in Fig.~\ref{fig:otherCOM_mass_vector}, i.e.\ it is obtained using the first approach, as presented in Sec.~\ref{sec:otherCOM1}. The sensitivity of the other planned lepton colliders is approximated with the second approach, see Sec.~\ref{sec:otherCOM2}. The second and third bars give estimates inspired by the case of circular colliders, CEPC~\cite{CEPCStudyGroup:2018ghi} and FCC-ee~\cite{Abada:2019zxq}, all without beam polarisation. None of these configurations with integrated luminosities of $5$\,ab$^{-1}$ at $250$\,GeV and $1.5$\,ab$^{-1}$ at 350\,GeV surpasses the sensitivity of the ILC's $2$\,ab$^{-1}$ of polarised data at $250$\,GeV, which would require $10$\,ab$^{-1}$ at $350$\,GeV to be reached with unpolarised beams. The polarisation is not only essential to reduce SM backgrounds, but also (as discussed in Sec.~\ref{sec:systematics_effect}) to control systematic uncertainties, which becomes more and more important with increasing size of the data set. The limited increase with higher integrated luminosities for unpolarised beams is also illustrated in Fig.~\ref{fig:lumi_scan}. The CLIC result at 380\,GeV with only 0.5\,ab$^{-1}$ shows again that a higher \com energy is favored over high integrated luminosities. This conclusion is further confirmed by the extension to higher \com energies in Fig.~\ref{fig:manhattan}. For the ILC setup at 1\,TeV and the CLIC configurations, only the data sets for $P(e^-)=(+80\%)$ (and in the case of the ILC $P(e^+)=(-20\%)$) are taken into account, as they dominate the sensitivity.

\section{Conclusions} \label{sec:conclusion}

The expected sensitivity of the ILC to WIMP dark matter was studied based on pair production of WIMPs in association with an ISR photon. In a full simulation of the ILD detector concept, the expected discovery and exclusion reach were evaluated for WIMP masses up to $250$\,GeV at a \com energy of $500$\,GeV. Dedicated data sets of the most important Standard Model background processes, namely neutrino pair production and Bhabha scattering, were produced with up to 4 detectable photons in the matrix element. This is the first WIMP study at the ILC with a thorough treatment of the systematic uncertainties and their correlations. 

The results are expressed in terms of the energy scale of the new interaction $\Lambda$ in the framework of effective operators. With an integrated luminosity of $4$\,ab$^{-1}$ shared between all four polarisation sign combinations, up to $\Lambda_{95}=3$\,TeV can be probed at the $95\%$ confidence level and a signal with $\Lambda_{5\sigma}=1$\,TeV could be discovered if the WIMP production is mediated by a vector operator. Polarised beams are essential to reduce the Standard Model background from neutrino pair production and, depending on the tested effective operator, can increase the signal, which leads to an improved sensitivity of $50\%$ or more. In addition, the impact of systematic uncertainties can be reduced significantly when combining data with different polarisation settings. In absence of polarisation, these benefits cannot be easily compensated for by increasing the luminosity. 
Furthermore, the hermeticity of the detector in the very forward region is crucial for this type of analysis. Results degrade by about $25\%$ if the detector acceptance starts only at $20$\,mrad rather than at $6-7$\,mrad as in case of the ILD detector.

The results of the full detector simulation study at 500\,GeV have been
extrapolated to other centre-of-mass energies. This shows that already
the first stage of the ILC at $\sqrt{s}=250$\,GeV can probe new physics scales of $\Lambda_{95}$ of up to $\approx 1.4$\,TeV with $2$\,ab$^{-1}$ of polarised data. With the upgrade to $\sqrt{s}=1$\,TeV, already a data set of $3.2$\,ab$^{-1}$ with $P(e^-,e^+)=(-80\%,+30\%)$ can probe scales of up to $\Lambda_{95}=4.8$\,TeV and $4.2$\,TeV for vector and axial vector operators, respectively.

This study demonstrates that the ILD concept meets the key requirements of a WIMP search in the mono-photon channel. It also shows the decisive role of beam polarisation and of higher \com energies for this aspect of the physics programme of future lepton colliders. 

\section{Acknowledgements}
This work was supported by the Deutsche Forschungsgemeinschaft (DFG) through the Collaborative Research Centre SFB 676 ``Particles, Strings and the Early Universe'', project B1.
We would like to thank the LCC generator working group and the ILD software working group for providing the simulation and reconstruction tools and producing the Monte Carlo samples used in this study.
This work has benefited from computing services provided by the ILC Virtual Organization, supported by the national resource providers of the EGI Federation and the Open Science GRID,
and of those of the German National Analysis Facility (NAF).

\printbibliography

@article{Abada:2019zxq,
      author         = "Abada, A. and others",
      title          = "{FCC-ee: The Lepton Collider}",
      collaboration  = "FCC",
      journal        = "Eur. Phys. J. ST",
      volume         = "228",
      year           = "2019",
      number         = "2",
      pages          = "261-623",
      doi            = "10.1140/epjst/e2019-900045-4",
      reportNumber   = "CERN-ACC-2018-0057",
      SLACcitation   = "%%CITATION = 00619,228,261;%%"
}

@article{CEPCStudyGroup:2018ghi,
    author = "Dong, Mingyi and others",
    editor = "Guimarães da Costa, João Barreiro and others",
    archivePrefix = "arXiv",
    collaboration = "CEPC Study Group",
    eprint = "1811.10545",
    primaryClass = "hep-ex",
    reportNumber = "IHEP-CEPC-DR-2018-02, IHEP-EP-2018-01, IHEP-TH-2018-01",
    title = "CEPC Conceptual Design Report: Volume 2 - Physics \& Detector"
}

@phdthesis{Karl:2019hes,
      author         = "Karl, Robert",
      title          = "{From the Machine-Detector Interface to Electroweak
                        Precision Measurements at the ILC --- Beam-Gas Background,
                        Beam Polarization and Triple Gauge Couplings}",
      school         = "Hamburg U.",
      year           = "2019",
      address        = "Hamburg",
      doi            = "10.3204/PUBDB-2019-03013",
      reportNumber   = "DESY-THESIS-2019-018",
      SLACcitation   = "%%CITATION = DESY-THESIS-2019-018;%%"
}

@article{CBDA1,
  title={Characterising {WIMPs} at a future $e^+ e^-$ linear collider},
  author={Bartels, Christoph and Berggren, Mikael and List, Jenny},
  journal={EPJ C},
  volume={72},
  number={11},
  pages={2213},
  year={2012},
  publisher={Springer},
  url={https://doi.org/10.1140/epjc/s10052-012-2213-9}
}

@article{TDRvol1,
  title={The {International Linear Collider} Technical Design Report - Volume 1: Executive Summary},
  author={Behnke, Ties and Brau, James E and Foster, Brian and others},
  journal={arXiv:1306.6327},
  year={2013}
}

@article{TDRvol2,
  title={The {International Linear Collider} technical design report - volume 2: Physics},
  author={Baer, Howard and Barklow, Tim and Fujii, Keisuke and others},
  journal={arXiv:1306.6352},
  year={2013}
}

@article{TDRvol3I,
  title={The {International Linear Collider Technical} Design Report - Volume 3. {I}: Accelerator in the Technical Design Phase},
  author={Adolphsen, Chris and Barone, Maura and Barish, Barry and others},
  year={2013},
  journal={arXiv:1306.6353},
}

@article{TDRvol3II,
  title={The {International Linear Collider} Technical Design Report - Volume 3. {II}: Accelerator Baseline Design},
  author={Adolphsen, Chris and Barone, Maura and Barish, Barry and others},
  journal={arXiv:1306.6328},
  year={2013}
}

@article{TDRvol4,
  title={The {International Linear Collider} Technical Design Report - Volume 4: Detectors},
  author={Behnke, Ties and Brau, James and Burrows, Philip and others},
  year={2013},
  journal={arXiv:1306.6329}
}

@article{Kilian:2007gr,
  title={{WHIZARD} - simulating multi-particle processes at {LHC} and {ILC}},
  author={Kilian, Wolfgang and Ohl, Thorsten and Reuter, J{\"u}rgen},
  journal={EPJ C-Particles and Fields},
  volume={71},
  number={9},
  pages={1--29},
  year={2011},
  publisher={Springer},
  url={https://doi.org/10.1140/epjc/s10052-011-1742-y}
}

@article{Moretti:2001zz,
  title={O'{Mega}: An Optimizing matrix element generator},
  author={Moretti, Mauro and Ohl, Thorsten and Reuter, J{\"u}rgen},
  journal={arXiv preprint hep-ph/0102195},
  year={2001}
}

@misc{circe2,
 title={Circe2 manual: $\kappa\iota\rho\kappa\eta$ Version 2.0: Beam Spectra for Simulating Linear Collider Physics},
 author={Ohl, Thorsten},
 url={https://whizard.hepforge.org/circe2.pdf}
}

@inproceedings{Mokka,
  title={Detector simulation with {MOKKA/GEANT4}: Present and future},
  author={De Freitas, P Mora and Videau, Henri},
  booktitle={International Workshop on Linear Colliders (LCWS 2002), Jeju Island, Korea},
  pages={26--30},
  year={2002}
}

@misc{Marlin,
 title={{MARLIN} - {M}odular {A}nalysis and {R}econstruction for the {LIN}ear collider},
 url={http://ilcsoft.desy.de/portal/software_packages/marlin/index_eng.html}
}

@article{Pandora,
  title={Particle flow calorimetry and the {PandoraPFA} algorithm},
  author={Thomson, MA},
  journal={Nucl. Instrum. Methods Phys. Res. Section A: Accelerators, Spectrometers, Detectors and Associated Equipment},
  volume={611},
  number={1},
  pages={25--40},
  year={2009},
  publisher={Elsevier},
  url={https://doi.org/10.1016/j.nima.2009.09.009}
}

@article{BorouXu,
  title={Improvement of photon reconstruction in {PandoraPFA}},
  author={Xu, Boruo},
  journal={arXiv:1603.00013},
  year={2016}
}

@article{OldBeamCalReco,
  title={Forward instrumentation for {ILC} detectors},
  author={Abramowicz, Halina and Abusleme, Angel and Afanaciev, Konstantin and others},
  journal={Journal of Instrumentation},
  volume={5},
  number={12},
  pages={P12002},
  year={2010},
  url={http://stacks.iop.org/1748-0221/5/i=12/a=P12002},
  publisher={IOP Publishing}
}

@article{perelstein,
  title={Dark matter search at a linear collider: effective operator approach},
  author={Chae, Yoonseok John and Perelstein, Maxim},
  journal={Journal of High Energy Physics},
  volume={2013},
  number={5},
  pages={138},
  year={2013},
  publisher={Springer},
  eprint={1211.4008}
}

@article{Barklow:2015tja,
  title={{ILC} operating scenarios},
  author={Barklow, T and Brau, J and Fujii, K and others},
  journal={arXiv:1506.07830},
  year={2015}
}

@article{Matsumoto:2016hbs,
  title={Effective theory of {WIMP} dark matter supplemented by simplified models: singlet-like Majorana fermion case},
  author={Matsumoto, Shigeki and Mukhopadhyay, Satyanarayan and Tsai, Yue-Lin Sming},
  journal={Physical Review D},
  volume={94},
  number={6},
  pages={065034},
  year={2016},
  doi = {10.1103/PhysRevD.94.065034},
  publisher={APS}
}

@article{ecalres,
  title={Response of the {CALICE Si-W} electromagnetic calorimeter physics prototype to electrons},
  author={Adloff, C and Karyotakis, Y and Repond, J and others},
  journal={Nucl. Instrum. Methods Phys. Res. Section A: Accelerators, Spectrometers, Detectors and Associated Equipment},
  volume={608},
  number={3},
  pages={372--383},
  year={2009},
  publisher={Elsevier},
  url={https://doi.org/10.1016/j.nima.2009.07.026}
}

@article{effoplep,
  title={{LEP} shines light on dark matter},
  author={Fox, Patrick J and Harnik, Roni and Kopp, Joachim and Tsai, Yuhsin},
  journal={Physical Review D},
  volume={84},
  number={1},
  pages={014028},
  year={2011},
  doi = {10.1103/PhysRevD.84.014028},
  publisher={APS}
}

@article{schulte,
      author        ={Schulte, Daniel},
      title         = {Beam-Beam Simulations with {Guinea-Pig}},
      month         ={Mar},
      year          = {1999},
      reportNumber  = {CERN-PS-99-014-LP},
      url           = {http://cds.cern.ch/record/382453},
}

@article{grah,
  title={Beam parameter determination using beamstrahlung photons and incoherent pairs},
  author={Grah, Ch and Sapronov, A},
  journal={Journal of Instrumentation},
  volume={3},
  number={10},
  pages={P10004},
  year={2008},
  publisher={IOP Publishing},
  url={https://doi.org/10.1088/1748-0221/3/10/P10004}
}

@article{Fujii:2017vwa,
      author         = "Fujii, Keisuke and others",
      title          = "{Physics Case for the 250 GeV Stage of the International
                        Linear Collider}",
      year           = "2017",
      eprint         = "1710.07621",
      archivePrefix  = "arXiv",
      primaryClass   = "hep-ex",
      reportNumber   = "DESY-17-155, KEK-PREPRINT-2017-31, LAL-17-059,
                        SLAC-PUB-17161",
      SLACcitation   = "%%CITATION = ARXIV:1710.07621;%%"
}

@techreport{antiDID,
  title={{IR} optimization, {DID} and anti-{DID}},
  author={Seryi, Andrei and Maruyama, Takashi and Parker, Brett and others},
  year={2006},
  institution={Stanford Linear Accelerator Center (SLAC)},
  url={https://www.osti.gov/biblio/876041}
}

@article{newL*,
  title={The design of the {ILD} forward region},
  author={Levy, Aharon},
  journal={arXiv:1701.01923},
  year={2017}
}

@misc{ilcsoft,
 title={{ILCSoft} web page},
 url={http://ilcsoft.desy.de/portal}
}

@article{geant4_1,
  title={Geant4 developments and applications},
  author={Allison, John and Amako, Katsuya and Apostolakis, Jea and others},
  journal={IEEE Transactions on Nuclear Science},
  volume={53},
  number={1},
  pages={270--278},
  year={2006},
  publisher={IEEE},
  doi={10.1109/TNS.2006.869826}
}

@article{geant4_2,
  title={Recent developments in {Geant4}},
  author={Allison, J and Amako, K and Apostolakis, J and others},
  journal={Nucl. Instrum. Methods Phys. Res. Section A: Accelerators, Spectrometers, Detectors and Associated Equipment},
  volume={835},
  pages={186--225},
  year={2016},
  publisher={Elsevier},
  url={https://doi.org/10.1016/j.nima.2016.06.125}
}

@article{geant4_3,
  title={GEANT4 - a simulation toolkit},
  author={Agostinelli, Sea and Allison, John and Amako, K al and others},
  journal={Nucl. Instrum. Methods Phys. Res. section A: Accelerators, Spectrometers, Detectors and Associated Equipment},
  volume={506},
  number={3},
  pages={250--303},
  year={2003},
  publisher={Elsevier},
  url={https://doi.org/10.1016/S0168-9002(03)01368-8}
}

@article{BCal_arxiv,
  title={High Energy Electron Reconstruction in the {BeamCal}},
  author={Sailer, Andre and Sapronov, Andrey},
  journal={arXiv:1702.06945},
  year={2017}
}

@article{bhwide,
title = "{BHWIDE} 1.00: O($\alpha$) {YFS} exponentiated{ Monte Carlo for Bhabha} scattering at wide angles for {LEP1/SLC and LEP2}",
journal = "Physics Letters B",
volume = "390",
number = "1",
pages = "298 - 308",
year = "1997",
issn = "0370-2693",
doi = "https://doi.org/10.1016/S0370-2693(96)01382-2",
author = "S. Jadach and W. P\l{}aczek and B.F.L. Ward"
}

@article{malik2015interplay,
  title={Interplay and characterization of dark matter searches at colliders and in direct detection experiments},
  author={Malik, Sarah A and McCabe, Christopher and Araujo, Henrique and others},
  journal={Physics of the Dark Universe},
  volume={9},
  pages={51--58},
  year={2015},
  publisher={Elsevier},
  url={https://doi.org/10.1016/j.dark.2015.03.003}
}

@article{busoni2014validity,
  title={On the validity of the effective field theory for dark matter searches at the {LHC}},
  author={Busoni, Giorgio and De Simone, Andrea and Morgante, Enrico and Riotto, Antonio},
  journal={Physics Letters B},
  volume={728},
  pages={412--421},
  year={2014},
  publisher={Elsevier},
  url={https://doi.org/10.1016/j.physletb.2013.11.069}
}

@article{alwall2009simplified,
  title={Simplified models for a first characterization of new physics at the {LHC}},
  author={Alwall, Johan and Schuster, Philip C and Toro, Natalia},
  journal={Physical Review D},
  volume={79},
  number={7},
  pages={075020},
  year={2009},
  publisher={APS},
  doi = {10.1103/PhysRevD.79.075020},
}

@article{dreiner2013illuminating,
  title={Illuminating dark matter at the {ILC}},
  author={Dreiner, Herbert K and Huck, Moritz and Kr{\"a}mer, Michael and others},
  journal={Physical Review D},
  volume={87},
  number={7},
  pages={075015},
  year={2013},
  publisher={APS},
  doi = {10.1103/PhysRevD.87.075015}
}

@article{beltran2010maverick,
  title={Maverick dark matter at colliders},
  author={Beltran, Maria and Hooper, Dan and Kolb, Edward W and others},
  journal={Journal of High Energy Physics},
  volume={2010},
  number={9},
  pages={37},
  year={2010},
  publisher={Springer},
  url={https://doi.org/10.1007/JHEP09(2010)037}
}

@article{bai2010tevatron,
  title={The {Tevatron} at the frontier of dark matter direct detection},
  author={Bai, Yang and Fox, Patrick J and Harnik, Roni},
  journal={Journal of High Energy Physics},
  volume={2010},
  number={12},
  pages={48},
  year={2010},
  publisher={Springer},
  url={https://doi.org/10.1007/JHEP12(2010)048}
}

@article{goodman2010constraints,
  title={Constraints on dark matter from colliders},
  author={Goodman, Jessica and Ibe, Masahiro and Rajaraman, Arvind and others},
  journal={Physical Review D},
  volume={82},
  number={11},
  pages={116010},
  year={2010},
  doi = {10.1103/PhysRevD.82.116010},
  publisher={APS}
}

@article{zheng2012constraining,
  title={Constraining the interaction strength between dark matter and visible matter: I. fermionic dark matter},
  author={Zheng, Jia-Ming and Yu, Zhao-Huan and Shao, Jun-Wen and others},
  journal={Nuclear Physics B},
  volume={854},
  number={2},
  pages={350--374},
  year={2012},
  publisher={Elsevier},
  url={https://doi.org/10.1016/j.nuclphysb.2011.09.009}
}

@article{yu2012constraining,
  title={Constraining the interaction strength between dark matter and visible matter: {II}. scalar, vector and spin-3/2 dark matter},
  author={Yu, Zhao-Huan and Zheng, Jia-Ming and Bi, Xiao-Jun and others},
  journal={Nuclear Physics B},
  volume={860},
  number={1},
  pages={115--151},
  year={2012},
  publisher={Elsevier},
  url={https://doi.org/10.1016/j.nuclphysb.2012.02.016}
}

@article{tsyslimit_hera,
  title={Search for leptoquark bosons in ep collisions at {HERA}},
  author={Aktas, Adil and others},
  collaboration={H1},
  journal={Physics Letters B},
  volume={629},
  number={1},
  pages={9--19},
  year={2005},
  publisher={Elsevier},
  url={https://doi.org/10.1016/j.physletb.2005.09.048}
}

@article{kahlhoefer2017review,
  title={Review of {LHC} dark matter searches},
  author={Kahlhoefer, Felix},
  journal={International Journal of Modern Physics A},
  volume={32},
  number={13},
  pages={1730006},
  year={2017},
  DOI={10.1142/S0217751X1730006X},
  publisher={World Scientific}
}

@article{Bock:2004xz,
      author         = "Bock, P.",
      title          = "{Computation of confidence levels for exclusion or
                        discovery of a signal with the method of fractional event
                        counting}",
      journal        = "JHEP",
      volume         = "01",
      year           = "2007",
      pages          = "080",
      doi            = "10.1088/1126-6708/2007/01/080",
      eprint         = "hep-ex/0405072",
      archivePrefix  = "arXiv",
      primaryClass   = "hep-ex",
      SLACcitation   = "%%CITATION = HEP-EX/0405072;%%"
}

@article{karl2017polarimetry,
  title={Polarimetry at the {ILC}},
  author={Karl, Robert and List, Jenny},
  journal={arXiv:1703.00214},
  year={2017}
}

@techreport{read2000modified,
  title={Modified frequentist analysis of search results (the {CL}$_{S}$ method)},
  author={Read, Alexander L},
  year={2000},
  institution={Cern},
  url={http://cds.cern.ch/record/451614/files/p81.pdf}
}

@article{poss2014luminosity,
  title={Luminosity spectrum reconstruction at linear colliders},
  author={Poss, St{\'e}phane and Sailer, Andr{\'e}},
  journal={EPJ C},
  volume={74},
  number={4},
  pages={2833},
  year={2014},
  publisher={Springer},
  url={https://doi.org/10.1140/epjc/s10052-014-2833-3}
}

@article{jelisavvcic2013luminosity,
  title={Luminosity measurement at {ILC}},
  author={Jelisav{\v{c}}i{\'c}, I Bo{\v{z}}ovi{\'c} and Luki{\'c}, S and Dumbelovi{\'c}, G Milutinovi{\'c} and others},
  journal={Journal of Instrumentation},
  volume={8},
  number={08},
  pages={P08012},
  year={2013},
  publisher={IOP Publishing},
  url={https://doi.org/10.1088/1748-0221/8/08/P08012}
}

@article{junk1999confidence,
  title={Confidence level computation for combining searches with small statistics},
  author={Junk, Thomas},
  journal={Nucl. Instrum. Methods Phys. Res. Section A: Accelerators, Spectrometers, Detectors and Associated Equipment},
  volume={434},
  number={2-3},
  pages={435--443},
  year={1999},
  publisher={Elsevier},
  url={https://doi.org/10.1016/S0168-9002(99)00498-2}
}

@phdthesis{thesis,
 title={Dark Matter at the {International} {Linear} {Collider}},
 author={Habermehl, Moritz},
 year={2018},
 school={University of Hamburg},
 url={https://bib-pubdb1.desy.de/record/417605?}
}

@article{Aghanim:2018eyx,
      author         = "Aghanim, N. and others",
      title          = "{Planck 2018 results. VI. Cosmological parameters}",
      collaboration  = "Planck",
      year           = "2018",
      eprint         = "1807.06209",
      archivePrefix  = "arXiv",
      primaryClass   = "astro-ph.CO",
      SLACcitation   = "%%CITATION = ARXIV:1807.06209;%%"
}

@article{Sirunyan:2018dsf,
      author         = "Sirunyan, Albert M and others",
      title          = "{Search for new physics in final states with a single
                        photon and missing transverse momentum in proton-proton
                        collisions at $\sqrt{s} =$ 13 TeV}",
      collaboration  = "CMS",
      journal        = "JHEP",
      volume         = "02",
      year           = "2019",
      pages          = "074",
      doi            = "10.1007/JHEP02(2019)074",
      eprint         = "1810.00196",
      archivePrefix  = "arXiv",
      primaryClass   = "hep-ex",
      reportNumber   = "CMS-EXO-16-053, CERN-EP-2018-248",
      SLACcitation   = "%%CITATION = ARXIV:1810.00196;%%"
}

@article{Sirunyan:2017jix,
      author         = "Sirunyan, A. M. and others",
      title          = "{Search for new physics in final states with an energetic
                        jet or a hadronically decaying $W$ or $Z$ boson and
                        transverse momentum imbalance at $\sqrt{s}=13\text{
                        }\text{ }\mathrm{TeV}$}",
      collaboration  = "CMS",
      journal        = "Phys. Rev.",
      volume         = "D97",
      year           = "2018",
      number         = "9",
      pages          = "092005",
      doi            = "10.1103/PhysRevD.97.092005",
      eprint         = "1712.02345",
      archivePrefix  = "arXiv",
      primaryClass   = "hep-ex",
      reportNumber   = "CMS-EXO-16-048, CERN-EP-2017-294",
      SLACcitation   = "%%CITATION = ARXIV:1712.02345;%%"
}

@article{Aaboud:2019yqu,
      author         = "Aaboud, Morad and others",
      title          = "{Constraints on mediator-based dark matter and scalar
                        dark energy models using $\sqrt s = 13$ TeV $pp$ collision
                        data collected by the ATLAS detector}",
      collaboration  = "ATLAS",
      journal        = "JHEP",
      volume         = "05",
      year           = "2019",
      pages          = "142",
      doi            = "10.1007/JHEP05(2019)142",
      eprint         = "1903.01400",
      archivePrefix  = "arXiv",
      primaryClass   = "hep-ex",
      reportNumber   = "CERN-EP-2018-334",
      SLACcitation   = "%%CITATION = ARXIV:1903.01400;%%"
}

@article{Fox:2011fx,
      author         = "Fox, Patrick J. and Harnik, Roni and Kopp, Joachim and
                        Tsai, Yuhsin",
      title          = "{LEP Shines Light on Dark Matter}",
      journal        = "Phys. Rev.",
      volume         = "D84",
      year           = "2011",
      pages          = "014028",
      doi            = "10.1103/PhysRevD.84.014028",
      eprint         = "1103.0240",
      archivePrefix  = "arXiv",
      primaryClass   = "hep-ph",
      reportNumber   = "FERMILAB-PUB-11-039-T",
      SLACcitation   = "%%CITATION = ARXIV:1103.0240;%%"
}

@article{Abdallah:2003np,
      author         = "Abdallah, J. and others",
      title          = "{Photon events with missing energy in e+ e- collisions at
                        s**(1/2) = 130-GeV to 209-GeV}",
      collaboration  = "DELPHI",
      journal        = "Eur. Phys. J.",
      volume         = "C38",
      year           = "2005",
      pages          = "395-411",
      doi            = "10.1140/epjc/s2004-02051-8",
      eprint         = "hep-ex/0406019",
      archivePrefix  = "arXiv",
      primaryClass   = "hep-ex",
      reportNumber   = "CERN-EP-2003-093",
      SLACcitation   = "%%CITATION = HEP-EX/0406019;%%"
}

@article{Abdallah:2008aa,
      author         = "Abdallah, J. and others",
      title          = "{Search for one large extra dimension with the DELPHI
                        detector at LEP}",
      collaboration  = "DELPHI",
      journal        = "Eur. Phys. J.",
      volume         = "C60",
      year           = "2009",
      pages          = "17-23",
      doi            = "10.1140/epjc/s10052-009-0874-9",
      eprint         = "0901.4486",
      archivePrefix  = "arXiv",
      primaryClass   = "hep-ex",
      reportNumber   = "CERN-PH-EP-2008-013",
      SLACcitation   = "%%CITATION = ARXIV:0901.4486;%%"
}

@article{Yokoya:1985xx,
      author         = "Yokoya, Kaoru",
      title          = "{Quantum Correction to Beamstrahlung Due to the Finite
                        Number of Photons}",
      journal        = "Nucl. Instrum. Meth.",
      volume         = "A251",
      year           = "1986",
      pages          = "1",
      doi            = "10.1016/0168-9002(86)91144-7",
      reportNumber   = "KEK-PREPRINT-85-53",
      SLACcitation   = "%%CITATION = NUIMA,A251,1;%%"
}

\end{document}